\documentclass[12pt]{article}
\usepackage{amsmath,amssymb,amsfonts}
\usepackage{hyperref}
\usepackage{cancel}
\usepackage[normalem]{ulem}
\usepackage{color}

\numberwithin{equation}{section}

\newcommand{\diff}{\mathrm{d}}
\newcommand{\ii}{\mathrm{i}}

\definecolor{deepblue}{rgb}{0,0,0.65}
\definecolor{deepred}{rgb}{0.7,0,0}
\definecolor{deepgreen}{rgb}{0,0.6,0}

\definecolor{blue-violet}{rgb}{0.54, 0.17, 0.89}
\definecolor{PineGreen}{cmyk}{0.92, 0, 0.59, 0.25}
\definecolor{OliveGreen}{cmyk}{0.64, 0, 0.95, 0.40}
\definecolor{RawSienna}{cmyk}{0, 0.72, 1, 0.45}
\definecolor{Gray}{cmyk}{0, 0, 0, 0.50}
\definecolor{MidnightBlue}{cmyk}{0.98, 0.13, 0, 0.43}
\definecolor{Orange}{cmyk}{0, 0.61, 0.87, 0}
\definecolor{LimeGreen}{cmyk}{0.50, 0, 1, 0}
\definecolor{Green}{cmyk}{1, 0, 1, 0}

\begin{document}

\title{Black holes with topological charges in Chern-Simons AdS$_5$ supergravity}
\author{Laura Andrianopoli $^{1,}$\thanks{laura.andrianopoli@polito.it}\;, Gaston Giribet $^{2,}$\thanks{gaston@df.uba.ar}\;,  \\
Dar\'{\i}o L\'opez D\'{\i}az $^{3,}$\thanks{dario.lopez@pucv.cl}\;  and Olivera Miskovic $^{3,}$\thanks{olivera.miskovic@pucv.cl} \bigskip\\
{\small $^1$ DISAT, Politecnico di Torino, Corso Duca
    degli Abruzzi 24, I-10129 Turin, Italy \&}\\
{\small Istituto Nazionale di  Fisica Nucleare (INFN) Sezione di Torino, Italy \&} \\
{\small Arnold - Regge Center, Via P. Giuria 1, 10125 Torino, Italy} \\
{\small $^2$ Departamento de Física, Universidad de Buenos Aires y  IFIBA-CONICET,}\\
{\small  Ciudad Universitaria, Pabellón 1, 1428 Buenos Aires, Argentina}\\
{\small $^3$ Instituto de F\'\i sica, Pontificia Universidad Cat\'olica de Valpara\'\i so,}\\
{\small Casilla 4059, Valpara\'{\i}so, Chile}\\}
\date{\today}
\maketitle

\begin{abstract}
 We study static black hole solutions with locally spherical horizons coupled to non-Abelian field in $\mathcal{N}=4$ Chern-Simons AdS$_5$ supergravity. They are governed by three parameters associated to the mass, axial torsion and amplitude of the internal soliton, and two ones to the gravitational hair. They describe geometries that can be a global AdS space, naked singularity or a (non-)extremal black hole. We analyze physical properties of two inequivalent asymptotically AdS solutions when the spatial section at radial infinity is  either a 3-sphere or a projective 3-space. An important feature of these 3-parametric solutions is that they possess a topological structure including two $SU(2)$ solitons   that wind nontrivially around the black hole horizon,  as characterized by the Pontryagin index.  In the extremal black hole limit, the solitons' strengths match and a soliton-antisoliton system unwinds. That limit admits both non-BPS and BPS configurations. For the latter, the pure gauge and non-pure gauge solutions preserve $1/2$ and $1/16$ of the original supersymmetries, respectively. In a general case, we compute conserved charges in Hamiltonian formalism, finding many similarities with standard supergravity black holes.
\end{abstract}

\newpage

\tableofcontents

\newpage

\section{Introduction}

Chern-Simons (CS) supergravity \cite{Zanelli:2005sa} provides an excellent theoretical laboratory to investigate many theoretical aspects of gravitational physics, including black holes. In particular, when formulated about five-dimensional anti-de Sitter (AdS$_5$) space, the theory gives a tractable model to study the effects of introducing higher-curvature terms in the context of the AdS/CFT correspondence. On the one hand, in contrast to other theories involving higher-curvature terms, CS theory is of second order. On the other hand, lacking of local degrees of freedom when linearizing about AdS$_5$, it happens to evades the standard causality bound arguments coming from AdS/CFT \cite{Brigante:2008gz}, cf. \cite{Brigante:2007nu, Hofman:2008ar, Hofman:2009ug}.

The theory is defined by considering the five-dimensional CS action for a supergroup that includes the four-dimensional conformal group $SO(2,4)$ as a part of its bosonic components. A simple example of such a supergroup is given by $SU(2,2|\mathcal{N})$, cf. \cite{Troncoso:1998ng}. This construction yields a gauge field theory that, when expressed in the appropriate variables, takes the form of a higher-curvature theory of gravity coupled to additional fields. The latter fields represent the matter content of the theory, and they typically consist in fermions and gauge fields that are associated to the other fermionic and bosonic generators of the supergroup, respectively. In spite of containing quadratic-curvature terms in the gravitational part of the action, the field equations of the theory are of second order, which follows from the fact that CS gravity is a particular case of Lovelock theory. This makes the theory free of ghosts and tractable analytically. In fact, many sectors of the solution space of the CS (super-)gravity theory can be explicitly worked out. This includes black holes \cite{Boulware:1985wk, Banados:1993ur, Zegers:2005vx, Deser:2005gr}, wormholes \cite{Dotti:2007az}, geometries with anisotropic scale invariance \cite{Adams:2008zk, AyonBeato:2009nh}, $pp$-waves, and many other interesting geometries \cite{Dotti:2006cp}. The existence of such a diversity in the configuration space is somehow associated to certain degree of degeneracy that the theory exhibits, which relates to its peculiar dynamical structure \cite{Chandia:1998uf, Miskovic:2006ei}.

AdS$_5$ black hole solutions in CS supergravity have been explicitly constructed, for instance, in references \cite{Canfora:2007xs, Giribet:2014hpa, Aros:2002rk,Brihaye:2013vsa}. In particular, in \cite{Giribet:2014hpa}, charged AdS$_5$ black hole solutions were studied for the case of the supergroup $SU(2,2|\mathcal{N})$, which is the minimal supergroup containing AdS$_5 \times U(1)$ configurations. For this model, asymptotically locally AdS$_5$ black hole solutions carrying $U(1)$ charge were explicitly found. It was shown that non-vanishing torsion is necessary for the $U(1)$ to be present, with the coupling of torsion being realized by an operator that resembles that of the universal axion of string theory. Here, we will generalize the construction of \cite{Giribet:2014hpa} by studying spherically symmetric black hole solutions coupled to non-Abelian field in $\mathcal{N}=4$ CS AdS$_5$ supergravity. These solutions are characterized by five different parameters, including the axial torsion coupling. In particular, we will construct BPS black hole configurations, along with other solutions. We will compute the topological charges of the solutions as well as the conserved charges, yielding the black hole mass spectrum.

The field content of CS supergravity differs from that of standard supergravity. In the former, all the fields, including the ones associated with the gravity sectory, namely the vielbein $e^a$, the spin connection $\omega^{ab}$, and the gravitino $\psi_s$, are gauge components of a Lie algebra-valued gauge connection. However, for the sake of simplicity and interpretational purpose, hereafter the term `gauge field' will mainly be reserved for the part of the CS gauge connection associated with an internal symmetry fields: the $SU(\mathcal{N})$ non-Abelian field $\mathcal{A}^{\Lambda}$ and the $U(1)$ gauge field $A$. The latter field, to which we will refer as $U(1)_q$, becomes strongly coupled in the case with $\mathcal{N} = 4$ supersymmetries, and thus we will mainly focus on solutions uncharged under it.

Despite the difference with standard supergravity, CS theory presents many similarities with it. For instance, as in standard supergravity, the Bogomol'nyi–Prasad–Sommerfield (BPS) bound appears as a no-force condition, a sort of balance between gravitational and gauge fields interactions. More precisely, the black holes we will construct here are dressed with a topological structure that can be thought of as the presence of two $SU(2)$ solitons winding nontrivially around the event horizon. This is captured by the Pontryagin index. In the limit in which the black holes become extremal, limit that includes both non-BPS as well as BPS configurations preserving $1/2$ and $1/16$ supersymmetries, the soliton-antisoliton system unwinds. This can be regarded as the no-force phenomenon and it is expressed by the topological charges of the solution. As said, the BPS black holes are {a} particular case of extremal configurations, so presenting vanishing Hawking temperature.

The paper is organized as follows: In Section 2, we review CS AdS$_5$ supergravity, we present our conventions and collect formulae that will be used in the rest of the paper. In Section 3, we present the general ansatz for the black hole solutions we want to construct. We discuss different  fibrations and topologies for the horizon, as well as the variety of torsion field configurations supporting the solutions. In Section 4, we construct black holes with non-Abelian charges, generalizing previous works and, in particular, the solutions studied in \cite{Giribet:2014hpa}. In Section 5, we study BPS solutions. We explicitly solve the Killing spinor equations and show that black holes preserving supersymmetries can be constructed analytically. In Section 6, we discuss the global properties of the solutions and we compute the topological charges associated to them. We compute the black hole mass spectrum and we perform a comparative analysis with other black hole solutions found in the literature. Section 7 contains our conclusions, while a series of appendices contain formulae and details of the calculations.

\section{Chern-Simons AdS supergravity in five dimensions}

Chern-Simons (\textbf{CS}) AdS supergravity in five dimensions is a gauge theory described by the action \cite{Chamseddine:1976bf}
\begin{eqnarray}
I[\mathbf{A}] &=&k \int\limits_{\mathcal{M}}L(\mathbf{A})= \frac{\ii k}{3} \int\limits_{\mathcal{M}}\left\langle \mathbf{AF}^2 -\frac{1}{2}\,\mathbf{A}^3 \mathbf{F}+\frac{1}{10}\,\mathbf{A}^{5}\right\rangle \,,  \notag \\
\diff L &=&\frac{\ii}{3}\,\left\langle \mathbf{F}\wedge \mathbf{F}\wedge \mathbf{F}\right\rangle =\frac 13 \,g_{MNK}\,\mathbf{F}^M \wedge \mathbf{F}^N \wedge  \mathbf{F}^K\,,  \label{L(A)}
\end{eqnarray}
 invariant under  $SU(2,2|\mathcal{N})$, which is the supersymmetric extension  (with $\mathcal{N}$ supersymmetry generators) of the AdS$_{5}$ isometry group $SO(2,4)\simeq SU(2,2)$. The  real number $k$ gives the gravitational constant. More precisely, $k^{-1/3}=\ell_P /\ell $ controls the Planck length $\ell_P$ relative to the AdS radius $\ell $. Hereafter, we omit writing the wedge product $\mathbf{\wedge }$ between the forms for short.

The supertrace is defined through the symmetric invariant tensor (antisymmetric for  fermionic indices) of rank three,
\begin{equation}
g_{MNK}=\ii\left\langle \mathbf{G}_M\mathbf{G}_{N}\mathbf{G}_{K}\right\rangle  \equiv \frac{\ii}{2} \, \mathrm{STr} \left( \left\{  \mathbf{G}_M,\mathbf{G}_{N} \right\} \mathbf{G}_{K} \right) \,,
\end{equation}
with the generators
\begin{equation}
\mathbf{G}_M=\left\{ \mathbf{J}_{AB},\mathbf{G}_{\Lambda },\mathbf{G}_1 ,\mathbf{Q}_{s}^{\alpha },\mathbf{\bar{Q}}_{\alpha }^{s}\right\} \,.
\end{equation}
Here, $ \mathbf{J}_{AB},\mathbf{G}_{\Lambda },\mathbf{G}_1 $ are the $15+\mathcal{N}^2$ generators of the bosonic part of the superalgebra, $SU(2,2)\times SU(\mathcal{N})\times U(1)_q$, $\mathbf{Q}_{s}^{\alpha },\mathbf{\bar{Q}}_{\alpha }^{s}$ being instead the $8\mathcal{N}$ odd generators. The index `$q$' in the Abelian group is to distinguish it from other Abelian subalgebras within $SU(\mathcal{N})$ that will play a role in our discussion. AdS generators are the $SO(2,4)\simeq SU(2,2)$ generators $\mathbf{J}_{AB}=-\mathbf{J}_{BA}$, with$\ A=(a,5)$
($a=0,1,2,3,4$) and $\mathbf{J}_{a5}=\mathbf{P}_{a}$. They can be represented in terms of the gamma matrices with the matrix elements $\left( \Gamma _{a}\right) _{\beta }^{\alpha }$, where $\alpha ,\beta =1,\ldots ,4$ are spinorial indices in the fundamental representation of $SU(2,2)$ as $\mathbf{J}_{AB}=\frac{1}{2}\,\Gamma _{AB}$, with $\Gamma _{a5}=\Gamma _{a}$ and $\Gamma _{ab}=\frac{1}{2}\,\left[ \Gamma _{a},\Gamma _{b}\right] $. On the other hand, the $SU(\mathcal{N})$ generators $\mathbf{G}_{\Lambda }$ can be represented by $\mathcal{N}\times \mathcal{N}$ matrices $\tau
_{\Lambda }$, with the matrix elements $\left( \tau _{\Lambda }\right) _{s}^{u}$, where $s,u=1,\ldots ,\mathcal{N}$. When $\mathcal{N}=1$, the internal generators
are absent. The dimension of the superalgebra $SU(2,2|\mathcal{N})$ is $\mathfrak{D} =\mathcal{N}^2+8\mathcal{N}+15$, and from that $\mathcal{N}^2+15$ generators are bosonic and $8\mathcal{N}$ fermionic. The off-shell number of bosonic and fermionic degrees of freedom coincides only when $\mathcal{N}=3$ or $\mathcal{N}=5$. 
Representation of all generators is given in Appendix \ref{Representation} and the properties of the gamma matrices in Appendix \ref{Gamma}.

The Lie algebra of the anti-Hermitian generators $\mathbf{G}_M$ defines the structure constants
\begin{equation}
[ \mathbf{G}_M,\mathbf{G}_{N}]=f_{MN}^{\ \ \ K}\,\mathbf{G}_{K}\,,\qquad M,N,\ldots =\left[ AB\right] ,\Lambda ,1,\binom{\alpha }{s},\binom{s}{\alpha }\,.
\end{equation}
The non-vanishing (anti)commutators are \cite{Chandia:1998uf}
\begin{eqnarray}
\left[ \mathbf{J}_{AB},\mathbf{J}_{CD}\right] &=&\eta _{AD}\,\mathbf{J}_{BC}-\eta _{BD}\,\mathbf{J}_{AC}-\eta _{AC}\,\mathbf{J}_{BD}+\eta _{BC}\,
\mathbf{J}_{AD}\,,  \notag \\
\left[ \mathbf{G}_{\Lambda _1 },\mathbf{G}_{\Lambda _2 }\right] &=&f_{\Lambda _1 \Lambda _2 }^{\ \ \ \ \ \Lambda _{3}}\,\mathbf{G}_{\Lambda _{3}}\,,  \notag \\
\left\{ \mathbf{Q}_{s}^{\alpha }\mathbf{,\bar{Q}}_{\beta }^{u}\right\} &=&
\frac{1}{4}\,\delta _{s}^{u}\,\left( \Gamma ^{AB}\right) _{\beta }^{\alpha}\,\mathbf{J}_{AB}-\delta _{\beta }^{\alpha }\,\left( \tau ^{\Lambda}\right) _{s}^{u}\,\mathbf{G}_{\Lambda }+\ii \,\delta _{\beta }^{\alpha}\,\delta _{s}^{u}\,\mathbf{G}_1 \,, \label{superA_1}
\end{eqnarray}
and the mixed bosonic-fermionic ones,
\begin{equation}
\begin{array}[b]{llll}
\left[ \mathbf{J}_{AB},\mathbf{Q}_{s}^{\alpha }\right] & =-\dfrac{1}{2} \,\left( \Gamma _{AB}\right) _{\beta }^{\alpha }\,\mathbf{Q}_{s}^{\beta }\,,\qquad & \left[ \mathbf{J}_{AB},\mathbf{\bar{Q}}_{\alpha }^{s}\right] & =\dfrac{1}{2}\,\mathbf{\bar{Q}}_{\beta }^{s}\,\left( \Gamma _{AB}\right)_{\alpha }^{\beta }\,, \medskip\\
\left[ \mathbf{G}_{\Lambda },\mathbf{Q}_{s}^{\alpha }\right] & =\left( \tau_{\Lambda }\right) _{s}^{u}\,\mathbf{Q}_{u}^{\alpha }\,, & \left[ \mathbf{G}_{\Lambda },\mathbf{\bar{Q}}_{\alpha }^{s}\right] & =-\mathbf{\bar{Q}}_{\alpha }^{u}\,\left( \tau _{\Lambda }\right) _{u}^{s}\,, \medskip\\
\left[ \mathbf{G}_1 ,\mathbf{Q}_{s}^{\alpha }\right] & =-\ii q\,\mathbf{Q}_{s}^{\alpha }\,, & \left[ \mathbf{G}_1 ,\mathbf{\bar{Q}}_{\alpha }^{s}\right] & =\ii q\,\mathbf{\bar{Q}}_{\alpha }^{s}\,, 
\end{array}\label{superA_2}
\end{equation}
where $q=\frac{1}{4}-\frac{1}{\mathcal{N}}$ is the $U(1)_q$-charge of fermions. We already see here that the case $\mathcal{N}=4$ is special; when $\mathcal{N}=4$, the $U(1)_q$ generator becomes a central charge of $SU(2,2|4)$. Our conventions related to the supersymmetry generators are
\begin{equation}
\mathbf{Q}=[\mathbf{Q}^{\alpha }]\,,\qquad \mathbf{\bar{Q}}=\mathbf{Q}%
^{\dagger }\Gamma _{0}=[\mathbf{\bar{Q}}_{\alpha }]\,,\qquad \{\Gamma
_{a},\Gamma _{b}\}=2\eta _{ab}\,.
\end{equation}
The metric $\eta _{ab}=diag(-,+,+,+,+)$ is mostly positive, and the $SO(2,4)$ embedding metric  is $\eta_{AB}= diag(-,+,+,+,+,-)$.

The fundamental field is a gauge connection 1-form, namely
\begin{eqnarray}
\mathbf{A} &=&A_{\mu }^M(x)\,\diff x^{\mu }\,\mathbf{G}_M =\frac{1}{2} \,\omega ^{AB}\mathbf{J}_{AB}+\mathcal{A}^{\Lambda }\mathbf{G}_{\Lambda }+\bar{\psi}_{\alpha }^{s}\mathbf{Q}_{s}^{\alpha }-\mathbf{\bar{Q}}_{\alpha}^{s}\psi _{s}^{\alpha }+A\,\mathbf{G}_1   \notag \\
&=&\frac{1}{\ell }\,e^{a}\mathbf{P}_{a}+\frac{1}{2}\,\omega ^{ab}\mathbf{J}_{ab}+\mathcal{A}^{\Lambda }\mathbf{G}_{\Lambda }+\bar{\psi}_{\alpha }^{s}%
\mathbf{Q}_{s}^{\alpha }-\mathbf{\bar{Q}}_{\alpha }^{s}\psi _{s}^{\alpha }+A\,\mathbf{G}_1 \,. 
\end{eqnarray}
In our notation the generators are anti-Hermitean, and therefore the same applies for the gauge field $\mathbf{A}=A^M\mathbf{G}_M$. The fermionic part of the gauge field, $\bar{\psi}\mathbf{Q} -\mathbf{\bar{Q}}\psi $, forms an anti-Hermitean bilinear combination because the fermions satisfy $ \left( \bar{\psi}\mathbf{Q}\right) ^{\dagger }=\mathbf{\bar{Q}}\psi$. 
The corresponding field-strength is
\begin{eqnarray}
\mathbf{F} &=&\diff \mathbf{A+A\wedge A=}\left( \diff A^M+\frac{1}{2}\,f_{NK}^{\ \ \ M}\,A^{N}\wedge A^K\right) \mathbf{G}_M  \notag \\
&=&\frac{1}{\ell }\,T^{a}\mathbf{P}_{a}+\frac{1}{2}\,F^{ab}\mathbf{J}_{ab}+F^{\Lambda }\mathbf{G}_{\Lambda }+\nabla \bar{\psi}^{s}\mathbf{Q}_{s}-\mathbf{\bar{Q}}^{s}\nabla \psi _{s}+F\,\mathbf{G}_1 \,,
\label{fs}\end{eqnarray}
with the components
\begin{equation}
\begin{array}{llll}
F^{ab} & =R^{ab}+\dfrac{1}{\ell^2 }\,e^{a}e^{b}-\dfrac{1}{2}\,\bar{\psi}^{s}\Gamma ^{ab}\psi _{s}\,, \qquad & F & =\diff A+\ii \bar{\psi}^{s}\psi _{s}\,,\medskip \\
F^{a5} & =\dfrac{1}{\ell}\,T^{a}+\dfrac{1}{2}\, \bar{\psi}^{s}\Gamma ^{a}\psi_{s}\,, & F_{s} & =\nabla \psi _{s}\,, \medskip\\
F^{\Lambda } & =\mathcal{F}^{\Lambda }+\bar{\psi}^{s}\left( \tau ^{\Lambda}\right) _{s}^{u}\psi _{u}\,, &  &
\end{array}  \label{F}
\end{equation}
where we defined the Riemann curvature 2-form $R^{ab}=\diff\omega ^{ab}+\omega^{ac}\wedge \omega _c ^{\ b}$, torsion tensor 2-form $T^{a}=\diff e^{a}+\omega _{\ b}^{a}\wedge e^{b}$ and the internal symmetry field strength $\mathcal{F}=\mathcal{F}^{\Lambda }\mathbf{G}_\Lambda=\diff\mathcal{A}+\mathcal{A}\wedge \mathcal{A}$.
The group covariant derivative acts on the spinors as
\begin{equation}
\nabla \psi _{s}=\left( \diff +\frac{1}{4}\,\omega ^{ab}\Gamma _{ab}+\frac{1}{2\ell }\,e^{a}\Gamma _{a}\right) \psi _{s}-\mathcal{A}^{\Lambda}\left( \tau _{\Lambda }\right) _{s}^{u}\,\psi _{u}+\ii qA\,\psi _{s}\,.
\label{spinorD}
\end{equation}

The symmetric invariant tensor $g_{MNK}$ has general form\footnote{Compared to the notation of \cite{Chandia:1998uf}, the invariant tensor is $g_{MNK}=\ii\Delta _{MNK}$.} \cite{Chandia:1998uf}  
\begin{equation}
\begin{array}[b]{llll}
g_{[AB][CD][EF]} & =\dfrac{1}{2}\,\epsilon _{ABCDEF}\,,\qquad  & g_{111} &
=\beta \,, \\
g_{1[AB][CD]} & =\alpha \,\eta _{[AB][CD]}\,, & g_{[AB]\left(
_{u}^{\alpha }\right) \left( _{\beta }^{s}\right) } & =\ii\nu _1 \,\left(
\Gamma _{AB}\right) _{\beta }^{\alpha }\delta _{u}^{s}\,, \\
g_{\Lambda _1 \Lambda _2 \Lambda _{3}} & =\sigma \,\gamma _{\Lambda_1 \Lambda _2 \Lambda _{3}}\,, & g_{\Lambda \left( _{u}^{\alpha }\right)
\left( _{\beta }^{s}\right) } & =\ii\nu _2 \,\delta _{\beta }^{\alpha }\left(
\tau _{\Lambda }\right) _{u}^{s}\,, \\
g_{1\Lambda _1 \Lambda _2 } & =\rho \,\gamma _{\Lambda _1 \Lambda _2 }\,,
& g_{1\left( _{u}^{\alpha }\right) \left( _{\beta }^{s}\right) } & =\lambda
\,\delta _{\beta }^{\alpha }\delta _{u}^{s}\,,
\end{array}
\label{inv tensor}
\end{equation}
where the Cartan-Killing metric of the AdS group is $\eta _{[AB][CD]}=\eta _{AC}\,\eta _{BD}-\eta _{AD}\,\eta _{BC}$ with $\eta _{AB}=\mathrm{diag}(\eta _{ab},-1)$.  The symmetric invariant tensors of $SU(\mathcal{N})$ that appear in the CS 5-form Lagrangian are $\gamma_{\Lambda _1 \Lambda _2 }=\mathrm{Tr}\left( \tau _{\Lambda _1 }\tau_{\Lambda _2 }\right) $ and $\ii\gamma _{\Lambda _1 \Lambda _2 \Lambda_{3}}=\mathrm{Tr}\left( \tau _{\Lambda _1 }\tau _{\Lambda _2 }\tau_{\Lambda _{3}}\right) $. Levi-Civita tensor in six-dimensional embedding space is defined by $\epsilon _{abcde5}=\epsilon _{abcde}$. In terms of the Lorentz indices, the bosonic components of the invariant tensor are
\begin{align}
g_{a[bc][de]}& =\frac{1}{2}\,\epsilon _{abcde}\,,  \notag \\
g_{1[ab][cd]}& ={\alpha }\,\left( \eta _{ac}\eta _{bd}-\eta _{ad}\eta_{bc}\right) \,,  \notag \\
g_{1ab}& =-{\alpha }\,\eta _{ab}\,.
\end{align}
The dimensionless coupling constants $\{\alpha ,\beta ,\nu ,\sigma ,\rho,\lambda \}$ are not independent,  $SU(2,2|\mathcal{N})$ being a simple supergroup. Invariance of the tensor ($D(\mathbf{A})g_{MNK}=0$) establishes the local supersymmetry when \cite{Chandia:1998uf}
\begin{equation}
\begin{array}[b]{llll}
\alpha  & =\dfrac{1}{4}\,, & \lambda  & =-\left( \dfrac{1}{4}+\dfrac{1}{\mathcal{N}} \right) \,,\medskip  \\
\beta  & =\dfrac{1}{4^2}-\dfrac{1}{\mathcal{N}^2} \,, \quad & \nu _1  & =\dfrac{1}{2}\,,\medskip  \\
\rho  & =\dfrac{1}{\mathcal{N}}\,, & \nu _2  & =-1\,.\medskip \\
\sigma  & =1\,, &  &
\end{array}  \label{susy}
\end{equation}
The only independent coupling,  that  defines the level of the CS action, is the gravitational constant $k$. 

On the other hand, the Cartan-Killing metric $g_{MN}=\left\langle \mathbf{G}_M\mathbf{G}_{N}\right\rangle$ of the supergroup is given by eq.~\eqref{CK} in Appendix \ref{Representation}. When $\mathcal{N}\neq 4$, the algebra $\mathfrak{su}(2,2|\mathcal{N})$ is
semi-simple because the Cartan-Killing metric is non-singular. When $\mathcal{N}=4$, the $U(1)$ generator becomes the center (it commutes with other supergenerators, as given by \eqref{superA_1}--\eqref{superA_1}) and the metric is such that $g_{1M}=0$. Thus, the supergroup becomes $\mathfrak{psu}(2,2|\mathcal{N})$. Its bosonic generators satisfy $g_{1MN}=-\frac{1}{4}\,g_{MN}$.

Using the invariant tensor \eqref{inv tensor}, the Lagrangian \eqref{L(A)} can be written in components in a covariant way. Focusing on the bosonic sector ($\psi _{s}=0$),  we get 
\begin{align}
L=L_{\mathrm{G}}(e,\omega)+L_{SU(\mathcal{N})}(\mathcal{A})+L_{U(1)}(A,e,\omega,\mathcal{A})\,,    
\end{align}
where, up to a boundary term, the gravitational Lagrangian has the form
\begin{align}
L_{\mathrm{G}}=\frac{1}{8\ell}\,\epsilon_{abcde}\left(R^{ab}R^{cd}e^{e}+\frac{2}{3\ell^2}\,R^{ab}e^{c}e^{d}e^{e}+\frac{1}{5\ell^4}\,e^{a}e^{b}e^{c}e^{d}e^{e} \right)\,.     
\end{align}
Rewritten in the tensorial notation with $L_{\mathrm{G}}=\diff^5x \sqrt{-g}\,\mathcal{L}_{\mathrm{G}}$, we recognize the gravitational Lagrangian density as the Einstein-Gauss-Bonnet AdS gravity \cite{Boulware:1985wk}
\begin{equation}
\mathcal{L}_{\mathrm{G}}=-\frac{1}{2\ell^3}\,\left[ \rule{0pt}{13pt}R-2\Lambda +\alpha _{\mathrm{GB}}\,(R^{\mu \nu \alpha \beta }R_{\mu \nu \alpha \beta
}-4R^{\mu \nu }R_{\mu \nu }+R^2 )\right] ,
\end{equation}
 with the fixed Gauss-Bonnet coupling constant $\alpha_{\mathrm{GB}} = \ell^2 /4$, known as the CS point. {The CS AdS supergravity can, therefore, be regarded as a supersymmetric extension of the Einstein-Gauss-Bonnet AdS gravity in the CS point.\footnote{{Supersymmetric extension of the Gauss-Bonnet term with $\mathcal{N}=1$ and $\mathcal{N}=2$ in four dimensions  has been constructed in \cite{Andrianopoli:2014aqa}.}}} The cosmological constant $\Lambda=-3/\ell^2<0$  also has different form in terms of the AdS radius compared to General Relativity\footnote{In General Relativity ($\alpha_{\mathrm{GB}} =0$) the cosmological constant in $D$ dimensions is $\Lambda=-\frac{(D-1)(D-2)}{2\ell^2}$, which corresponds to $\Lambda=-6/\ell^2$ when $D=5$.}. Note that the limit $\alpha_{\mathrm{GB}} \to \ell^2 /4$ is not continuous because the CS action has different local symmetries and a number of bulk degrees of freedom than the Einstein-Gauss-Bonnet AdS gravity with a generic coupling $\alpha_{\mathrm{GB}} \neq \ell^2 /4$.
 
The five-dimensional analog of the Newton constant $G_{\mathrm{N}}=\ell^3 /8 \pi k$ is read from the term $R/16\pi G_{\mathrm{N}}$. In terms of the Planck length introduced after eq.~\eqref{L(A)},  we have $G_{\mathrm{N}} \propto \ell_P^3 $. The normalization of the Lagrangian is such that the dimensionless CS level $k$ is proportional to the ratio $\ell^3/G_{\mathrm{N}}$.

The gauge field Lagrangians, on the other hand, are given by 
\begin{eqnarray}
L_{SU(\mathcal{N})}&=&-\frac{\ii}{3}\,\mathrm{Tr}\left( \mathcal{AF}^2 -\frac{1}{2}\,\mathcal{A}^3 \mathcal{F}+\frac{1}{10}\,\mathcal{A}
^{5}\right) \,,  \notag \\
L_{U(1)_q}&=&\frac{\beta }{3}\,AF^2 -\frac{1}{4\ell ^2 }\left( T^{a}T_{a} -\frac{\ell^2 }{2}\,R^{ab}R_{ab}-R^{ab}e_{a}e_{b}\right) A+\frac{1}{\mathcal{N}}\, \mathcal{F}^{\Lambda }\mathcal{F}_{\Lambda }A\,,\label{estaeq}
\end{eqnarray}
where the interaction is included in the $U(1)_q$ part and $\mathcal{A}=\mathcal{A}^\Lambda \tau_\Lambda$.

The CS coupling for the $U(1)$ field $AF^2 $ is $\beta $ given by \eqref{susy}, which vanishes for $\mathcal{N}=4$ and becomes negative for $\mathcal{N}>4$. If, in spite of this, we consider the large $\mathcal{N}$ limit, we find that the non-Abelian coupling $\mathcal{F}^{\Lambda }\mathcal{F}_{\Lambda }A$ vanishes as  $\rho \sim 1/\mathcal{N}$, while the one of $U(1)$ goes as $\beta \sim -1/\mathcal{N}^2$ and it also vanishes. 
Therefore, in the large $\mathcal{N}$ limit of the CS AdS$_5$ supergravity, the bosonic part of the action contains the gravitational and $SU(\mathcal{N})$ kinetic terms and the gravitational coupling with $U(1)_q$ field. Gravitational dynamics of this theory has been discussed in \cite{Giribet:2014hpa}. In addition,
the whole supersymmetric part of the action remains finite in this limit, both the kinetic term of the fermions and their interaction with the gauge fields.

On the other hand, in the case $\mathcal{N} \to 4$, as mentioned before, the coefficient $\beta $, which appears in the first term of the Lagrangian $L_{U(1)_q}$ in (\ref{estaeq}), vanishes. This can, thus, be regarded as a point where the ${U(1)_q}$ field becomes strongly coupled. This will be relevant for our discussion of BPS configurations.

The field equations have the form
\begin{equation}
\delta A^M:\qquad \mathcal{E}_M=g_{MNK}\,F^{N}F^K=0\,,
\end{equation}
written here as $4$-forms. The bosonic part of these equations in components reads 
\begin{eqnarray}
\delta e^{a} &:&\qquad \mathcal{E}_{a}=\frac 18\,\epsilon
_{abcde}\,F^{bc}F^{de}-\frac{1 }{2\ell }\,T_{a}F\,,  \notag \\
\delta \omega ^{ab} &:&\qquad \mathcal{E}_{ab}=\frac{1}{2\ell }\,\epsilon
_{abcde}\,F^{cd}T^{e}+\frac 12 \,F_{ab}F\,,  \notag \\
\delta A &:&\qquad \mathcal{E}=\frac{1 }{8}\,\left( F^{ab}F_{ab}-\frac{2}{\ell ^2 }\,T^{a}T_{a}\right) +\beta \,FF+\frac{1}{\mathcal{N}}\, \mathcal{F}^{\Lambda }\mathcal{F}_{\Lambda }\,,  \notag \\
\delta \mathcal{A}^{\Lambda } &:&\qquad \mathcal{E}_{\Lambda } =\frac{2}{\mathcal{N}} \,\mathcal{F}_{\Lambda }F+\gamma _{\Lambda \Lambda _1 \Lambda _2 }\mathcal{F}^{\Lambda _1 }\mathcal{F}^{\Lambda _2 }\,.  \label{EOM}
\end{eqnarray}
The first set of the above equations are the Einstein field equations augmented with higher-curvature modifications and a coupling to torsion tensor. $F^{ab}$ in those equations correspond to the so-called AdS-curvature $2$-form (see \eqref{F}), which vanishes for solutions that have constant curvature locally.   

\section{Static spherically symmetric black hole ansatz}

We are interested in static, spherically symmetric black hole solutions of the supergravity theory described above. Consider the local coordinates $x^{\mu }=(t,r,y^m)$ on the spacetime manifold $\mathcal{M}\simeq \mathbb{R}\times \Gamma$, such that $\Gamma$ corresponds to the four-dimensional spatial manifold $t=const$ with coordinates $(r,y^m)$, and $\Sigma $ defines the three-dimensional transversal section $t,r=const$ with local coordinates $y^m$. The corresponding coordinates on the tangent bundle will be labeled by $a=(0,1,i)$, where $i=2,3,4$.

The most general static, spherically symmetric  ansatz is given by the metric 
\begin{equation}
\diff s^2 =-N^2(r) f^2 (r)\,\diff t^2 +\frac{\diff r^2 }{f^2 (r)}+r^2  \diff \Omega ^2 \,,  \label{metric}
\end{equation}
where $f(r)$ is the metric function whose positive roots define the horizons location, and $N(r)$ is the lapse function. We choose the transversal line element $r\diff\Omega $ such that $\Omega $ describes a three-dimensional space of constant curvature locally,
\begin{equation}
\diff \Omega ^2 =\gamma _{mn}(y)\,\diff y^m\diff y^{n}=\eta _{ij}\, \tilde{e}^{i}\tilde{e}^j\,,\qquad \gamma _{mn}=\eta _{ij}\,\tilde{e}_m^{i}\tilde{e}_{n}^j\,.  \label{Omega}
\end{equation}
The difference between the submanifold $\Sigma$, with the induced metric $r^2\gamma_{mn}(y)$, and the submanifold $\Omega$ endowed with the (intrinsic) metric $\gamma_{mn}(y)$, is the radial fibration. When $\Omega$ is maximally symmetric, its intrisic radius can be rescaled to $1$, $-1$ or $0$. Here, we allow non maximally symmetric constant curvature configurations with more than one scale factor, and not all the radii can be rescaled to the unit size at the same time. This is why we keep the size of $\Omega$ arbitrary in general.  
Possible geometries of the $\Sigma$ section discussed here have topologies with  $\Omega=\mathcal{M}^1 \times
\mathcal{M}^1 \times \mathcal{M}^1 $, $\mathcal{M}^1 \times \mathcal{M}^2 \ $ or $\mathcal{M}^3 $, where each of maximally symmetric spaces $\mathcal{M}^{n}\in \{\mathbb{R}^{n},\mathbb{S}^{n},\mathbb{H}^{n}\}$ can be locally flat, spherical or hyperbolic, i.e., have zero, positive or negative intrinsic curvature constructed from the metric of the submanifold $\mathcal{M}^{n}$. For example, we can consider $\Sigma $ to be a compact space $\mathbb{H}^3/\Upsilon $, with $\Upsilon$  being a discrete subgroup of the isometries of the $3$-hyperbolic space. On $\Omega$, the curvature $\tilde{R}^{ij}=\frac{1}{2}\, \tilde{R}_{mn}^{ij}\,\diff y^m\wedge \diff y^{n}$ is associated to the metric $\gamma _{mn}$. Note that in general the transverse space is not maximally symmetric. For example, $\mathbb{S}^1 \times \mathbb{S}^2 $ geometry breaks the maximal isometry group $SO(4)$ of $\Sigma$ to $SO(2)\times SO(3)$.

The five independent spherically symmetric components of the torsion tensor are\footnote{In order to have static spherically symmetric geometry, we impose the ansatz only on the metric and torsion (field strength associated to $\mathbf{P}_{a}$), and not on the gauge fields. We will see that this is sufficient to obtain BPS solutions.}
\begin{equation}
T_{ttr}(r)=0\,,\ T_{rtr}(r)=0\,, \ T_{ntm}=\tilde{\chi}(r)\gamma _{nm}=0\,,\  T_{nrm}=\chi (r)\gamma _{nm},\ T_{nmk}=\phi (r)\, \tilde{\epsilon}_{nmk}\,. \label{torsion}
\end{equation}
We restrict the torsion tensor only to the cases where the first three components vanish because other solutions involving these components have been discussed in \cite{Giribet:2014hpa}.
 In addition, when $\phi = 0$ and some of other torsional components are turned on, the Schwarzschild-like black hole solution was found in \cite{Aros:2006qc}. In our case, we will keep $\phi$ non-vanishing, as it will turn out that this is an important field for construction of BPS states for the black holes.
Therefore, for the solutions considered here at most the trace $\chi $ and the axial torsion $\phi $ could contribute to the dynamics of the system.

On the transversal section $\Sigma$, $\epsilon _{nmk}$ is a constant Levi-Civita tensor and $\tilde{\epsilon}_{nmk}$ is a $\Sigma$-covariant Levi-Civita tensor,  
\begin{equation}
\tilde{\epsilon}_{nmk}=\sqrt{\gamma }\,\epsilon _{nmk}\,,\qquad \epsilon_{nmk}\equiv \epsilon _{nmktr}\,.
\end{equation}
This summarizes our 6D, 5D and 3D Levi-Civita conventions (all  with mostly plus signature conventions) as
\begin{eqnarray}
\epsilon _{abcde5} &=&\epsilon _{abcde}\,,\qquad \epsilon _{01ijk5}=\epsilon _{01ijk}=\epsilon _{ijk}\,,  \notag \\
\epsilon ^{abcde5} &=&-\epsilon ^{abcde}\,,\qquad \epsilon^{01ijk5}=-\epsilon ^{01ijk}=\epsilon ^{ijk}\,, \\ \label{LeviCivita}
\epsilon _{012345} &=&\epsilon ^{012345}=+1\,.  \notag
\end{eqnarray}

As regarding $U(1)_q \times SU(\mathcal{N})$ fields, their form is determined by equations of motion. Namely, the BPS states are solitonic solutions, which are extended objects, and therefore they do not necessarily possess spherical symmetry. However, for simplicity, we will consider only $U(1)_q$ field components  
\begin{equation}
A_{t}(r)\,,\qquad A_r(r)=0\,,
\end{equation}
which are the ones that are compatible with the metric isometries.
The radial component is set to zero because it can always be written as total differential by redefinition of the radial coordinate. The above ansatz is suitable when the spacetime topology  does not contain a $\mathbb{S}^1$ factor which could, in principle, allow for a non-trivial winding of the $U(1)_q$ field around the circle.\medskip

Let us single out three types of transversal sections \eqref{Omega}:\medskip

\quad (\textit{i}) Planar geometry allows the topologies $\mathbb{S}%
^1 \times \mathbb{S}^1 \times \mathbb{S}^1 $, $\mathbb{R}^2 \times
\mathbb{S}^1 $ and $\mathbb{R}^3 $, and is given by the flat metric
\begin{equation}
\gamma _{nm}=\delta _{mn}\,,\qquad \tilde{R}^{ij}=0\,,\qquad \kappa =0\,.
\label{Omega-planar}
\end{equation}

\quad (\textit{ii}) The 3-sphere $\mathbb{S}^3 $ of  radius $a$ has constant positive curvature 
\begin{equation}
\tilde{R}^{ij}=\kappa \,{\tilde{e}}^{i}\wedge {\tilde{e}}^j\,,\qquad
\kappa =\frac{1}{a^2 }>0\,,  \label{S3}
\end{equation}
and $\gamma _{nm}$ depends on three angular variables. Changing the topology to the  hyperbolic one can be done by setting $\kappa <0$. The planar case ($\kappa=0$) has been included in the previous point \eqref{Omega-planar}.

\quad (\textit{iii}) Non maximally symmetric transversal section is $\mathcal{M}^1 \times \mathcal{M}^2 $. For example, if the boundary topology is $\mathbb{S}^1 \times \mathbb{S}^2 $, the maximal number (six) of isometries is broken to four isometries. The coordinates can be chosen as $y^m=(z,\theta ,\varphi )$ and \eqref{Omega} is given by
\begin{equation}
\diff \Omega ^2 =a_1\,\diff z^2 +a_2\,\left( \diff \theta
^2 +\sin ^2 \theta \,\diff \varphi ^2 \right) \,.
\label{Omega-2sphere}
\end{equation}
Here, the $z\simeq z+2\pi $ direction is compactified on the circle $\mathbb{S}^1 $ of  radius $a_1$, and $a_2$ is the radius of the sphere $\mathbb{S}^2 $, the curvature has only one non-vanishing component $\tilde{R}_{\theta\varphi }^{\theta \varphi }=1/a_2^2$. 

It is worth mentioning that the classification above refers to local properties of the $3$-spaces, and that locally equivalent spaces that are constructed from the ones above through identifications are also included.  

Under these assumptions, and considering here only the cases (\textit{i})--(\textit{ii}) (the case (\textit{iii}) will be discussed elsewhere), we can write out the following vielbein and spin connection
1-forms,
\begin{equation}
\begin{array}{lll}
e^{0}=Nf\,\diff t\,,  & e^1 =\dfrac{\diff r}{f}\,,\qquad  &
e^{i}=r\,\tilde{e}^{i}\,,\medskip  \\
\omega ^{01}=f(Nf)'\,\diff t\,, \qquad & \omega ^{1i}=\nu \,\tilde{e}^{i},
& \omega ^{ij}=\tilde{\omega}^{ij}-\varphi \,\epsilon ^{ijk}\,\tilde{e}_{k}\,,
\end{array}
\label{e,w}
\end{equation}
where we introduced the definitions 
\begin{equation}
\nu \equiv \frac{f\left( \chi -r\right) }{r}\,,\qquad \varphi \equiv \frac{\phi }{2r^2 }\,. \label{nu_chi}
\end{equation}
The torsion becomes
\begin{equation}
T^{i}=\frac{f+\nu }{f}\,\diff r\wedge \tilde{e}^{i}+r\varphi \,\epsilon
_{\ jk}^{i}\,\tilde{e}^j\wedge \tilde{e}^k\,,  \label{T}
\end{equation}
and the AdS curvature reads
\begin{equation}
\begin{array}{llll}
F^{01} & =\left[ \dfrac{1}{\ell ^2 }-\left( \rule{0pt}{13pt}f(Nf)'\right)'\right] \,\diff t\wedge \diff r\,,\; \; & F^{1i} & =\left( \nu'+\dfrac{r}{\ell ^2 f}\right) \diff r\wedge \tilde{e}^{i}+\varphi \nu \,\epsilon _{\ jk}^{i}\, \tilde{e}^j\wedge \tilde{e}^k\,,
\\
F^{0i} & =f\left[ \nu (Nf)' +\dfrac{r}{\ell ^2 }\right] \,\diff t\wedge \tilde{e}^{i}\,, & F^{ij} & =-\varphi' \epsilon _{\ k}^{ij}\,\diff r\wedge \tilde{e}^k+\Xi \,\tilde{e}^{i}\wedge \tilde{e}^j\,.
\end{array}
\label{AdS comp}
\end{equation}
To simplify the expressions, we denoted
\begin{equation}
\Xi(r) \equiv \kappa +\frac{r^2 }{\ell ^2 }-\varphi ^2 -\nu ^2 \,.  \label{Xi}
\end{equation}
The first two equations of motion \eqref{EOM} can be written in components as 
\begin{align}
\mathcal{E}_{0}& =\frac{1}{2}\left[ \Xi \left( \nu'+\frac{r}{\ell ^2 f}\right) -2\nu \varphi \varphi'\right] \epsilon _{ijk}\,\diff r\,\tilde{e}^{i}\,\tilde{e}^j\tilde{e}^k\,,  \notag \\
\mathcal{E}_1 & =-\frac{\Xi \,f}{2}\,\left[ \nu (Nf)'+\frac{r}{\ell^2}\right] \,\epsilon _{ijk}\,\diff t\,\tilde{e}^{i}\tilde{e}^j\tilde{e}^k\,,  \notag \\
\mathcal{E}_i& =\left[ \frac{\Xi }{2}\left( \frac{1}{\ell ^2 }-\left(\rule{0pt}{13pt} f(Nf)'\right)'\right)
+f\left( \nu (Nf)'+\frac{r}{\ell ^2 }\right) \left(
\nu'+\frac{r}{\ell ^2 f}\right) -\frac{r\varphi F_{tr}}{2\ell }\right] \epsilon _{ijk} \,\diff t\,\diff r\,\tilde{e}^j\tilde{e}^k\,,  \notag \\
\mathcal{E}_{01}& =\frac{1}{2\ell }\,\left( \frac{f+\nu }{f}\,\Xi -2r\varphi \varphi'\right) \epsilon _{ijk}\,\diff r\,\tilde{e}^{i}\tilde{e}^j\tilde{e}^k\,,  \notag \\
\mathcal{E}_{0i}& =0\,,  \notag \\
\mathcal{E}_{1i}& =\left[ -\frac{f+\nu }{\ell }\left( \nu (Nf)'+\frac{r}{\ell ^2 }\right) +\frac 12\, \nu \varphi F_{tr}\right] \epsilon _{ijk}\,\diff t\,\diff r\,\tilde{e}^j\tilde{e}^k\,,  \notag \\
\mathcal{E}_{ij}& =\left[ -\frac{2r\varphi }{\ell }\left( \frac{1}{\ell ^2 }-\left( \rule{0pt}{13pt}f(Nf)'\right)'\right) +\frac 12\, F_{tr}\,\Xi \right] \diff t\,\diff r\,\tilde{e}_i\tilde{e}_j\,.  \label{EOM_AdS}
\end{align}
The last two equations in \eqref{EOM} give restrictions to the Abelian $A$ and internal $\mathcal{A}^{\Lambda }$ fields. In particular, while the  $U(1)_q$-gauge field in the static ansatz has vanishing Pontryagin density, $FF=0$, it is possible to have nontrivial contributions  to $\mathcal E$ from the non-Abelian internal symmetry sector and from gravity. Substituting our ansatz \eqref{e,w} for the metric and spin connection, we get for that equation of motion 
\begin{equation}
\mathcal{E}=\frac 12\,\epsilon _{ijk}\left( \varphi\nu \nu'-\frac{ r\varphi}{\ell^2}-\Xi\varphi'\right) \diff r\tilde{e}^{i}\tilde{e}^j\tilde{e}^k+\frac{1}{\mathcal{N}}\, \mathcal{F}^{\Lambda }\mathcal{F}_{\Lambda }\,.
\label{EOM_SU(4)_1}
\end{equation}
The last equation that determines the dynamics of the internal symmetry gauge field, $\mathcal{E}_{\Lambda }=0$, in
components read
\begin{eqnarray}
\diff t  &:&\quad 0=\gamma _{\Lambda \Lambda _1 \Lambda _2 }\mathcal{F}_{rm}^{\Lambda_1 }\mathcal{F}_{nl}^{\Lambda _2 }\,\tilde{\epsilon}^{mnl}\,, \notag  \\
\diff r  &:&\quad 0=\gamma _{\Lambda \Lambda _1 \Lambda _2 }\mathcal{F}_{tm}^{\Lambda_1 }\mathcal{F}_{nl}^{\Lambda _2 }\,\tilde{\epsilon}^{mnl}\,,\notag   \\
\diff y^m  &:&\quad 0=\left[ \frac{1}{\mathcal{N}}\, F_{tr}\,\mathcal{F}_{\Lambda nl}+
\,\gamma _{\Lambda \Lambda _1 \Lambda _2 }\left( \mathcal{F}_{tr}^{\Lambda
_1 }\mathcal{F}_{nl}^{\Lambda _2 }-2\mathcal{F}_{tn}^{\Lambda _1 }
\mathcal{F}_{rl}^{\Lambda _2 }\right) \right] \tilde{\epsilon}^{mnl}\,.
\label{EOM_SU(4)_2}
\end{eqnarray}

The equations \eqref{EOM_AdS}--\eqref{EOM_SU(4)_2} are valid for any $\mathcal{N}$. As already noticed, the case $\mathcal{N}=4$ has vanishing coupling $\beta =0$, but the chosen ansatz is such that $FF=0$, thus the only term involving this coupling, $\beta FF$, vanishes for any $\mathcal{N}$.

As usual in higher-curvature gravity, the solution space of the equations \eqref{EOM_AdS}--\eqref{EOM_SU(4)_2} contains several independent branches with well-defined
geometries (in the sense that the metric does not contain arbitrary functions of the radial coordinate). Our goal is to find one involving the internal non-Abelian fields that, at the same time, admits unbroken supersymmetries. To this end, let us review the known solutions and discuss the existence of the BPS states within them:

\subparagraph{Global AdS space.}

Maximally symmetric, global AdS spacetime, with $F^{AB}=0$, is a particular solution of CS supergravity. It amounts to setting
\begin{equation}
f^2 =\kappa +\frac{r^2 }{\ell ^2 }\,,  \label{global AdS}
\end{equation}
in the metric \eqref{metric}. It can exist with or without the addition of the $U(1)\times SU(\mathcal{N})$ matter fields.

Without matter, the Killing spinors in the global
AdS space have been constructed in \cite{Aros:2002rk} based on the Killing spinor equation within a class of supergravities. They are completely determined by the Killing spinors of the transverse section $\Sigma$, and for maximally symmetric $\Sigma$ they admit the maximum number of supersymmetries. Killing spinors have been classified for complete, connected, irreducible Riemannian manifolds of Euclidean signature in \cite{Wang,Bar:1993gpi,Baum}, and their geometric construction in homogeneous space-times has been performed in \cite{Alonso-Alberca:2002wsh}.
In particular, the spinors on $\Sigma\simeq \mathbb{S}^{n}$ have been constructed in \cite{Lu:1998nu}. 

\subparagraph{Global AdS space with the $U(1)_q\times SU(4)$ fields.}

In \cite{Miskovic:2006ei}, the Killing spinors and the $\frac{1}{16}$-BPS state were constructed in the case of $\mathcal{N}=4$, when the spacetime is globally AdS$_{5}$, such that
\begin{equation}
F^{AB}=0\,,\qquad F\neq 0\,,\qquad \mathcal{F}^{\Lambda }\neq 0\,.
\end{equation}
The space has the planar $\mathbb{S}^1 \times \mathbb{S}^1 \times \mathbb{S}^1 $ topology \eqref{Omega-planar} with $\kappa =0$, such that the isometry group is broken as $SO(4) \to U(1)\times U(1)\times U(1)$, and the $SU(4)$ group is also broken to $U(1)\times U(1)\times U(1)$. Because of $\pi _1 (U(1)) =\mathbb{Z}$, the non-trivial central charge and the BPS state are characterized by three integers, one for each coordinate $y^m\simeq
y^m+2\pi $.

Explicitly, the isomorphism $SU(4)\simeq SO(6)$ enables to represent the internal symmetry generators $\mathbf{G}_{\Lambda }=\left\{ \mathbf{T}_{IJ}\right\}$ with the $4\times 4$ blocks $\tau_\Lambda=\left\{ \tau_{IJ}\right\}$ in terms of an independent set of the gamma matrices $\tau_{IJ}=\frac 12 \, \tilde{\Gamma}_{IJ}$ ($I,J=1,...,6$). The cancellation of different
components of $\mathcal{F}^{IJ}$ is achieved along the Cartan subalgebra  $\mathbf{T}_{12}\times \mathbf{T}_{34}\times \mathbf{T}_{56}$, with two
non-vanishing Abelian components of the field strength (subgroup of $SU(4))$, and also non-vanishing $U(1)_q$ field. 

The Abelian fields appearing in this solution are not spherically symmetric (for example, $F_{rm},F_{mn}\neq 0$) and they describe a soliton. We expect similar feature to hold for the Abelian and internal non-Abelian fields in the case of black holes.

\subparagraph{Dimensionally continued black hole.}

Dimensionally continued black holes are exact solutions of CS AdS gravity in odd dimensions $D\geq 5$ when the torsion vanishes \cite{Banados:1993ur}; the BTZ black hole can be regarded as the case $D=3$. They are characterized by the mass $M$ and, when coupled to the Maxwell field, by the electric charge $Q$. In pure CS gravity, $Q=0$, and
\begin{equation}
f^2 (r)=\frac{r^2 }{\ell ^2 }-M\,,\qquad T^{a}=0\,.  \label{DCBH}
\end{equation}
The spacetime topology is the one of the unit sphere, $\Sigma \simeq \mathbb{S}^{D-2}$. The AdS space corresponds to a state\ of  mass $M=-1$.
The black holes solutions are separated by a gap of naked singularities $-1/a=-1<M<0$ from the black hole solutions with $M\geq 0$. This behavior of the solutions is similar to the one in $D=3$ dimensions (i.e. BTZ black holes), therefore the name `dimensionally continued black holes'.

In sum, the solutions behave in the following way:
\begin{itemize}
\item[ ] \qquad 
\begin{tabular}[t]{ll}
$M>0$ & Black holes \\
$M=0$ & Black hole ground state \\
$-\frac{1}{a^2 }<M<0$ \qquad & Naked singularities \\
$M=-\frac{1}{a^2 }$ & Global AdS space
\end{tabular}
\end{itemize}
In the list above, we introduced the radius $a$ of the sphere $\mathbb{S}^{D-2}$ (compared to \cite{Banados:1993ur}); recall, one can rescale $a=1$. The scalar curvature behaves near $r=0$ as
\begin{equation}
R\simeq \left( \frac{1}{a^2 }+M\right) \frac{\left( D-2\right) \left(
D-3\right) }{r^2 }-\frac{D\left( D-1\right) }{\ell ^2 }+\cdots \,.
\label{singularity}
\end{equation}
We see that the singularity at $r=0$ is absent only in the global AdS space. The black hole ground state with $M=0$ is the smallest black hole and it has the horizon with zero radius. 

Black hole solutions  with zero horizon area (and then vanishing black hole entropy at the supergravity level), are familiar in string theory, where they are called ``small black holes''. They are typically associated to string states and require, for finding the entropy, a quantum attractor mechanism taking into account the presence of higher curvature terms \cite{small1,small2,small3}.

The BPS states exist only for the global AdS space ($M=-1$), as
discussed in the previous paragraph. In particular, the zero mass black hole is not supersymmetric \cite{Aros:2002rk}.

\subparagraph{$U(1)_q$-neutral black hole with axial torsion.}

In absence of torsion, CS AdS gravity is in fact a special case of Lovelock AdS gravity, whose exact spherically symmetric, static solutions are dimensionally continued black holes. They do not admit BPS states if $M\neq -1$. Addition of the axial torsion ($\phi \neq 0$, $\chi =0$) has been discussed in \cite{Canfora:2007xs}. The solution is%
\begin{equation}
f^2 =\frac{r^2 }{\ell ^2 }+br-M\,,\qquad \phi =2Cr^2 \,,\qquad \kappa
=C^2 \,,
\end{equation}
where $b$ is gravitational hair, an integration constant of gravitational origin different than mass, charge or angular momentum. This kind of
constants is known to appear in CS AdS$_{5}$ gravity \cite{Canfora:2007xs,Giribet:2014hpa}, but it can also be found in New Massive Gravity in three dimensions \cite{Oliva:2009ip}, or conformal gravity \cite{Riegert:1984zz} in arbitrary dimension. In \cite{Canfora:2007xs}, the torsion field stabilises the ground state black hole solution. Namely, it becomes a $\frac{1}{2}$-BPS state when $b=0$, for the spacetime with the (locally parallelized) $\mathbb{S}^3 $ topology, such that the transversal section curvature is $\tilde{R}^{ij}=C^2 \,\tilde{e}^{i}\wedge \tilde{e}^j$.

\subparagraph{$U(1)_q$-charged black hole.}

Black hole solutions in the AdS$_{5}\times U(1)_q$ sector with $\mathcal{A}^{\Lambda }=0$ were discussed in \cite{Giribet:2014hpa} for arbitrary
coupling constants (not necessarily supersymmetric). This branch of solutions is completely independent on the previous one, in the sense that the latter is not a special case of the former, although both branches do share common particular sub-cases. In case of the planar geometry \eqref{Omega-planar} (where $\tilde{\omega}^{ab}=0$) and the torsion tensor with $\chi \phi \neq 0$, the general solution reads
\begin{equation}
\begin{array}{ll}
f=\sqrt{\dfrac{r^2 }{\ell ^2 }+br-\mu }\,,\medskip & \varphi =C\,,\qquad
\nu =\varepsilon h\,,\qquad \varepsilon =\pm 1\,, \\
h=\sqrt{\dfrac{r^2 }{\ell ^2 }-C^2 }\,,\quad r\geq \ell C\,,\quad & \chi
=r\left( 1+\dfrac{\varepsilon h}{f}\right) \,. \\
A_{t}=A_{0t}-\dfrac{1}{2C\ell \alpha }\left( \dfrac{r^2 }{\ell ^2 }+\dfrac{br}{2}+\varepsilon fh\right) \,, &
\end{array}
\label{f,h}
\end{equation}
where $A_{0t}$ is an integration constant. Note that $\mu =M$ because $\kappa =0$. The  potential behaves as
\begin{equation}
A_{t}=A_{\infty }-\frac{1+\varepsilon  }{2\alpha C\ell }\left( \frac{r^2 }{\ell ^2 }+\frac{br}{2}\right) +\mathcal{O}\left( \frac{1}{r}\right) \,,  \label{At}
\end{equation}
thus the electromagnetic field becomes divergent in the asymptotic sector, unless $\varepsilon =-1$.

The non-vanishing $U(1)_q\times $AdS$_{5}$ field strength 2-forms \eqref{AdS comp} have the form
\begin{equation}
\begin{array}{llll}
F^{0i} & =\left( \dfrac{r(f-h)}{\ell ^2 }-\dfrac{bh}{2}\right) \,\diff t\wedge \diff y^{i}\,, & F^{1i} & =-\dfrac{r\left( f-h\right) }{\ell ^2 hf}\,\diff r\wedge \diff y^{i}-Ch\,\epsilon _{\ nm}^{i}\,\diff y^{n}\wedge \diff y^m\,, \\
F & =A'_t\,\diff r\wedge \diff t\,, & F^{i5} & =\dfrac{1}{\ell }\,T^{i}=\dfrac{f-h}{\ell f}\,\diff r\wedge \diff y^{i}+\dfrac{r}{\ell }\,C\,\epsilon _{\ nm}^{i}\,\diff y^{n}\wedge \diff y^m\,.
\end{array}
\label{FAdS,FU(1)}
\end{equation}
Note that the AdS curvature is divergent, $F_{rm}^{1i}\to \infty $, on two surfaces: on the horizon $r=r_{+}$ (when $f=0$) and at $r=\ell C$ (when $h=0$). It turns out that these points introduce singularities in the torsional curvature invariants, but not in the Riemann curvature invariants.

This solution does not admit Killing spinors in general.

\section{Black hole solutions with the non-Abelian field}

In this section we will add an $SU(\mathcal{N})$ field to the solution. We consider the case where the transversal space is isomorphic to the 3-sphere. Then equations \eqref{EOM_AdS}--\eqref{EOM_SU(4)_2} have at least two branches of  solutions, distinguished by the vanishing or not of the function 
\begin{equation}
  \Xi(r) = \kappa +\frac{r^2 }{\ell ^2 }-\varphi ^2 -\nu ^2   \,,
\end{equation}
introduced in eq.~\eqref{Xi}
(see, for instance, $\mathcal{E}_1 $). 
Recall that, geometrically, in case of a constant axial torsion $\varphi=C$,  $\Xi$ becomes the AdS field strength on the 3-sphere, as we can see from the last of eqs.~\eqref{AdS comp}.

We will analyze  separately the two cases $\Xi =0$ and $\Xi \neq 0$. From eqs.~\eqref{AdS comp} and \eqref{Xi}, it can be seen that this parameter geometrically represents the `magnetic' components of the AdS field strength caused by presence of non-axial torsion, similarly as $\varphi'$ are `electric' components (in radial foliation) due to existence of the axial torsion. In the case considered in this section, the electric field-like contribution is vanishing, thus $\Xi =0$ and $\Xi \neq 0$ distinguish, respectively, absence and presence of the AdS magnetic field on the spatial boundary of the spacetime.

\subsection{Branch with \texorpdfstring{$\Xi=0$}{Xi =0}}
\label{Xi-neutral}

We assume that $\varphi \neq 0$ because it defines the axial torsion. Then the condition $\Xi =0$ determines the function $\nu $ in terms of $\varphi $ up to a global sign $\varepsilon =\pm 1$ by means of \eqref{Xi}. One can show that taking $\nu =0$ is in contradiction with $\mathcal{E}_{01}$, and $\nu \neq 0$ leads to $\varphi ^{\prime }=0$ in order to satisfy $\mathcal{E}_{0}$, so that the solution is
\begin{equation}
\nu =\varepsilon h=\varepsilon \sqrt{\frac{r^2 }{\ell ^2 }+\kappa -C^2 }\,,\qquad \varphi =C\,, \label{solchi0}
\end{equation}
where $C$ is an integration constant. The remaining independent equations not involving the internal non-Abelian field, $\mathcal{E}_{ij}$ and $\mathcal{E}_i$, become
\begin{equation}
-2\alpha\ell CA'_t=(f+\nu) (Nf+\nu)'\,,\qquad f(Nf)' =\frac{r}{\ell ^2 }+\frac{b}{2}\,.
\label{splitting_point}
\end{equation}
We can distinguish three cases to determine the unknown functions $A_t$, $N$ and $f$.

\subparagraph{1)}

When $A'_t \neq 0$, the solution is
\begin{equation}
A'_t=-\frac{f+\nu }{2\alpha \ell Cf}\left( \frac{f+\nu }{\nu }\frac{r}{\ell ^2 }+\frac{b}{2}\right) \,,\qquad f(Nf)' =\frac{r}{\ell ^2 }+\frac{b}{2}\,. \label{2 of 3}
\end{equation}
These two equations are not enough to solve three radial functions.
In absence of internal non-Abelian fields, $N(r)$ remains arbitrary. In that case, a radial evolution of the spacetime geometry is not determined by initial conditions. Such ``free geometry'' has been observed in  Lovelock gravity \cite{Wheeler:1985qd}, higher-order curvature theory that can have multiply degenerate vacua and accidental symmetries that turn fields into non dynamical ones. Similar feature has been noticed in CS AdS$_5$ supergravity with maximal number of spherically symmetric torsional components \cite{Giribet:2014hpa} and in New Massive Gravity in \cite{Gabadadze:2012xv}. In this work, we will not consider these degenerate geometries.

When internal non-Abelian fields are added, there is an additional  interaction with the metric which could help to determine $N$. The only equation with such interaction is (see last line in \eqref{EOM_SU(4)_2})
\begin{equation}
\rho A'_{t}\,\mathcal{F}_{\Lambda nl}=\sigma \,\gamma _{\Lambda \Lambda _1 \Lambda _2 }\left( 2\mathcal{F}_{tn}^{\Lambda _1 } \mathcal{F}_{rl}^{\Lambda _2 } -\mathcal{F}_{tr}^{\Lambda _1 } \mathcal{F}_{nl}^{\Lambda _2 }\right) \tilde{\epsilon}^{mnl}\,.
\end{equation}
Therefore, to avoid a free geometry, the above equation together with \eqref{2 of 3} has to fully determine $A_t$, $f$ and $N$. We expect that a $U(1)_q$-charged black hole with internal non-Abelian fields would exist only when $\mathcal{N} \neq 4$.

When $\mathcal{N}=4$, the $U(1)_q$ field becomes non dynamical (a central extension in the algebra) and the black hole is $U(1)_q$-neutral.

\subparagraph{2)}

When $A_{t}=0$ and $(Nf+\nu)'=0$, the solution reads
\begin{equation}
A_{t}=0\,, \quad Nf=-\left( \frac{\varepsilon b_{0}\ell }{2}+\nu \right) \,,\quad f=-\left( 1+\frac{b\ell ^2 }{2r}\right) \nu \,,
\end{equation} 
or explicitly 
\begin{eqnarray}
f(r) &=&\left( 1+\frac{b\ell ^2 }{2r}\right) \sqrt{\frac{r^2 }{\ell ^2 }+\kappa -C^2 }\,,  \notag \\
N(r) &=&\frac{\frac{b_{0}\ell }{2}+\sqrt{\frac{r^2 }{\ell ^2 }+\kappa -C^2 }}{\left( 1+\frac{b\ell ^2 }{2r}\right) \sqrt{\frac{r^2 }{\ell ^2 }+\kappa
-C^2 }}\,,  \label{Nf1}
\end{eqnarray}
where $b_{0}$ is another integration constant, conveniently normalized. Thus we have free parameters $\{C,b_{0},b\}$. The black hole horizon is determined by $f(r_h)=0$, $N(r_h)\neq 0$.

The gravitational hair modifies the metric through $N\neq 1$, since it behaves asymptotically as
\begin{equation}
N\sim 1+\frac{\left( b_{0}-b\right) \ell ^2 }{2r}\,.
\end{equation}
The metric describes asymptotically AdS space with gravitational hair,
\begin{eqnarray}
(fN)^2 &\sim &\frac{r^2 }{\ell ^2 }+b_{0}r+\kappa -C^2 +\frac{b_{0}^2 \ell ^2 }{4}\,,  \notag \\
f^2  &\sim &\frac{r^2 }{\ell ^2 }+br+\kappa -C^2 +\frac{b^2 \ell ^2 }{4}\,.
\end{eqnarray}
 
\subparagraph{3)}

Another nonequivalent $\Xi $-neutral solution with $A_t=0$ is obtained when we take $f+\nu =0$ in \eqref{splitting_point}, which yields
\begin{equation}
A_{t}=0\,, \quad f=-\nu=-\varepsilon h \,,\quad N=1-\dfrac{\varepsilon b_{0}\ell }{2\nu }+\dfrac{b\ell }{2\nu }\,\mathrm{\ln }\left( \nu +\dfrac{r}{\ell }\right) .  \label{Nf2}
\end{equation}
The expressions in this case involve the log terms in the metric functions $N$ and $fN$ when $b\neq 0$, and the geometry does not describe any longer usual AAdS space. It has faster fall-off when $\varepsilon =-1$ than when $\varepsilon = 1$. In particular, when $b=0$, the space is AAdS, but still it possesses the gravitational hair $b_{0}\neq 0$ that modifies its asymptotics. When $b=b_0=0$, we recover the solution with $N=1$. 
\bigskip

In the last two cases, 2) and 3), the internal non-Abelian field that satisfies equations \eqref{EOM_SU(4)_1} and \eqref{EOM_SU(4)_2} allows the static
configuration
\begin{equation}
\mathcal{A}^{\Lambda }=\mathcal{A}^{\Lambda }(r,y)\,,  \label{static}
\end{equation}
with the internal non-Abelian field strength without support on the transversal section,
\begin{equation}
\mathcal{F}^{\Lambda }=\mathcal{F}_{tr}^{\Lambda }\,\diff t\wedge \diff r+\mathcal{F}_{rm}^{\Lambda }\,\diff r\wedge \diff y^m\,.
\label{SU(4) particular}
\end{equation}

An interesting case of the $\Xi$-neutral solution corresponds to $b=b_0=0$,  in which case the  branches \textbf{2)} and \textbf{3)} coincide. Then, without the internal non-Abelian field, it becomes a $\frac{1}{2}$-BPS state when $\kappa =C^2$ \cite{Canfora:2007xs}. We will show in Section \ref{Neutral} that there is also a BPS state that includes a pure gauge internal non-Abelian field.

\subsection{Branch with \texorpdfstring{$\Xi\neq 0$}{Xi = 0} \label{LS}}

To solve the equations of motion, we express $N$ from $\mathcal{E}_1 $ and require that the spacetime has
non-vanishing axial torsion. Then $\mathcal{E}_{1i}$ implies $A_{t}=0$ and $\mathcal{E}_{ij}$ enables to find $f$. The solution is
\begin{equation}
A_t=0\,, \quad f=-\nu \left( 1+\frac{b\ell ^2 }{2r}\right) \,,\quad (Nf)' =-\frac{r}{\nu \ell ^2 }\,.
\end{equation}
On the other hand, $\mathcal{E}_{01}$ and $\mathcal{E}_{0}$ present a system of equations in $\varphi $ and $\nu $,
\begin{equation}
\frac{f+\nu }{rf}\,\Xi =2\varphi \varphi'\,,\qquad \left( f\nu' +\frac{r}{\ell ^2 }\right) \frac{\Xi}{\nu f} =2\varphi \varphi'\,. \label{2_phi_phi'}
\end{equation}
Recall that the function $\Xi (r)$ is given by eq.~\eqref{Xi}. Comparing  the last equations leads to a differential equation in $\nu ^2 $ that can be straightforwardly solved. The general solution is 
\begin{eqnarray}
\nu  &=& - \frac{f}{1+\frac{b\ell ^2 }{2r}}\,,\qquad f=\sqrt{\frac{r^2 }{\ell ^2 }+br+\kappa -\mu }\,,\qquad N=1-\frac{b_{0}\ell }{2f}\,.  \notag \\
\varphi ^2  &=&\left[ C^2 -\frac{b\ell ^2 }{2}\frac{\mu +\frac{b^2 \ell ^2 }{4}}{r+\frac{b\ell ^2 }{2}}+\frac{b^2 \ell ^{4}}{4}\frac{\left( r+ \frac{b\ell ^2 }{3}\right) \left( \mu +\frac{b^2 \ell ^2 }{4}-\kappa \right) }{\left( r+\frac{b\ell ^2 }{2}\right) ^3 }\right] \left( 1+\frac{b\ell ^2 }{2r}\right) ,  \label{varphi2}
\end{eqnarray}
with the integration constants $\{C,b_{0},b,\mu \}$.
 We will choose a sign of the axial torsion so that $\mathrm{sgn}(\varphi )=\mathrm{sgn}(C)$.

When $b\neq 0$, the gravitational hair causes a non-trivial behaviour of the axial torsion that tends to a constant value on the boundary at $r\to\infty$,
\begin{eqnarray}
\varphi ^2  &\to &C^2 \,,  \notag \\
\Xi  &\to &\mu -C^2 +\frac{b^2 \ell ^2 }{4}\,, \label{Branch-Xi-neq-0}
\end{eqnarray}
whereas the second gravitational hair parameter $b_{0}$ produces $N\neq 1$, such that asymptotically
\begin{equation}
N\to 1 - \frac{b_{0}\ell ^2 }{2 r}\,.
\end{equation}

With addition of the gravitational hair, the spacetime acquires additional singular terms at $r=0$, as the Riemann scalar behaves near the singularity as
\begin{align}
\mathring{R}\simeq & \frac{6(M+\kappa) }{r^2} - \frac{3b\,(8\sqrt{-M} + 3 b_0\ell) }{(2\sqrt{-M} +b_0 \ell)r } +\cdots \,,
\end{align} 
where for $b_0=0$ we recover eq.~\eqref{R_ring}. Non-vanishing hair parameter $b_0 \neq 0$ favors naked singularities ($M < 0$), global AdS space ($M=-1$) and the ground state black hole ($M=0$). 

However, we have to set $b_0=0$ in order to have finite $N$ on the horizon.

Regarding the internal non-Abelian field contribution to the field equations \eqref{EOM}, the equation $\mathcal{E}$ implies that the non-Abelian Pontryagin term has to satisfy 
\begin{equation}
\rho \mathcal{F}^{\Lambda }\mathcal{F}_{\Lambda }=\alpha \,\left( b\varphi \,
\frac{1-\frac{\ell ^2 \nu ^2 }{r^2 }}{1+\frac{b\ell ^2 }{2r}}+2\Xi
\varphi ^{\prime }\right) \epsilon _{ijk}\diff r\tilde{e}^{i}\tilde{e}%
^j\tilde{e}^k\,.  \label{FF-SU(4)}
\end{equation}
In the general case with $b\neq 0$, $\varphi \neq const.$, the above equation requires both $\mathcal{F}_{rm}^{\Lambda }$ and $\mathcal{F}_{mn}^{\Lambda }$ to be different from zero.

On the other hand, the solution with no gravitational hair (that is, with both $b=0$ and $b_{0}=0$, in which case $\varphi=const.$ and $N=1$) allows for a simpler but still non trivial solution, where the internal non-Abelian field strength takes value only on the transverse space $\Sigma$. In this special case, $f+\nu =0$ and $\Xi $ becomes a constant `charge', with the solution 
\begin{eqnarray}
A_{t} &=&0\,,\qquad \Xi =\mu -C^2 \neq 0\,,  \notag \\
f &=&-\nu \;=\;h =\; \sqrt{\dfrac{r^2 }{\ell ^2 }-M}\,,\quad M=\mu -\kappa \geq 0\,,  \label{Laura} \\
\varphi  &=&C\,,\qquad b=0\,.  \notag
\end{eqnarray}
An additional feature of this particular solution is that it allows the torsion component $C$ to decouple completely from the functions $\nu (r)$ and $f(r)$,  leading to a $U(1)_q $-neutral and $\Xi $-charged black hole. Note that, unlike \eqref{f,h}, now the metric function $f=h$ does not depend on the constant $C$. Furthermore, in this case the r.h.s. of \eqref{FF-SU(4)} is zero so we can choose $\mathcal{F}_{rm}^{\Lambda }=0$ to solve it. Then it is straightforward to prove that the internal non-Abelian field equations \eqref{EOM_SU(4)_1} and \eqref{EOM_SU(4)_2} are solved. In consequence,  $\mathcal{F}$ has support only on $\Sigma $, namely
\begin{equation}
\mathcal{F=}\frac{1}{2}\,\mathcal{F}_{mn}\,\diff y^m\wedge \diff
y^{n}\neq 0\,.  \label{F_Laura}
\end{equation}

\subsection{The space of parameters of the black hole}

All three solutions mentioned above, \eqref{Nf1} and \eqref{Nf2} (with $\Xi =0$), and \eqref{varphi2} (with $\Xi \neq 0$), have the space of the parameters characterized by, at most, four free parameters $b_0$, $b$, $\mu $ and $C$. The metric function is such that the horizons are defined by
\begin{equation}
f(r_{\pm })=0\quad \Rightarrow \quad \frac{r^2 _{\pm }}{\ell ^2 }+Lr_{\pm }-M=0\,,  \label{f(r)}
\end{equation}
where the mass parameter $M$ and the value of $L$ depend on the solution ($M=C^2 -\kappa $, $L=0$ when $\Xi =0$, and $M=\mu -\kappa $, $L=b$ when $\Xi \neq 0$). The geometry has at most two horizons $r_{-}$ and $r_{+}\geq r_{-}$ of the form
\begin{equation}
r_{\pm }=-L\ell ^2 \pm 2\ell p\,.\qquad L\leq 0\,.
\end{equation}
The event horizon $r_{h}=r_{+}$ exists when the \textit{extremality parameter} $p$ is real: 
\begin{equation}
p^2 =\left( \frac{r_{+}-r_{-}}{4\ell }\right) ^2 =M +\frac{L^2\ell ^2 }{4}\geq 0\,.  \label{extremality}
\end{equation}%
The black hole becomes extremal in the limit when the two horizons coincide, $p=0$.
Thus, the extrThus, the extremality parameter is a measure of the deviation of the black hole solution

emality parameter is a measure of the deviation of the black hole solution
from its ground state (the extremal case).

Indeed, similarly to \eqref{singularity}, in $D=5$ the Riemannian scalar curvature of the solution near $r=0$ behaves as
\begin{equation}
\mathring{R}\simeq \frac{6(M+\kappa) }{r^2 } -\frac{12L}{r}-\frac{20}{\ell ^2 }+\cdots \,, \label{R_ring}
\end{equation}
showing  a singularity when $M+\kappa\neq 0$ or $L\neq 0$.

When $p^2 \geq 0$, the black hole temperature is
\begin{equation}
    T=\frac{1}{4\pi}\,\left[\rule{0pt}{13pt} N(f^2)'\right]_{r_+}=\frac{N(r_+)}{2\pi}\, \left( \frac{r_+}{\ell^2}+\frac L2 \right),
\end{equation}
and it becomes zero in the extremal case $p=0$.

The cases where $p^2<0$, $p^2\neq -1$ correspond  to \emph{naked singularities}, where an event horizon hiding the singularity at $r=0$ cannot be defined.

In the next section we will focus on the solution \eqref{Laura},  with $\mathcal{N}=4$ supercharges. 

\section{BPS states with the internal non-Abelian field}

In this section we are interested in solutions with a standard asymptotically AdS space-time (without gravitational hair ($b=b_0=0$)) in the pseudo-Riemannian sector, and  with a gauge field in the internal symmetry sector. In that way, the torsional degrees of freedom can modify the asymptotic behavior of the supercurvature, but its pseudo-Riemannian part has the usual asymptotically AdS form.

We will focus in particular on the case   $\mathcal{N}=4$, that is with  gauge supergroup $SU(2,2|4)$. Then the $U(1)_q$ generator becomes a central extension, making the corresponding field $A_\mu$ non dynamical\footnote{As mentioned before, when $\mathcal{N}=4$ the coefficient $\beta $ vanishes and the $U(1)_q$ field becomes strongly coupled.}.
Moreover, as we will explicitly discuss after eq.~\eqref{ks}, in this case the odd generators of the superalgebra  become $U(1)_q$-neutral, so that the $U(1)_q$ charge should not play a role while looking to BPS solutions preserving a fraction of the $\mathcal{N}=4$ supersymmetries. 
However, the $U(1)_q$ field could have a non-trivial effect on topology \cite{Miskovic:2006ei}, but typically not in the spherically symmetric ansatz.
It was shown in the previous section that for spherically symmetric
solutions, the time component $A_{t}$ of the $U(1)_{q}$ gauge field either vanishes ($A_{t}=0$), or it can produce non-physical configurations which
require further investigation. The last statement means that, when $\Xi =0$, it can become divergent on the boundary, as in \eqref{At}, and when  $\Xi \neq 0$, it allows free geometries (with the metric containing arbitrary functions).
It is worthwhile emphasizing that divergent fields on the asymptotically AdS boundary are not a problem as long as energy and other conserved charges are finite. For simplicity, we  set $A_t=0$.

We seek for bosonic solutions of the field equations ($\psi _{s},\bar{\psi}^{s}=0$) that are left invariant under (globally defined) local supersymmetry transformations $\epsilon _{s},\bar{\epsilon}^{s}\neq 0$. In general, gauge transformations with the local parameter
\begin{equation}
\mathbf{\Lambda }=\Lambda ^M\,\mathbf{G}_M=\frac{1}{2}\,\Lambda ^{AB}\mathbf{J}_{AB}+\Lambda^{\Lambda }\mathbf{G}_{\Lambda }+\bar{\epsilon}_{\alpha }^{s}\mathbf{Q}_{s}^{\alpha } -\mathbf{\bar{Q}}_{\alpha
}^{s}\epsilon _{s}^{\alpha } +\Lambda^1\, \mathbf{G}_1 \,,
\end{equation}
act on the gauge field as
\begin{equation}
\delta \mathbf{A}=D\mathbf{\Lambda }=\diff \mathbf{\Lambda }+[\mathbf{A}, \mathbf{\Lambda }]\,.
\end{equation}
In particular, the supersymmetry transformations for any $\mathcal{N}$ read
\begin{eqnarray}
\delta _{\epsilon }\psi _{s} &=&\nabla \epsilon _{s}=\left( \diff +\frac 14\,\omega ^{ab}\Gamma _{ab}+\frac{1}{2\ell }\,e^{a}\Gamma _{a}+\ii qA \mathbf{G}_1 \right) \epsilon _{s}-\mathcal{A}^{\Lambda }(\tau _{\Lambda})_{s}^{u}\epsilon _{u}\,,  \notag \\
\delta _{\epsilon }\bar{\psi}^{s} &=&\nabla \bar{\epsilon}^{s}=\bar{\epsilon}^{s}\left( \overset{\leftarrow }{\diff }-\frac{1}{2\ell }\,e^{a}\,\Gamma_{a}-\frac{1}{4}\,\omega ^{ab}\Gamma _{ab}+\ii qA\mathbf{G}_1 \right) +\mathcal{A}^{\Lambda }\bar{\epsilon}^{u}(\tau _{\Lambda })_{u}^{s}\,.\label{ks}
\end{eqnarray}
In our case $\mathcal{N}=4$ and the fermions are $U(1)_q$-neutral, $q=\frac{1}{4}-\frac{1}{\mathcal{N}}=0$, thus
the $U(1)_q$ field drops out of the covariant derivative. In consequence, the bosonic BPS states satisfy the Killing spinor equation
\begin{equation}
\delta _{\epsilon }\psi _{s}=\nabla \epsilon _{s} =(\diff +\mathbf{\hat A} -\mathcal{A}) \epsilon_s =0\,,
\label{KS Eq}
\end{equation}
where we denoted the AdS gauge field by
\begin{equation}
\mathbf{\hat A}=\frac 12\,\omega ^{ab}\mathbf{J}_{ab} +\frac{1}{\ell }\,e^{a}\mathbf{P}_{a}=\frac{1}{4}\, \omega ^{ab}\Gamma _{ab}+\frac{1}{2\ell }\,e^{a}\Gamma _{a}\,.
\end{equation}

\subparagraph{Integrability condition.}

Applying once more the covariant derivative $\nabla$ over \eqref{KS Eq}, we obtain the integrability condition of the Killing spinor equation, 
\begin{equation}
\nabla \nabla \epsilon _{s}= \mathbf{\hat F}\,\epsilon _{s}-\mathcal{F}_{\;s}^{u}\epsilon _{u}^{\alpha }=0
\quad \Rightarrow \quad \mathbf{\hat F} _{\ \beta }^{\alpha }\epsilon_{s}^{\beta }=\mathcal{F}_{\;s}^{u}\epsilon _{u}^{\alpha }\,,
\label{ConsistencyKS}
\end{equation}
where the AdS curvature reads
\begin{equation}
\mathbf{\hat F}=\frac 12\,F^{ab}\mathbf{J}_{ab}+\frac{1}{\ell }\,T^{a}\mathbf{P}_{a} =\frac{1}{4}\,F^{ab}\Gamma _{ab}+\frac{1}{2\ell }\,T^{a}\Gamma _{a}\,.  \label{F-AdS def}
\end{equation}
Unlike in standard supergravity, this condition  --linear in the field strength-- cannot be expressed as a combination of the equations of motion which are quadratic in the field strength, but it has to be satisfied on-shell. We will evaluate it for the asymptotically AdS black hole, that means, without the gravitational hair. At first we will
include all branches with (vanishing or non-vanishing) $\Xi$-charge,
\begin{equation}
b=b_0=0\,,\qquad\Xi =\mu -C^2\,.
\end{equation}
Recall that we also have the extremality parameter,
\begin{equation}
p^2 =\mu -\kappa \geq 0\,,
\end{equation}
where the negative values of $p^2$ do not correspond to black holes, and it is convenient to introduce the third (dependent) parameter
\begin{equation}
\tilde\Xi=\Xi -p^2=\kappa -C^2\leq \Xi\,. \label{new parameter}
\end{equation}
We will see later that, geometrically, the parameters $\Xi$ and $\tilde{\Xi}$ characterize the on-shell field strength \eqref{fs} with support on $\Sigma$, restricted to $SU(2)$ subgroups. In the extremal case ($p^2 =0$), the two parameters coincide, $\Xi =\tilde\Xi$.

Common features of all $U(1)_q $-neutral solutions without hair found in the previous section are the following conditions on the fields
\begin{equation}
A_{t}=0\,,\quad N=1\,,\quad \varphi =C\,,\quad \nu =-f\,,\quad \nu \nu'=ff'=\frac{r}{\ell ^2 }\,. \label{common}
\end{equation}
In particular, in the extremal limit $p\to 0$ we have
\begin{equation}
\Xi \to \tilde\Xi \,, \qquad\mbox{that is }\quad \mu \to \kappa\,,\qquad \mbox{so that }\quad    f\to \frac  r\ell\,.
\label{ex}
\end{equation}
Using the index decomposition $a=(0,1,i)$ in \eqref{F-AdS def} and plugging the above fields in the components of the AdS curvature \eqref{T} and \eqref{AdS comp}, we find the general form of the AdS curvature on-shell,
\begin{equation}
\mathbf{\hat F}=C\left(\frac{r}{\ell}\, \mathbf{P}_i -f\,\mathbf{J}_{1i}\right) \epsilon _{\ jk}^{i}\, \tilde{e}^j\tilde{e}^k+\frac{1}{2}\,\Xi \, \tilde{e}^{i}\tilde{e}^j\,\mathbf{J}_{ij}\,, \label{F common}
\end{equation}
where only the components on $\Sigma$ remain different from zero, and in the representation of the gamma matrices they have the form
\begin{equation}
\mathbf{\hat F}_{mn}=C\tilde{\epsilon}_{\ mn}^{i}\Gamma _i\left( \frac{r}{\ell }+f\,\Gamma _1 \right) +\frac{1}{2}\,\Xi \,\Gamma _{ij}\tilde{e}_m^{i}\tilde{e}_{n}^j\,.  \label{FAdS}
\end{equation}
The consistency \eqref{ConsistencyKS} of the Killing spinor equation with the AdS curvature \eqref{FAdS} reads
\begin{equation}\label{intbps}
0=(\mathbf{\hat F}_{mn}-\mathcal{F}_{mn}) \epsilon \,,\quad \mathcal{F}_{rm}\epsilon =0\,,\quad \mathcal{F}_{tm}\epsilon =0\,,\quad \mathcal{F}_{tr}\epsilon =0\,.
\end{equation}
The equations in \eqref{intbps} can be satisfied in the extremal limit, in which the conditions  \eqref{ex} hold, by taking the Killing spinors to be the eigenvectors of the $\Gamma _1 $ matrix:
\begin{equation}
\Gamma _1 \epsilon =\lambda \epsilon \,,\qquad \lambda =\pm 1\,.
\end{equation}
We then obtain the general on-shell conditions on the internal non-Abelian field $\mathcal{F}_{\mu \nu }=\mathcal{F}_{\mu \nu }^{\Lambda }\mathbf{G}_{\Lambda }$ and the spinor parameter $\epsilon $:
\begin{eqnarray}
\mathcal{F}_{mn}\epsilon  &=&\left[ (1+\lambda) \,\frac{rC}{\ell }\,\tilde{\epsilon}_{\ mn}^{i}\Gamma _i+\frac{1}{2}\,\Xi \,\Gamma
_{ij}\tilde{e}_m^{i}\tilde{e}_{n}^j\right] \epsilon \,,\notag \\
\mathcal{F}_{rm}\epsilon  &=&0\,,\qquad \mathcal{F}_{tm}\epsilon =0\,,\qquad \mathcal{F}_{tr}\epsilon =0\,,  
\label{SU(4) consistency}
\end{eqnarray}
that have to be satisfied in order for
the extremal black hole (with $p^2=0$ and  $f=\frac{r}{\ell }$) to be compatible with the BPS condition \eqref{KS Eq}. 

The first term in \eqref{FAdS} cancels out when acting on $\epsilon$ only if $\lambda=-1$. In the other  case $\lambda=+1$, $p^2=0$, then $\mathcal{F}_{ij}$ has an extra contribution proportional to $2C\frac r\ell \,\Gamma_{ij}$.

The above conditions can be used to identify the subgroup of $SU(4)$ to which $\mathcal{F}$ is restricted.
Let us now proceed with finding the Killing spinors.

\subparagraph{Static Killing spinor equation.}

Let us write out in components the Killing spinor equation \eqref{KS Eq},
\begin{equation}
0=\diff \epsilon +\mathbf{\hat A}\epsilon -\mathcal{A}\epsilon \,,
\label{KSeq}
\end{equation}
for all static AdS black holes previously discussed, satisfying \eqref{common}. 

The on-shell AdS gauge field in these cases, corresponding to the AdS field strength \eqref{F common}, is
\begin{eqnarray}
\mathbf{\hat A} &=&\left(\frac r\ell\, \mathbf{J}_{01} +f\,\mathbf{P}_{0}\right) \frac{\diff t}{\ell } +\dfrac{\diff r}{\ell f}\,\mathbf{P}_1  +\left( \frac{r}{\ell }\,\mathbf{P}_i-f\,\mathbf{J}_{1i}\right) \tilde{e}^{i}+\frac{1}{2}\,\left( \tilde{\omega}^{ij}-C\,\epsilon ^{ijk}\,\tilde{e}_{k}\right) \mathbf{J}_{ij}  \notag \\
&=&\dfrac{r}{2\ell ^2 }\,\Gamma _{0}\left( 1+\Gamma _1 \right) \diff t+\dfrac{\diff r}{2r}\,\Gamma _1 +\frac{r}{2\ell }\,\Gamma _i\left(
1+\Gamma _1 \right) \tilde{e}^{i}+\frac{1}{4}\,\left( \tilde{\omega}
^{ij}-C\,\epsilon ^{ijk}\,\tilde{e}_{k}\right) \Gamma _{ij}\,, \label{AdS-Xineq0}
\end{eqnarray}
where in the second line we applied $p^2=0$ and therefore $f=\frac r\ell.$ We also assume a static configuration of the spinor,
\begin{equation}
\partial _{t}\epsilon _{s}^{\alpha }=0\,.
\end{equation}
From \eqref{KSeq}, the Killing spinor satisfies a system of partial differential equations taken along $\diff x^{\mu }$ that, applying $\Gamma _1 \epsilon =\lambda \epsilon $, acquire the form
\begin{eqnarray}
\diff t &:&\quad 0=(\lambda+1)\,\dfrac{r}{2\ell ^2 }\,\Gamma _0 \epsilon -\mathcal{A}_t\, \epsilon \,,  \notag \\
\diff r &:&\quad 0=\partial _r\epsilon +\dfrac{\lambda}{2r}\, \epsilon -\mathcal{A}_r\,\epsilon \,,  \label{KS_Eq_components} \\
\diff y^m &:&\quad 0=\partial _m\epsilon +\left[ (1+\lambda)\,\frac{r}{2\ell }\,\Gamma
_i \tilde{e}_m^{i}+\frac{1}{4}\,\left( \tilde{\omega}_m^{ij}-C\,\epsilon ^{ijk}\,\tilde{e}_{km}\right) \Gamma _{ij}
\right] \epsilon -\mathcal{A}_m\epsilon \,. \notag
\end{eqnarray}
Recall that $\tilde{\omega}^{ij}$ is the intrinsic (torsioneless)  spin connection of $\mathbb{S}^3$ satisfying
 $\tilde{T}^{i}=\diff \tilde{e}^{i}+\tilde{\omega}^{ij}\wedge
\tilde{e}_j=0$,  so that it is a known function of $\tilde{e}^k$. In contrast, the spacetime torsion $T^a$ and the supertorsion $F^{a5}$, defined in \eqref{F}, are non-vanishing.

A solution of \eqref{KS_Eq_components} depends on the explicit form of the black hole, the eigenvalue $\lambda$ and the particular subgroup of $SU(4)$ considered for the internal non-Abelian field. 

As respect to the spinors that are the positive eigenstates of the $\Gamma_1$-matrix, for them the first equation in \eqref{SU(4) consistency} acquires an extra term due to $\lambda=+1$, that yields a contribution to $\mathcal{F}_{ij}$  proportional to $2C\frac r\ell \,\Gamma_{ij}$. However, solving \eqref{KS_Eq_components} for this configuration shows that it produces a constant flux of the non-Abelian field with $\mathcal{F}_{tr}\neq 0$, that is inconsistent with our requirement  of static solutions, and in particular with the second line of \eqref{SU(4) consistency}. Thus, we will disregard this solution for the given ansatz of the non-Abelian field, and focus only on the $\lambda=-1$ case.

\subsection{Broken symmetry of the non-Abelian field}

The fact that the BPS state has to satisfy eq.~\eqref{ConsistencyKS} suggests that the compact subgroup\footnote{Looking at $SO(2)\times SO(4)\subset SO(2,4)$ would involve closed timelike curves.} $SO(4)\subset SO(2,4)$  of the spacetime  isometry group has to be  `tuned' with the internal symmetry group $SU(4)$.
This in particular implies that, on the BPS solution, the internal symmetry $SU(4)\simeq SO(6)$ (we do not consider $U(1)_q$ any longer), will be broken at least to its  subgroup $SO(4)\subset SO(6)$.
Setting to zero all components of  $\mathcal{F=F}^{\Lambda }\mathbf{G} _{\Lambda }$ that correspond to the broken symmetries also simplifies solving the field equations of CS AdS$_5$ gravity that are quadratic in the field strengths.

The result depends on both the subgroup of $SU(4)$ and on the topology of the spatial boundary $\Sigma$, whose metric is $\gamma _{mn}$ given by \eqref{Omega}.
The existence of a topological charge, associated 
to a non trivial embedding, with winding number $n\in \mathbb{Z}$, of the  internal symmetry group $G$ in the asymptotic transverse space with  topology $\mathbb{S}^k$, depends on the $k$-th homotopy group $\pi _{k}(G)$.  In our case, since the transverse space $\Sigma$ is three dimensional, $k=1,2,3$.  Using the table from \cite{Nakahara}, we identify the existence of possible  winding numbers for the following relevant cases: 
\begin{equation*}
\begin{array}{clll}
\mathbb{S}^1 \simeq U(1):\medskip  & \pi _1 (\mathbb{S}^1 )=\mathbb{Z}\,,
& \pi _2 (\mathbb{S}^1 )=0\,, & \pi _{3}(\mathbb{S}^1 )=0\,, \\
\mathbb{S}^2 \simeq \frac{SO(3)}{SO(2)}:\medskip  & \pi _1 (\mathbb{S}^2 )=0\,, & \pi _2 (\mathbb{S}^2 )=\mathbb{Z}\,, & \pi _{3}(\mathbb{S}^2 )=\mathbb{Z}\,, \\
\mathbb{S}^3 \simeq \frac{SO(4)}{SO(3)}\simeq SU(2):\quad& \pi _1 (\mathbb{S}^3 )=0\,, & \pi _2 (\mathbb{S}^3 )=0\,, & \pi _{3}(\mathbb{S}^3 )=\mathbb{Z}\,,
\end{array}
\end{equation*}
and, correspondingly, for the isometry groups:
$$\pi _1 (U(1))=\mathbb{Z}\,,\quad
\pi_{3}(SU(2))=\mathbb{Z}\,,\quad \pi_{3}(SO(4))=\mathbb{Z} + \mathbb{Z}\,.
$$
The table presented above has much richer content when we also include quotient spaces of the type $\mathbb{S}^k/\Upsilon$, with $\Upsilon$ a discrete subroup of {\it Isom}$(\mathbb{S}^k)$, such as ``lens spaces'' $\mathbb{S}^k/\mathbb{Z}_s$ \cite{Brody}. 
In particular, for the 3-sphere with isometries $\Upsilon=\mathbb{Z}_2$, the identification of its antipodal points leads to the topological space $\mathbb{RP}^3\equiv\mathbb{S}^3/\mathbb{Z}_2$ that is the real projective  3-space, whose nontrivial homotopy groups are 
$$\pi _1 (\mathbb{RP}^3)=\mathbb{Z}_2\,,\quad
\pi_2 (\mathbb{RP}^3)=0\,,\quad
\pi_{3}(\mathbb{RP}^3)=\mathbb{Z}\,,
$$
the group $\pi_1$ being different compared to the one of $\mathbb{S}^3$.

In the case discussed in this paper, $\Sigma $  is the maximally symmetric 3-sphere $\mathbb{S}^3$ or its identifications, and the corresponding (locally) isometry group  is $SO(4)$, while the internal symmetry is $SU(4)$, which is locally equivalent to $SO(6)$. 

Non-trivial configurations then require in particular  the internal symmetry group $SU(4)$ to be broken to  $SO(4)$ or subgroups thereof.  There are  two nonequivalent {\it maximal  subgroups} of $SU(4)$ containing $SO(4)\simeq SU(2)_{+}\times SU(2)_{-}$ as a semi-simple factor (see for example \cite{Slansky:1981yr}).
One  corresponds to  $SU(4)\to SO(4)\simeq SU(2)_{+}\times SU(2)_{-}$ irreducibly. In terms of $SO(6)$, it  corresponds to the decomposition $SO(6)\to SO(3)\times SO(3)$.
The other decomposition is  $SU(4)\to SU(2)_{+}\times SU(2)_{-}\times U(1)_c $ that in terms of $SO(6)$ reads  $SO(6) \to SO(4)\times SO(2)_c$.
The two decompositions are distinguished by the corresponding nonequivalent  branchings of the irreducible representations of $SU(4)$. In particular, for the former one, the fundamental representation ${\bf 4}$ of $SU(4)$ (the one pertaining to the odd generators) decomposes into the $({\bf 2,2}) $ of  $SU(2)_{+}\times SU(2)_{-}$, while for the latter it goes into the $({\bf 2,1})_{+1} + ({\bf 1,2})_{-1} $ of  $SU(2)_{+}\times SU(2)_{-}\times U(1)_c $, so that the odd generators are $U(1)_c$-charged. More information about these two decompositions can be found in  Appendix \ref{two_branchings}.

As we are going to discuss in the following text, our solution corresponds to choosing the second of the branchings above, where the Killing spinor decomposes into a $U(1)_c$-charged Dirac spinor on $\Sigma$.
\bigskip

To implement explicitly the breaking of internal symmetry on the solution, it is convenient to represent the $SU(4)$ generators in terms of $SO(6)$ antisymmetric matrices $\mathbf{T}_{IJ}=-\mathbf{T}_{JI}$, where  the indices labeling the generators run as $I,J=0,\ldots,5$ for convenience (the Lie group is compact), and expand the $SU(4)$ gauge field as
\begin{equation}
\mathcal{A}=\mathcal{A}^{\Lambda }\mathbf{G}_{\Lambda }=\frac 12 \,\mathcal{A}^{IJ}\mathbf{T}_{IJ}\,.
\end{equation}
We will set to zero the components of $\mathcal{A}^{IJ}$ corresponding to the broken generators. It depends on $\Xi$ and $\lambda$ which ones are broken but, in all cases, they have to describe a soliton on the geometry whose transversal section is $\mathbb{S}^3 $. Because we know that, due to $\pi _{3}(SU(2))=\mathbb{Z}$, the $SU(2)$ group can have a non-trivial topological structure with winding around the 3-sphere, let us assume from the very beginning  that the $SU(4)$ is broken to a subgroup containing at least an $SU(2)$ factor, to be identified on the solution with the diagonal $\widehat{SU}(2)_D\subset SU(2,2)$ tangent space symmetry on $\mathbb{S}^3$  (a hat over $SU(2)$ is to distinguish it from  $SU(2)$ factors in the internal symmetry group). Generators of the diagonal subgroup $SU(2)_D\subset SU(4)$ are neither of the generators of $SU(2)_{\pm}$, but their linear combination.\footnote{Sometimes the term `diagonal subgroup' is used for an Abelian subgroup of diagonal matrices, but this is not the case here.}
Let us consider the following decomposition, which corresponds to the second of the branchings mentioned above.  Using $I=(0,1,\hat\imath,5)$, we write
\begin{equation*}
\begin{array}{ccccc}
SO(6) \simeq SU(4) & \to  & SU(2)_+ \times SU(2)_- \times U(1)_c &  \to  & SU(2)_D \times U(1)_c\,, \\
\mathbf{T}_{IJ} & \to  & \{\mathbf{T}_{+\hat\imath}\,,\quad \mathbf{T}_{-\hat\imath}\,, \quad \mathbf{T}_c \}  & \to  & \{ \mathbf{T}_{\hat\imath}\,, \quad \mathbf{T}_c \}\,,
\end{array}
\end{equation*}
The $U(1)_c$ subgroup is generated by any commuting generator, for instance $\mathbf{T}_{15}$.  Furthermore, $\mathbf{T}_{\hat\imath}$ given by \eqref{dualT} are the generators of the diagonal subgroup $SU(2)_D\subset SU(2)_{+}\times SU(2)_{-}$, written in the ``dual'' form
\begin{equation}
\mathbf{T}_{\hat\imath} = \frac 12 \, \epsilon _{\hat\imath}^{\  \hat\jmath \hat{k}} \,\mathbf{T}_{\hat\jmath \hat{k}}\,. \label{dualT}
\end{equation}
On the other hand, with the above choice
the `left' and `right' $SU(2)_\pm$ generators are
\begin{equation}
\mathbf{T}_{\pm \hat\imath}=\frac 12 \left( \pm\mathbf{T}_{0\hat\imath}- \mathbf{T}_{\hat\imath}\right)\,,\quad (\hat\imath=2,3,4),  \label{Tipm}
\end{equation}
and they close two independent $\mathfrak{su}(2)$ algebras,
\begin{equation}
\left[ \mathbf{T}_{\pm \hat\imath},\mathbf{T}_{\pm \hat\jmath}\right] =\epsilon _{\hat\imath \hat\jmath}^{\ \ \hat{k}}\,\mathbf{T}_{\pm \hat{k}}\,,\qquad \left[\mathbf{T}_{+\hat\imath}, \mathbf{T}_{-\hat\jmath}\right] =0\,. \label{2independent}
\end{equation}

Therefore,  the gauge field in the maximal subgroup of the $SU(4)$ gauge group, consistent with the isometries of the 3-sphere is
\begin{equation}
\mathcal{A}=\mathcal{A}_{+}^{\hat{\imath}}\mathbf{T}_{+\hat{\imath}} +\mathcal{A}_{-}^{\hat{\imath}}\mathbf{T}_{-\hat{\imath}}+\mathcal{A}^{15}\mathbf{T}_c \,.\quad   \label{U(1)xSU(2)}
\end{equation}
Note that
\begin{equation} 
\mathcal{A}^{0\hat\imath}=\frac 12\,\left( \mathcal{A}_{+}^{\hat\imath}-\mathcal{A}_{-}^{\hat\imath}\right) \,,\qquad \mathcal{A}^{\hat\imath\hat\jmath}=-\frac 12\,\epsilon _{\ \ \hat k}^{\hat\imath\hat\jmath}\left( \mathcal{A}_{+}^{\hat k}+\mathcal{A}_{-}^{\hat k}\right) \,, \label{redef}
\end{equation}
and the diagonal condition $\mathcal{A}_{+}^{\hat{\imath}}=\mathcal{A}_{-}^{\hat{\imath}}\equiv -\mathcal{A}^{\hat{\imath}}$ 
identifies the  subgroup with  the non-vanishing $SU(2)$ field $\mathcal{A}^{\hat{\imath}}$,
\begin{equation}
SU(2)_{D}:\qquad \mathcal{A}^{0\hat{\imath}}=0\,,\quad \mathcal{A}^{\hat{\imath}\hat{\jmath}}\equiv \epsilon _{\ \ \hat{k}}^{\hat{\imath}\hat{\jmath}}\mathcal{A}^{\hat{k}}\,,\label{asu2d}
\end{equation}
corresponding to the components along the generators $\mathbf{T}_{\hat\imath}$ that close the $\mathfrak{su}(2)$ algebra,
\begin{equation}
[\mathbf{T}_{\hat{\imath}},\mathbf{T}_{\hat{\jmath}}]=-\epsilon _{\hat{\imath}\hat{\jmath}}^{\ \ \hat{k}} \, \mathbf{T}_{\hat{k}}\,. \label{diagonalA}
\end{equation}
In this text we will focus on the $SU(2)_D\times U(1)_c$-valued solution of the form
\begin{equation}
\mathcal{A}=\mathcal{A}^{\hat{\imath}}  \mathbf{T}_{\hat{\imath}}+\mathcal{A}^{15}\mathbf{T}_c\,.  \label{SU(2)xU(1)}
\end{equation}

With this choice at hand, we can focus on solving the Killing spinor equation construction of the particular BPS states.

In view of the above discussion and the definition \eqref{dualT}, from now on it will be more convenient to work with the `dual' AdS generators of $\widehat{SU}(2)_D$ and an associated `dual' spin connection on three-dimensional manifold $\Sigma$,
\begin{equation}
 \mathbf{J}_i \equiv\frac 12 \, \epsilon_i^{\ jk} \mathbf{J}_{jk} \,, \qquad   \tilde{\omega}_i \equiv\frac 12 \, \epsilon_i^{\ jk} \tilde{\omega}_{jk}\,.
\label{3Ddual}
\end{equation}

In addition, to simplify the expressions, from here on, we will rescale out an explicit appearance of the 3-sphere radius $a$  as
\begin{equation}
t\to at\,,\qquad r\to \frac{r}{a}\,,\qquad \mu \to
\frac{\mu }{a^2 }\,,\qquad C\to \frac{C}{a}\,, \label{rescaling}
\end{equation}
what is equivalent to setting $a=1$.

\subsection{\texorpdfstring{$\Xi$}{Xi}-charged BPS black hole}

When $\Xi=\mu -C^2 \neq 0$, the solution without gravitational hair is given by eqs.~\eqref{Laura}. Since the internal non-Abelian field has non-vanishing components $\mathcal{F}_{mn}$, this combined with the integrability condition \eqref{SU(4) consistency}  and the rescaling \eqref{rescaling} leads to the $U(1)_q$-neutral solution of the form
\begin{eqnarray}
\varphi  &=&C\;\neq \;0\,, \qquad N=1\,, \notag \\
f &=&-\nu \;=\;\sqrt{1 +\frac{r^2 }{\ell ^2 }-\mu }\,,  \notag \\
\mathcal{F} &\mathcal{=}&\frac{1}{2}\,\mathcal{F}_{mn}\,\diff
y^m\wedge \diff y^{n}\neq 0\,,  \label{Xi-charged}
\end{eqnarray}
where $\mathcal{A}$ is yet to be determined. This solution allows for a non trivial embedding of an $SU(2)$ subgroup of the $SU(4)$ internal symmetry group on the three-dimensional section $\Sigma $, since it identifies the action on the Killing spinor of the $\widehat{SU}(2)_D$ generator $\mathbf{J}_i$ on the tangent space of $\Sigma $ with the one on the non vanishing generators of $SU(4)$. This suggests to choose $\mathcal{F}$ as a field strength for a  diagonal subgroup $SU(2)_D$.

When $\lambda =-1$, the Killing spinor equation \eqref{KS_Eq_components} becomes
\begin{eqnarray}
0 &=&\mathcal{A}_{t}\epsilon \,,  \notag \\
0 &=&\partial _r\epsilon -\frac{1}{2r}\,\epsilon -\mathcal{A}_r\,\epsilon \,,  \notag \\
0 &=&\tilde{D}_m\epsilon -\frac{1}{4}\,C\,\epsilon ^{ij}_{\ \ k}\,\tilde{e}^k_m\Gamma _{ij}\epsilon -\mathcal{A}_m\,\epsilon \,,  \label{KS no AdS}
\end{eqnarray}
where $\tilde{D}_m\epsilon =\left( \partial_m+\frac 14 \, \epsilon^{ij}_{\ \ k} \tilde{\omega}_m^k\Gamma _{ij}\right) \epsilon $ is the covariant derivative on the 3-sphere acting on the time-independent Killing spinor $\epsilon$, and the internal non-Abelian field is $\mathcal{A}_{\mu}=\mathcal{A}_{\mu}^{\Lambda}\mathbf{G}_{\Lambda}$. The first equation has as a particular solution
\begin{equation}
\mathcal{A}_{t}^{\Lambda}=0\,,
\end{equation}
and the second one can be solved by
\begin{equation}
\mathcal{A}_r^{\Lambda}=0\,,\qquad 
\epsilon(r,y) =\sqrt{\frac{r}{\ell }}\,\eta (y)\,,
\end{equation}
where $\eta (y)$ is an arbitrary spinor on the 3-sphere such that $\Gamma_1 \eta =-\eta $. 

To solve the third equation for the Killing spinor in \eqref{KS no AdS}, we finally impose the $SU(2)_D$ symmetry on the transversal part of the internal gauge field, with the generators $\mathbf{G}_\Lambda = \{ \mathbf{T}_{\hat\imath}, \mathbf{T}_c \}$, as given by eq.~\eqref{SU(2)xU(1)}. 
Replacing the known quantities, the last equation becomes
\begin{equation}
\partial _m\eta +\left(\tilde{\omega}_m^{i}-C\,\tilde{e}^i_m\right) \mathbf{J} _i\eta = (\mathcal{A}^{\hat\imath}\mathbf{T}_{\hat\imath} +\mathcal{A}^{15}\mathbf{T}_c )\,\eta \,.  \label{important}
\end{equation}
It can be solved by identifying, on the solution, the tangent space $\widehat{SU}(2)_D$ symmetry on $\Sigma$ (and the corresponding indices $ i,j,...$)  with the internal $SU(2)_D$ symmetry (labeled with indices $\hat\imath,\hat\jmath,...$), 
\begin{equation}
\mathcal{A}_m^{\hat\imath}=\tilde{\omega}_m^{i}-C\,\tilde{e}_m^{i}\,,\qquad \left( \mathbf{J}_i\right) _{\beta }^{\alpha
}\,\epsilon _{s}^{\beta }=\left(\mathbf{T}_{\hat\imath}\right) _{s}^{u}\,\epsilon
_{u}\,.  \label{iso}
\end{equation}

This solidary identification between the internal and spacetime symmetries is reminiscent of the t'Hooft-Polyakov monopole \cite{tHooft:1974kcl,Polyakov:1974ek}, where the adjoint indices of the Higgs field are identified with those of the 3-space, which is also known as embedding  of  the  gauge  group  in  the  spin connection \cite{Wilczek,Charap:1977ww}. 
Note that $\mathcal{A}_m^{\hat\imath}$ transforms as an $SU(2)_D$ gauge connection. 
Due to the identification of the generators \eqref{iso} when acting on the spinor, we also identify the corresponding indices,
\begin{equation}
    i \equiv \hat\imath \in \{2,3,4\}\,, \label{i}
\end{equation} 
and stop writing the hats above the $SU(4)$ counterpart from now on, unless we want to emphasize the difference. The signature on the identified sections is $(+,+,+)$ in both groups, so \eqref{i} can be performed consistently.

The equation \eqref{iso} then implies the following particular solution for the spinor,
\begin{equation} \label{Eq-Spinor-A15}
\partial _m\eta _{s}=\mathcal{A}^{15}_m(\mathbf{T}_c )_{s}^{u}\,\eta_u\quad \Rightarrow \quad \eta (y)=\mathrm{e}^{\Omega (y)\mathbf{T}_c }\eta_0 \,, 
\end{equation}
where $\mathcal{A}^{15}=\diff \Omega (y)$ is a locally pure gauge $U(1)_c $ configuration because it does not propagate, {as $\mathcal{F}^{15}=0$ is required by the integrability condition \eqref{ConsistencyKS}. Also}, $\eta_0 $ is a constant spinor. When $\Omega$ vanishes, the Killing spinor $\eta$ is constant on the sphere.

The anti-Hermitean generator $\mathbf{T}_c $ acts on the spinor, in the representation of gamma matrices, as $\tau_{15}=-\frac{\ii}{2}\,\tilde{\Gamma}_1 $. Similarly to usual gamma matrices whose explicit representation is given in Appendix \ref{Gamma}, the $(\tilde{\Gamma}_1 )_{s}^{u}$ matrix can be represented by a diagonal matrix $\tilde{\Gamma}_1 =\sigma _{3}\otimes \mathbb{I}_2 $, and it has the eigenvalues $\pm 1$. Therefore, we can write $\eta _{s}$ as a sum of  spinors $\eta _{(\pm)s}$ belonging to positive and negative eigenspaces of $\tilde{\Gamma}_1 $,
\begin{equation}\label{tildeg}
\eta_s(y)=\mathrm{e}^{-\frac{\ii}{2}\,\Omega(y)}\eta _{(+)s}+\mathrm{e}^{\frac{\ii}{2}\,\Omega(y)}\eta _{(-)s}\,,\qquad (\tilde{\Gamma}_1)^u_s\, \eta _{(\pm)u}=\pm \eta _{(\pm)s}\,,
\end{equation}
in agreement with the $SU(4)\to SU(2)_+\times SU(2)_-\times U(1)_c$ decomposition of the spinor charges, as in eq.~\eqref{spindec}.

In this solution, $\eta_0$ belongs to a negative eigenspace of  the $\Gamma_1$ matrix and it satisfies the projective relations \eqref{iso}. Thus, the final form of the Killing spinor is
\begin{equation}
\epsilon =\sqrt{\frac{r}{\ell }}\, \mathrm{e}^{\Omega (y)\mathbf{T}_c }\,\eta_0 \,,\quad \Gamma_1\eta_0 =-\eta_0\,,\quad \mathbf{J}_i\,\eta_{0s} =\left(\mathbf{T}_i\right)_s^u\,\eta_{0u}\,. \label{eta0}
\end{equation}

The corresponding internal field-strength has constant curvature on the transversal section, 
\begin{equation}
\mathcal{F}=\frac{1}{2}\,\tilde{\Xi}\, \epsilon_{ij}^{\ \ k} \,\tilde{e}^{i}\tilde{e}^j\mathbf{T}_{k}\,. \label{F-SU(4)BPS}
\end{equation}
Similarly as $\Xi$ corresponds to the magnitude of the AdS curvature on $\Sigma$, we see that $\tilde{\Xi}$ represents, for the given black hole solution, the magnitude of the internal $SU(2)_D$ field strength \eqref{fs} restricted to the transversal section $\Sigma$. Let us recall that this parameter is defined as $\tilde{\Xi}\equiv \Xi-p^2$, and for  black-hole solutions it satisfies $\tilde{\Xi}\leq \Xi$, so that the BPS condition \eqref{SU(4) consistency}, that holds in the extremal limit $p\to 0$, leads to
\begin{equation}
\tilde{\Xi}=\Xi \quad \Leftrightarrow \quad p^2 =0\quad \Leftrightarrow
\quad \mu =1 \,.\label{extremal}
\end{equation}
Note that, contrary to the  $\Xi$-neutral case found in Subection \ref{Xi-neutral} which is pure gauge, and where the BPS equation is solved by the vacuum state in Section \ref{Neutral}, here we find a non trivial solution to the BPS equation, corresponding to the extremal black hole condition $\mu =1$, where the axial torsion charge $C$ remains completely arbitrary, up to topological considerations that will be discussed in Subsection \ref{Winding-3sphere}.

Let us emphasize that the internal geometry of the three-sphere can be described in terms of torsionless quantities such that
\begin{equation}
\mathbb{S}^3 : \quad \tilde{R}^{ij}= \tilde{e}^{i}\wedge \tilde{e}^j\,,\quad \tilde{T}^{i}=0\,,  \label{RT of S3}
\end{equation}
while the torsion and curvature of the transversal section $\Sigma $ (with $\left. e^{i}\right\vert _{\Sigma }=r\tilde{e}^{i}$) are
\begin{equation}
\Sigma : \quad R^{ij}=\left( \Xi -\frac{r^2 }{\ell ^2 }\right) \,\tilde{e}^{i}\wedge \tilde{e}^j\,,\quad T^{i}=rC\epsilon _{\ jk}^{i}\tilde{e}^j\wedge \tilde{e}^k\,.
\end{equation}

In sum, using \eqref{iso}, the full expression for the BPS state gauge field is 
\begin{eqnarray}
\mathbf{A}_{\mathrm{BPS}} &=&\frac{r}{\ell ^2 }\,\left( \mathbf{J}_{01}+\mathbf{P}_{0}\right) \,\diff t+\frac{\diff r}{r}\,\mathbf{P}_1 + \frac{r}{\ell }\,\tilde{e}^i\left(
\mathbf{P}_i-\mathbf{J}_{1i}\right) \notag
\\ && + (\tilde{\omega}^i-C\,\tilde{e}^i) \, (\mathbf{J}_i+\mathbf{T}_i) + \diff \Omega(y)\, \mathbf{T}_c\,,  \label{A_BPS}
\end{eqnarray}
and the corresponding BPS state gauge curvature is 
\begin{equation}
\mathbf{F}_{\mathrm{BPS}}=\frac{rC}{\ell }\,\epsilon_{ij}^{\ \ k} \tilde{e}^{i}\tilde{e}^j\,\left(\mathbf{P}_{k}-\mathbf{J}_{1k}\right) +\frac{1}{2}\,\Xi \, \epsilon_{ij}^{\ \ k} \, \tilde{e}^i\tilde{e}^j\,(\mathbf{J}_k+\mathbf{T}_k)\,.  \label{F-BPS}
\end{equation}  
This solution is a $1/16$-BPS state because of the projective conditions on $\eta_0$ given in \eqref{eta0}, leaving finally only one (Dirac) supercharge unbroken.
A detailed counting of these supersymmetries and an explicit construction of the Killing spinor in a particular representation is given in Appendix \ref{Counting}.

To study both BPS and non-BPS solutions and their charges, we will look at a more generic solution with $b=0$, not necessarily BPS and  not necessarily extremal ($p^2 \geq 0$), characterized by the arbitrary, independent charges $\Xi $ and $\tilde{\Xi}$, conveniently defined by 
\begin{equation}
    \Xi = \mu-C^2 \,, \qquad \tilde\Xi \in \mathbb{R}\,, \label{newXi}
\end{equation}
such that, in a non-BPS state, $\widehat{SU}(2)_D$ and $SU(2)_D$ components of the gauge field do not have the same integration constants. The gauge field and the corresponding fields strengths are 
\begin{eqnarray}
\mathbf{A} &=&\left( \dfrac{r}{\ell }\,\mathbf{J}_{01}+f\,\mathbf{P}_{0}\right) \frac{\diff t}{\ell }+\dfrac{\diff r}{\ell f}\,\mathbf{P}_1 +\tilde{e}^{i}\left(\frac{r}{\ell }\,\mathbf{P}_i-f\,\mathbf{J}_{1i}\right)  +(\tilde{\omega}^i-C\,\tilde{e}^i)\, \mathbf{J}_i +\mathcal{A}^i \mathbf{T}_i
+\diff \Omega \, \mathbf{T}_c\,,  \notag \\
\mathbf{F} &=&C\,\epsilon_{ij}^{\ \ k} \tilde{e}^{i}\tilde{e}^j\,\left( \frac{r}{\ell }\,\mathbf{P}_{k}-f\,\mathbf{J}_{1k}\right) +\frac{1}{2}\,\epsilon_{ij}^{\ \ k} \, \tilde{e}^i\tilde{e}^j\,\left(\Xi\,\mathbf{J}_k+\tilde{\Xi}\,\mathbf{T}_k\right)\,,  
\label{Laura solution}
\end{eqnarray}
where $\mathcal{A}^i=\mathcal{A}^i_m(y)\diff y^m$ is any solution with constant  $\mathcal{F}^i$. It is asymptotically AdS in the Riemannian sector, but the field strength $\mathbf{F}$ does not vanish on the boundary, neither in the AdS sector (because of $\Xi\neq 0$) or the internal non-Abelian sector (when $\tilde{\Xi}\neq 0 $). 

A particular solution for the internal non-Abelian field that admits a BPS limit is 
\begin{equation}
    \mathcal{A}^i=\tilde{\omega}^{i}-B\,\tilde{e}^{i}\,, \qquad  \tilde{\Xi}=1-B^2 \leq 1\,, \label{internal soliton}
\end{equation}
where $B$ is the amplitude of the soliton. The upper bound on $\tilde\Xi(B)$ is chosen such that the field strength of this internal field on $\Sigma$ is always bounded from above.  

Since, with the generalized solution \eqref{internal soliton} for the internal soliton now we have, instead of \eqref{new parameter}, that the parameters are related by
\begin{equation}
\Xi =\tilde\Xi+p^2+B^2-C^2\,,  \label{new relation}
\end{equation}
the extremality condition \eqref{extremal} leads to
\begin{equation}
p^2=0\quad \Rightarrow \quad \Xi=\tilde\Xi+B^2-C^2\,,
\label{extremal2}
\end{equation}
and the BPS condition requires equality of the $\widehat{SU}(2)_D$ and $SU(2)_D$  field strengths and the amplitudes of two solitons, namely,
\begin{equation}
\Xi =\tilde{\Xi}\,,\qquad B= C\,.
\label{extremalityBC}
\end{equation}
On the other hand, the extremal configurations, defined by eq.~\eqref{extremal2}, allow also solutions with $\Xi=\tilde\Xi$ and $B=\pm C$.
Note, however, that only one of the two choices $B=\pm C$ corresponds to a BPS state, since the Killing spinor equation, associated with a supersymmetric black hole solution, requires $B=C$, as summarized by
\begin{equation}
   \mathrm{BPS}:\qquad  \mu=1\,, \qquad B=C\,, \qquad \Xi =\tilde\Xi\,, \label{BPSstate}
   \end{equation}
   and corresponding to the values  \eqref{A_BPS} and \eqref{F-BPS} of the gauge field and of its field strength.

The other choice, $B =-C$, still corresponds to an extremal solution, but it does not preserve any supersymmetry:
\begin{equation}
   \text{non-BPS}:\qquad  \mu=1\,, \qquad B =-C\,, \qquad \Xi =\tilde\Xi\,. \label{nonBPSstate}
\end{equation}

We then find that for this class of black hole solutions, in general the parameter space of extremal solutions ($\mu=1$ and $B,C$ arbitrary) is larger than the parameter space of BPS solutions (satisfying \eqref{BPSstate}). 
This is similar to standard supergravity, where the space of extremal solutions is also larger than the space of BPS solutions. \bigskip

\subparagraph{Global properties of the Killing spinor.}

Having obtained local solution at hand, we still have to
analyse global properties of the Killing spinor \eqref{Eq-Spinor-A15}--\eqref{eta0} on $\Sigma$,  which is locally a 3-sphere. This is generally related to the existence of a generalized spin structure on the manifold \cite{Hawking:1977ab} that insures consistent coupling of spinors to bosonic fields. Keeping aside detailed analysis of spin structures, here we will discuss examples of two interesting topological spaces $\Sigma$: global $\mathbb{S}^3 $ that is a simply connected space, and the real projective 3-space, $\mathbb{RP}^3 \simeq \mathbb{S}^3/\mathbb{Z}_2$, obtained from the (parallelizable) 3-sphere by identification of its antipodal points, which is not simply connected.
In absence of torsion and $SU(2)$ gauge field,   $\mathbb{S}^3 $ has one and $\mathbb{RP}^3$ two inequivalent spin structures \cite{Dabrowski:1986en}.

Understanding of global behaviour of locally $\mathbb{S}^3 $ topologies is clearer if we reduce the problem to three dimensions, i.e., introduce a 3-dimensional Killing bispinor $\tilde{\eta}$ on the sphere. We use the representation of $\Gamma $-matrices \eqref{Repr5D} given in Appendix \ref{Gamma} and express the left chiral spinors, satisfying $\Gamma _1 \eta _{s}=-\eta _{s}$,  in terms of the bispinor $\tilde{\eta}_{s}$ as
\begin{equation}
\eta_{s}=\left(
\begin{array}{c}
0 \\
\tilde{\eta}_{s}
\end{array}
\right) \,.
\end{equation}
Then it is interesting to notice that the local 3-sphere admits constant spinor solutions.
This is not possible on $\mathbb{S}^3 $ for spinors neutral with respect to internal gauge symmetries and on a torsionless sphere,  while constant Killing spinors are allowed when the topology of the space is $\mathbb{RP}^3 $.\footnote{In absence of additional fields ($C=0$, $\mathcal{A}=0$), for each $s$, the bispinor $\tilde{\eta}_s$ is a solution of the three-dimensional Killing spinor equation  $\tilde{D}\tilde{\eta}=\frac{\ii}{2}\,\tilde{e}^{i}\gamma _i\tilde{\eta}$  discussed in \cite{Lu:1998nu}. Using the 3D $\gamma $-matrices duality identity \eqref{id}, it can also be written as $\diff\tilde{\eta}+\frac{i}{2}\,(\tilde{\omega}^{i}-\tilde{e}^{i})\gamma _i\tilde{\eta}=0$. The spinors become constant in $\mathbb{RP}^3 $ space, $\tilde{\eta}=\tilde{\eta}_{0}$, in the frame with  $\tilde{\omega}^{i}=\tilde{e}^{i}$.  On the sphere $\mathbb{S}^3 $, the constant spinors are not solutions of this equation.}

In our case, this result can be generalized and globally well-defined constant spinors obtained in both $\mathbb{S}^3 $ and $\mathbb{RP}^3 $ in the presence of both torsion and gauge symmetry. 

In bicomponent notation, eq.~\eqref{important} becomes
\begin{equation}
\diff\tilde{\eta}=\left[ -\frac{\ii}{2}\,\left( \tilde{\omega}^{i}-C\,\tilde{e}^{i}\right) \gamma _i+\mathcal{A}\right] \,\tilde{\eta}\,,
\end{equation}
where the matrix $\gamma _i$ acts on the spinorial and an anti-Hermitean matrix $\mathcal{A}$ on the group indices of the bispinior. A general solution is  \cite{Howe:1995zm}
\begin{equation}
\tilde{\eta}=\mathcal{P}\,\mathrm{e}^{\int_{\mathcal{C}(y)}\left[ -\frac{\ii}{2}(\tilde{\omega}^{i}-C\tilde{e}^{i})\gamma_i+\mathcal{A}\right] }\tilde{\eta}_{0}\,,
\label{path}
\end{equation}
where  $\mathcal{P}$ denotes path ordering\footnote{ $\mathcal{P}$ orders a composition of paths in 1-parameter representation of $y^m(\tau)$ according to the value of the parameter $\tau$, e.g., $\mathcal{P}(\mathcal{C}\circ \mathcal{C}')=\mathcal{C}'\circ \mathcal{C}$ if $\tau ' <\tau$.} and $\mathcal{C}(y)$ is a path from a given point $y_0 \in \Sigma$ defined by $\tilde{\eta}(y_{0})=\tilde{\eta}_{0}$, to an arbitrary point $y\in \Sigma $.  

Globally well-defined Killing spinors  on $\Sigma$ can be obtained if, after the parallel transport 
along a closed path, the following holonomy condition  on the bosonic connections  is satisfied,
\begin{equation}
\mathcal{P}\,\mathrm{e}^{\oint_{\gamma }\left[ -\frac{\ii}{2}(\tilde{\omega}^{i}-C\tilde{e}^{i})\gamma _i+\mathcal{A}\right] }\eta _{0}=\lambda (\gamma )\eta _{0}\,, \label{loop}
\end{equation}
where  $\gamma$ is a loop connecting $y_0$ and $y$ and back to $y_0$ (it can be obtained by choosing two different paths $\mathcal{C},\mathcal{C}': y_{0}\rightarrow y$, and defining $\gamma =\mathcal{C}^{-1}\circ \mathcal{C}'$), and $\lambda (\gamma )=\pm 1$ is a sign depending on the choice of the spin structure.\\

{\it The 3-sphere manifold $\Sigma$}:\; 
We have shown that, when the spinor is charged by presence of additional fields, such as torsion and internal fields $\mathcal{A}$, then the holonomy condition \eqref{loop} on the bosonic connections ensures that the spinor is globally well-defined. This is in particular true when the manifold is $\mathbb{S}^3$.

While the case $C=0$, $\mathcal{A}=0$ does not admit the Killing spinors in CS AdS$_5$ supergravity \cite{Canfora:2007xs}, it is interesting that the torsion field itself ($C\neq 0$, $\mathcal{A}=0$) possesses non-trivial holonomy on the 3-sphere. Let us choose the Hopf's coordinates $y^m=(\theta ,\varphi _1 ,\varphi _2 )$ on $\mathbb{S}^3$, where $\theta \in \left[ 0,\frac{\pi }{2}\right]$ and $\varphi _1 ,\varphi _2  \in [0,2\pi]$, such that the vielbein and the connection are given by eqs.~\eqref{tilde e,w} in Appendix \ref{Sphere}. Then a closed path that is a circle $\gamma (y)=\mathbb{S}^1 (\theta )$ leads to the exponential factor $\mathrm{e}^{\ii \pi C\sigma _{3}}$ in \eqref{loop}, using the representation of gamma matrices given in Appendix \ref{Gamma}. As we will see later, $C$ is related to an integer $n_2$ according to $n_2 =C(C^2 -2)$, implying that this closed path will not always give zero due to $C\neq 0$. 
In this case, globally well-defined Killing spinors can be obtained when the holonomy of  the torsion contribution   is canceled by the holonomy of the internal connection $\mathcal{A}$.
The fact that topology and topological charges play an important role in construction of BPS states reminds on the extremal Reissner-Nordstr\"{o}m black hole, where the electric charge is topological and associated with a nontrivial central charge in $\mathcal{N}=2$ supergravity \cite{Gibbons:1982fy,Kallosh:1992ii}
and allows for globally well-defined Killing spinors.\\

{\it The projective 3-space  manifold $\Sigma$}:\;
When the antipodal points on the 3-sphere are identified, the manifold acquires the fixed points that change the topology of the manifold and two different spin structures are possible \cite{Dabrowski:1986en}. 
In this case, in the presence of both torsion and internal gauge field, Killing spinors can be defined when the holonomy of $\tilde{\omega}^{i}\gamma_i$ cancels the one of the other contributions, up to a sign, depending on the choice of the spin structure.
 
In $\mathbb{RP}^3$ space, the spin connection and vielbein can be chosen to globally obey  \cite{Canfora:2007xs}
\begin{equation}
\tilde{\omega}^{i}=\tilde{e}^{i}\,,  \label{e=w}
\end{equation}
as a consequence of the fact that isometries of the 3-sphere are locally described by $SO(4)\simeq SO(3)_{+}\times SO(3)_{-}$, which allows to set consistently one of two independent connections, say $SO(3)_{+}$, to zero. This parametrization describes the parallelized $\mathbb{RP}^3$, where the 1-form $\tilde e^i=\tilde\omega^i$ satisfies the Maurer-Cartan equation of the $SO(3)_-$ algebra: 
\begin{equation}
\left\{
\begin{array}{ll}
\tilde{R}^{i} & =\diff\tilde{\omega}^{i}-\frac{1}{2}\,\epsilon _{\ jk}^{i}\,\tilde{\omega}^j\tilde{\omega}^k=\frac{1}{2}\,\epsilon _{\ jk}^{i}\tilde{e}^j\tilde{e}^k\,, \medskip\\
\tilde{T}^{i} & =\diff\tilde{e}^{i}-\epsilon _{\ jk}^{i}\,\tilde{\omega}^j\tilde{e}^k=0\,,
\end{array}
\right. \quad \Leftrightarrow \quad \diff\tilde{e}^{i}-\epsilon _{\ jk}^{i}\,
\tilde{e}^j\tilde{e}^k=0\,.  \label{MC for e}
\end{equation}
\\
To parameterise the projective 3-space, we use the Euler angles adapted to parallelized $\mathbb{RP}^3$,  $y^m=(\psi ,\theta ,\varphi )$, with $\psi,\varphi \in [0,+2\pi]$ and $\theta \in [0,\pi]$. The vielbein is then given by the expression 
\eqref{e_RP3}.  More details are available in Appendix \ref{RProjective}.

\subsection{\texorpdfstring{$\Xi$}{Xi}-neutral BPS black hole} \label{Neutral}

When the black hole is $\Xi $-neutral, then the mass parameter is always positive, $\mu =C^2 \geq 0$, and the $U(1)_q $-neutral solution of Subsection \ref{Xi-neutral} with the 3-sphere radius rescaled to 1 through \eqref{rescaling}, is given by 
\begin{eqnarray}
\varphi  &=&C\;\neq \;0\,,\qquad N=1\,,  \notag \\
f &=&-\nu \;=\;\sqrt{1+\frac{r^2}{\ell^2}-C^2}\,,  \notag \\
\mathcal{F} &\mathcal{=}&\frac{1}{2}\,\mathcal{F}_{tr}\,\diff t\wedge \diff r+\frac 12\,\mathcal{F}_{rm}\,\diff r\wedge \diff y^m\,.  \label{non-charged}
\end{eqnarray}
Formally, this solution can be obtained from the $\Xi$-charged one \eqref{Xi-charged} by taking the limit $\Xi \to 0$, that is, by setting $\mu=C^2$, even if it belongs to a completely independent branch.

Proceeding the same as in the previous subsection, we break the internal symmetry as $SU(4)\to SU(2)_{D} \times U(1)_c$ and we find that the configuration can be a BPS state only in the extremal case
\begin{equation}
p^2=1-C^2=0\,, \qquad \tilde\Xi=0\,,
\end{equation}
that describes a pure gauge non-Abelian field,
\begin{equation}
 \mathcal{F}=0\,,\quad \mathcal{A}=g\diff g^{-1}+ \diff \Omega(y) \mathbf{T}_c\,, \qquad  g \in SU(2)_D\,.
\end{equation}
The solutions satisfying $\diff \mathcal{A}=-\mathcal{A}^2 $ are purely topological (without the bulk degrees of freedom).
Globally well-defined Killing spinors have the form
\begin{equation}
\epsilon =\sqrt{\frac{r}{\ell }}\,\mathcal{P}\,\mathrm{e}^{\int_{\mathcal{C}(y)} \left[ -\frac 12\,\epsilon _i^{\ jk}\,(\tilde{\omega}^{i}-C\tilde{e}^{i})\,\Gamma
_{jk}+\mathcal{A}\right] }\eta _{0}\,,\qquad \Gamma
_1 \eta _{0s}=-\eta _{0s}\,,
\end{equation}
where $\eta _0^{\alpha }$ is a constant spinor and \eqref{e=w} also applies. 
Taking closed paths $\gamma$ leads in general to
\begin{equation}
\mathcal{P}\,\mathrm{e}^{\oint_{\mathcal{\gamma }}\left[ -\frac{1}{2}\,\epsilon _i^{\ jk}(\tilde{\omega}^{i} -C\tilde{e}^{i})\Gamma _{jk}+\mathcal{A}\right] }\eta _{0}=\lambda (\gamma )\eta _{0}\,,
\end{equation}
with $\lambda=\pm1$.

Note that, since $\mathbf{\hat{F}}\eta _{0}=\mathcal{F}\eta _{0}$ is trivially satisfied on $\Sigma $ because the corresponding field strengths have $\Xi=\tilde{\Xi}=0$, there is no need to further project the spinor and we have
\begin{eqnarray}
\mathbf{A}_{\mathrm{BPS}} &=&\frac{r\diff t}{\ell^2}\,\left( \mathbf{J}_{01}+\mathbf{P}_{0}\right) +\frac{\diff r}{\ell f}\,\mathbf{P}_1 +\frac{r}{\ell }\,\tilde{e}^{i}\left( \mathbf{P}_i-\mathbf{J}_{1i}\right) +
(\tilde{\omega}^{i}-C\tilde{e}^{i})\, (\mathbf{J}_i +\mathbf{T}_i)+ \diff \Omega \mathbf{T}_c\,,  \notag \\
\mathbf{F}_{\mathrm{BPS}} &=&\frac{rC}{\ell }\,\,\epsilon _{ij}^{\ \ k}\, \tilde{e}^i\tilde{e}^j\,\left( \mathbf{P}_k-\mathbf{J}_{1k}\right)\,, \label{BPS-neutral}
\end{eqnarray}
that is a $\frac{1}{2}$-BPS state. In the non-BPS case, it becomes
\begin{eqnarray}
\mathbf{A} &=&\left(\frac{r}{\ell }\, \mathbf{J}_{01}+f\,\mathbf{P}_{0}\right) \frac{\diff t}{\ell }+\frac{\diff r}{\ell f}\,\mathbf{P}_1 +\tilde{e}^{i}
\left( \frac{r}{\ell }\,\mathbf{P}_i-f\,\mathbf{J}_{1i}\right) +(\tilde{\omega}^{i}-C\tilde{e}^{i})\, \mathbf{J}_i +\mathcal{A}^i \mathbf{T}_i
+ \diff \Omega \mathbf{T}_c\,, \notag \\
\mathbf{F} &=&C\,\epsilon _{ij}^{\ \ k}\,\tilde{e}^i\tilde{e}^j\,\left(\frac{r}{\ell }\,\mathbf{P}_k-f\,\mathbf{J}_{1k}\right),  \label{neutral}
\end{eqnarray}
where the internal field $\mathcal{A}^i=\tilde{\omega}^{i}-B \,\tilde{e}^{i}$ has the constant amplitude $B$. It is a pure gauge field when $\tilde{\Xi}=1-B ^2=0$, corresponding to two different configurations $B =\mp 1$ with the gauge field $\mathcal{A}_{\pm}^i=\tilde{\omega}^{i}\pm \tilde{e}^{i}$. Interestingly, on $\mathbb{RP}^3$, it is explicit that one of these fields  identically vanishes ($\mathcal{A}_{-}^i=0$), while the other one ($\mathcal{A}_{+}^i$) has nontrivial topology.

Note that the above expression, with two independent parameters $B$ and $C$, is different from the $\Xi $-charged solution \eqref{Laura solution} that depends on three independent parameters $\mu $, $B$ and $C$.


\section{Global properties of the solutions}

In this section we study physical properties of the solutions, i.e., their charges. Apart from Noether charges which are consequence of continuous symmetries in a theory, there are also topological charges which arise due to topological properties of the solutions defined on spacetime manifold.

\subsection{Topological properties \label{Winding-3sphere}}

In this subsection we focus on the topological properties of the solutions \eqref{Xi-charged} which, in the limit $\Xi =\tilde{\Xi}=0$ and $\mu=C^2 $, also include the $\Xi $-neutral ones \eqref{non-charged}. More precisely, we analyze a static configuration $\mathbf{A|}_{\Gamma }$ on the four-dimensional spatial manifold $\Gamma $, whose boundary at the infinity,
$\partial \Gamma =\Sigma$, is a compact topological space of  unit radius. We will discuss the cases $\mathbb{S}^3 $ and $\mathbb{S}^3 /\mathbb{Z}_2 =\mathbb{RP}^3 $.

Restricted to $t,r=const$, the solution \eqref{Laura solution} has non-trivial topology on $\Gamma $ and it takes the form
\begin{eqnarray}
\mathbf{A|}_{\Sigma } &=&\tilde{e}^{i}\left( \frac{r}{\ell }\,\mathbf{P}_i-f\,
\mathbf{J}_{1i}\right) +\hat{\mathcal{A}}^{i}\mathbf{J}_i+\mathcal{A}^{i}\mathbf{T}_i +\diff \Omega\, \mathbf{T}_c\,,  \notag \\
\mathbf{F|}_{\Sigma } &=&C\,\epsilon _{ij}^{\ \ k}\,\tilde{e}^i\tilde{e}^j\left( \frac{r}{\ell }\,\mathbf{P}_k-f\,\mathbf{J}_{1k}\right) +
\frac{1}{2}\,\epsilon _{ij}^{\ \ k}\tilde{e}^{i}\tilde{e}^j\,\left( \Xi\,
\mathbf{J}_{k}+\tilde{\Xi}\,\mathbf{T}_{k}\right)\,,  \label{solutionSigma}
\end{eqnarray}
where at the end we also have to take the boundary limit, $r\to \infty $. The  $\widehat{SU}(2)_D$-part of the AdS field $\hat{\mathcal{A}}^i$, 
with the axial torsion $C$, has the form
\begin{equation}
\hat{\mathcal{A}}^{i}=\tilde{\omega}^i-C\,\tilde{e}^i\,, \label{e=e}
\end{equation}
and the internal $SU(2)_{D}$ field $\mathcal{A}^{i}$, with the amplitude $B$, is given by 
\begin{equation}
\mathcal{A}^i=\tilde{\omega}^{i}-B \,\tilde{e}^{i}\,.   \label{A=A}
\end{equation}
The torsional soliton $\hat{\mathcal{A}}^{i}$ and the internal symmetry soliton  $\mathcal{A}^{i}$ have the same angular dependence and different amplitudes, determined by the integration constants $C$ and $B$.

The AdS field, in addition, has the term along the generators $\frac{r}{\ell }\,\mathbf{P}_i-f\,\mathbf{J}_{1i}$ that does not belong to the $\widehat{SU}(2)_D$ sector, which is a reason why $\hat{\mathcal{F}}^{i}\neq \mathcal{F}^{i}$ even in the extremal case. Namely,  the internal field strength for $SU(2)_{D}$,
\begin{equation}
\mathcal{F}^{i}=\left(\diff\mathcal{A}+\,\mathcal{A}\wedge \mathcal{A}\right) ^{i}=\diff\mathcal{A}^{i}-\frac 12 \, \epsilon _{\ jk}^{i}\,\mathcal{A}^j\wedge \mathcal{A}^k\,, \label{defF}
\end{equation}
has components
\begin{eqnarray}
\mathcal{F}^i &=&\frac 12 \,\tilde{\Xi}\,\epsilon^{ijk} \tilde{e}_j \tilde{e}_k \,,\qquad \tilde{\Xi}=1-B ^2\,,
\end{eqnarray}
while the $\widehat{SU}(2)_D$ field strength is also defined by \eqref{defF} for the field $\hat{\mathcal{A}}^{i}$, leading to
\begin{eqnarray}
\mathcal{\hat{F}}^{i}&=&
\tilde{R}^{i}-C\,\tilde{T}^{i}-\frac{C^2 }{2}\,\epsilon _{\ jk}^{i}\,\tilde{e}^j\wedge \tilde{e}^k
=\frac{1}{2}\,(1-C^2) \,\epsilon _{\ jk}^{i}\tilde{e}^j\tilde{e}^k\,,
\end{eqnarray}
after using \eqref{RT of S3}.  Thus, we can write
\begin{eqnarray}
\mathcal{\hat{F}}^{i} &=&\frac{1}{2}\,\left( 1-C^2 \right) \,\epsilon _{\ jk}^{i}\tilde{e}^j\tilde{e}^k\neq \frac{1}{2}\,\Xi \,\epsilon _{\ jk}^{i}\tilde{e}^j\tilde{e}^k\,, \notag \\
\mathcal{F}^{i} &=&\frac 12\,(1-B ^2 ) \,\epsilon _{\ jk}^{i}\tilde{e}^j\tilde{e}^k=\frac{1}{2}\,\tilde{\Xi}\,\epsilon _{\ jk}^{i}\tilde{e}^j\tilde{e}^k\,, \label{FF}
\end{eqnarray}
where we point out again that, above, only a part of the $\Xi$-charge appears in the AdS field strength because $\hat{\mathcal{F}}^{i}$ is the $\widehat{SU}(2)_D$ component of the AdS field $\hat{\mathcal{A}}^{i}$, and not the $\widehat{SU}(2)_D$ field strength associated to the full AdS field.\footnote{To obtain the usual constant $\Xi =\mu -C^2 $ in the AdS field strength starting from \eqref{FF}, one has to add a contribution from the non-$\widehat{SU}(2)_D$ term stemming from $[\mathbf{P}_i,\mathbf{P}_j]=\mathbf{J}_{ij}=-[\mathbf{J}_{1i},\mathbf{J}_{1j}]$.}  Note that $\hat{\mathcal{A}}^{i}$ and $\mathcal{A}^{i}$ are pure gauge solutions when $C=\pm 1$, $B =\pm 1$, but to determine whether they corresponds to trivial topology with identically vanishing gauge fields, or describe topological solitons, we have to compute their topological charges, which also depend on the global properties of $\Sigma$. 


The above static solution possesses topological charges obtained from the topological current $J_{\mathrm{top}}$ which satisfies $\left\langle \mathbf{F}^{2}\right\rangle |_{\Gamma }=\diff(^*J_{\mathrm{top}})$ locally, and it is also conserved. Globally, it is associated to the Pontryagin topological invariant that takes values on $\partial \Gamma =\Sigma $, whose corresponding charge  is known as the Pontryagin number. It is worth of pointing out that
the topological charge does not have an origin in a symmetry of the action, so the charge exists if the global properties of the solution are such that it is non-vanishing. In our case, as $SU(2,2|4)$ is broken to $SO(2,4)\times SU(2)_{D}\times U(1)_c $, there are several nontrivial topological charges, not necessarily independent.

First we define the total Pontryagin number computed in $SU(2,2|4)$, which is a real quantity ($P\in \mathbb{R}$) \cite{Jackiw:1977hi} 
\begin{equation}
SU(2,2,|4):\quad P=\Omega _{\Sigma }\int\limits_{\Gamma }\left\langle
\mathbf{F}^{2}\right\rangle =\Omega _{\Sigma }\int\limits_{\Sigma
}\left\langle \mathbf{AF}-\frac{1}{3}\,\mathbf{A}^{3}\right\rangle \,.
\label{P}
\end{equation}
Here $\Omega_{\Sigma }$ is a normalization constant that depends on the topology of $\Sigma$, and which will be determined later.
The supertrace $\langle \cdots \rangle$ is defined using the Cartan-Killing metric, $g_{MN}=\left\langle \mathbf{G}_m\mathbf{G}_{N}\right\rangle $ given in \eqref{CK(N)} in Appendix \ref{Representation}.

On the other hand, due to the non-vanishing third homotopy group $\pi_{3}(SU(2))=\pi_{3}(SO(3))=\mathbb{Z}$, the Pontryagin indices $n_1$ and $n_2$ associated to the solution \eqref{solutionSigma} restricted to  subgroups $SU(2)_{D}$ and $\widehat{SU}(2)_{D}$, respectively, are integers of the form
\begin{eqnarray}
SU(2)_{D} &:&\quad n_1 =\Omega _{\Sigma }\int\limits_{\Gamma }\left\langle \mathcal{F}^2 \right\rangle =\Omega _{\Sigma }\int\limits_{\Sigma}\left\langle \mathcal{AF}-\frac{1}{3}\,\mathcal{A}^3 \right\rangle ,  \notag \\
\widehat{SU}(2)_{D} &:&\quad n_2 =\Omega _{\Sigma } \int\limits_{\Gamma }\left\langle \mathcal{\hat{F}}^2 \right\rangle =\Omega _{\Sigma }\int\limits_{\Sigma}\left\langle \hat{\mathcal{A}}\mathcal{\hat{F}}-\frac{1}{3}\,\hat{\mathcal{A}}^3 \right\rangle ,\label{n1n2}
\end{eqnarray}
when properly normalized. They describe how many times $SU(2)$ and $\widehat{SU}(2)$ wind around the 3-sphere at the infinity.  Therefore, the Pontryagin index for $\widehat{SU}(2)_{D}\times SU(2)_{D}$ is also an integer, given by the total winding number
\begin{equation}
n=n_1+n_2\in \mathbb{Z}\,,  \label{n}
\end{equation}
because the invariant tensor is such that two terms decouple.

In eq.~\eqref{n1n2}, the normalization becomes relevant because it rescales the numbers $n_1 $ and $n_2 $ to integers. It depends on topology. Similarly as in \cite{Jackiw:1977hi}, the normalization in \eqref{n1n2} has been chosen so that a topologically non-trivial non-Abelian solution which tends to a pure gauge on the boundary ($\mathcal{F}\to 0$, $\mathcal{A}\to g\diff g^{-1}$) has the winding number $|n_1|=1$, and similarly for $n_2 $. A reason is that $SU(2)$ is geometrically a 3-sphere with invariant volume element proportional to $\mathrm{Tr}(g\diff g^{-1}\wedge g\diff g^{-1}\wedge g\diff g^{-1})$. For our solution, the pure gauge corresponds to $\tilde{\Xi}=0$. For details of the proof, see for example eqs.~(3.18)--(3.20) in Section 3.2 of \cite{ToplogicalSolitons}.
The result for the two topologies considered in
this text is
\begin{equation}
\Omega _{\mathbb{S}^3 }=\frac{1}{2\mathrm{Vol}(\mathbb{S}^3 )}\,, \qquad \Omega _{\mathbb{RP}^3 } = \frac{1}{8\mathrm{Vol}(\mathbb{RP}^3 )}\,.   \label{normalization}
\end{equation}
A difference with respect to Yang-Mills theory on the flat background considered in \cite{Jackiw:1977hi} is that, here, the dynamics is governed by the CS action which allows also for non pure-gauge geometries with finite energy in AdS space possessing the axial torsion.

For completeness, we also define the Pontryagin number for $SO(2,4)$,
\begin{equation}
SO(2,4): \quad \hat{P}=\Omega _{\Sigma }\int\limits_{\Gamma}\left\langle \mathbf{\hat{F}}^2 \right\rangle =\Omega _{\Sigma }\int\limits_{\Sigma}\left\langle \mathbf{\hat{A}\hat{F}}-\frac{1}{3}\,\mathbf{\hat{A}}^3 \right\rangle \,,  \label{hatP}
\end{equation}
which is also a real quantity (not integer).

For evaluation of conserved quantities, we need a coordinate system that parametrizes the manifold $\Sigma $. For the 3-sphere, the Hopf's coordinate frame $y^m=(\theta ,\varphi _1 ,\varphi _2 )$ with $\theta
\in \left[ 0,\frac{\pi }{2}\right] $ and $\varphi _1 ,\varphi _2 \in \left[
0,2\pi \right] $ has been introduced in Appendix \ref{Sphere}, whose vielbein and spin connection are given by eqs.~\eqref{tilde e,w}. Similarly, the projective 3-plane, due to identification of its points, satisfies the condition $\tilde{\omega}^{i}=\tilde{e}^{i}$, and the manifold can be described by the Euler coordinates $y^m=(\psi ,\theta ,\varphi )$, where $\psi ,\varphi \in \left[ 0,2\pi \right] $ and $\theta \in \left[ 0,\pi \right] $. In this case, the vielbein has the form \eqref{e_RP3}.

The constant $\Omega _{\Sigma }$ depends on the volume
\begin{equation}
\mathrm{Vol}(\Sigma)=\frac{1}{3!} \int \epsilon_{ijk}\,\tilde{e}^i\wedge \tilde{e}^j\wedge \tilde{e}^k=\int \diff^3 y\,\tilde{e}\,. \label{Vol_RP3}
\end{equation}
Using the vielbein \eqref{tilde e,w}, the volume of the 3-sphere is, as usual, $\mathrm{Vol}(\mathbb{S}^3)=2\pi^2$. On the other hand, the volume of the projective space depends on the choice of the geodesic length on the manifold, but it is always bounded from above \cite{Chavel-Chern}. For the metric choice given in Appendix \ref{RProjective}, it becomes a half of the volume of a 3-sphere, $\mathrm{Vol}(\mathbb{RP}^3)=\frac 12 \mathrm{Vol}(\mathbb{S}^3)=\pi^2$, as shown in eq.~\eqref{VolumeRP3}.

As an illustration, we write explicitly the form of the internal soliton $\mathcal{A}^{i}$ (and the same holds for the axial soliton $\hat{\mathcal{A}}^{i}$, with the replacement $B \rightarrow C$)  for two topologies. On the
3-sphere, it has the form (see eq.~\eqref{tilde e,w})
\begin{equation}
\mathbb{S}^3 :\quad \left\{
\begin{array}{ll}
\mathcal{A}^2  & =-B \,\diff\theta \,, \\
\mathcal{A}^3  & =-\sin \theta \,(B \,\diff\varphi _1 +\diff \varphi _2 ) \,, \\
\mathcal{A}^{4} & =-\cos \theta \,(B \,\diff \varphi _2 +\diff \varphi _1 )\,.
\end{array}
\right.
\end{equation}
The integration constant $B $ determines the amplitude of oscillation of particular components of $\mathcal{A}_m^{i}$. The fields $\hat{\mathcal{A}}^{i}$, $\mathcal{A}^{i}$ become degenerate when $C=\pm 1$, $B =\pm 1$ because then they depend on two coordinates only $(\theta ,\varphi _1 \pm \varphi _2 )$. These values of charges correspond to the pure gauge
configurations.

On the projective 3-plane, the internal soliton  reads (see eq.~\eqref{e_RP3})
\begin{equation}
\mathbb{RP}^3 :\quad \left\{
\begin{array}{ll}
\mathcal{A}^2  & =\frac{1-B }{2}\,\left( \sin \theta \,\diff \varphi -\sin
\varphi \cos \theta \,\diff \psi \right) \,, \medskip \\
\mathcal{A}^3  & =-\frac{1-B }{2}\,\left( \cos \theta \,\diff \varphi
+\sin \varphi \sin \theta \,\diff \psi \right) \,,\medskip  \\
\mathcal{A}^{4} & =\frac{1-B }{2}\,\left( \diff \theta +\cos \varphi \,\diff \psi \right) \,.
\end{array}
\right.   \label{nonAbelian A}
\end{equation}

To evaluate the topological charges $n_1$, $n_2$, $\hat{P}$  and $P$, we use the traces constructed in Appendix \ref{SU(4)}. In our case  $\mathbf{A}=\mathbf{\hat{A}} +\mathcal{A}$ along the generators $\mathbf{G}_m =\{\mathbf{J}_{AB},\mathbf{T}_{IJ}\}$, and the AdS and internal symmetry contributions decouple because, as shown in Appendix \ref{Representation} and Appendix \ref{SU(4)} (see eqs.~\eqref{CK(N)} and \eqref{CK-SU(4)})
\begin{eqnarray}
\left\langle \mathbf{J}_{AB}\mathbf{J}_{CD}\right\rangle  &=&\frac{1}{4}\, \mathrm{Tr}(\Gamma _{AB}\Gamma _{CD})=-\eta _{[AB][CD]}\,,  \notag \\
\left\langle \mathbf{T}_{IJ}\mathbf{T}_{KL}\right\rangle  &=&-\mathrm{Tr}\left( \tau _{IJ}\tau _{KL}\right) =\delta _{[IJ][KL]}\,,  \notag \\
\left\langle \mathbf{J}_{AB}\mathbf{T}_{IJ}\right\rangle  &=&0\,,
\label{CKmetric}
\end{eqnarray}
that implies
\begin{equation}
P=\hat{P}+n_1\,.
\end{equation}
We observe that two $SU(2)$ subgroups, generated by $\mathbf{T}_i$ and $\mathbf{J}_i$, have the Cartan-Killing metrics of the opposite sign as a consequence of the supersymmetry, namely 
\begin{equation}
\left\langle \mathbf{T}_i\mathbf{T}_j\right\rangle =\delta _{ij}\,,\qquad \left\langle \mathbf{J}_i\mathbf{J}_j\right\rangle =-\delta _{ij}\,,
\end{equation}
in agreement with \eqref{SU(2)all}. 

\subparagraph{$SU(2)_{D}$ soliton.}

To understand better the geometry of the full solution, let us focus first on the internal non-Abelian sector, by decoupling the gravitational sector by means of $\mu=0$ and $C=0$. This is possible by looking at the fixed gravitational background $\mathbf{\hat{A}}$ corresponding to the AdS space in spherical foliation, where $t,r=const$ is locally the 3-sphere that satisfies $\mathbf{\hat{F}}=0$. The gravitational part then vanishes both in the action and in the equations of motion. The $\mathcal{A}^{i}(y)$ field is a solitonic solution of five-dimensional CS theory for $SU(2)_D\times U(1)_c$ gauge group, defined on the manifold $\mathbb{R}\times \Gamma $, with the $U(1)_q$ field being zero on-shell. 
The $U(1)_c$ field does not contribute to the Pontryagin number because $\pi_1(\mathbb{S}^3)=0$ and, as respect to the projective space, although its fundamental group is non-trivial, $\pi_1(\mathbb{RP}^3)=\mathbb{Z}_2$,it is finite-dimensional, thus it cannot describe a usual winding of a soliton. Direct computation is in agreement with this observation.

Using the auxiliary expressions for the non-Abelian field, 
\begin{eqnarray}
\left\langle \mathcal{AF}\right\rangle  &=&\frac 12 \,\tilde{\Xi}\,\epsilon _{ijk}\, \tilde{e}^{i}\tilde{e}^j\mathcal{A}^k\,,   \notag \\
\left\langle \mathcal{A}^3 \right\rangle  &=&\frac{1}{2}\,\left\langle
\left[ \mathcal{A},\mathcal{A}\right] \mathcal{A}\right\rangle =-\frac{1}{2}\,\epsilon _{ijk}\mathcal{A}^{i}\mathcal{A}^j\mathcal{A}^k\,,
\end{eqnarray}
the integrals \eqref{eee} for the sphere and \eqref{Vol_RP3} for the projective plane, we arrive to the result for a non-Abelian soliton 
\begin{equation}
n_1 (\mathbb{S}^3 )=B (2-B ^2)\,,\qquad n_1 (\mathbb{RP}^3 )=\frac 14 \, (1-B )^2(2+B )\,,  \label{non-Abelian P}
\end{equation}
where the one for the 3-sphere has also been computed in \cite{Sochifi}.
Note that the winding number of the sphere is an odd function under the reflection transformation $B \rightarrow -B $, which only changes a direction of the winding of the soliton around it, $n_1 \rightarrow -n_1 $. The identification of the antipodal points on the 3-sphere breaks this reflection symmetry. Instead,  the transformation $B \rightarrow -B $ on $\mathbb{RP}^3 $ produces a shift $n_1\rightarrow -n_1+1$. The constant $\tilde{\Xi}=1-B ^2$, however, remains invariant under the reflection in both cases.

The non-Abelian solitons are classified by the Pontryagin index $n_1$, which relates the topological charge $n_1$ and the soliton of  amplitude $B$. Because the topological charge is an integer, this restricts the allowed
values of $B$. Explicitly, we find that $B$ is quantized by $n_1$ in the following way.\medskip

On the sphere, we get:
\begin{equation*}
\begin{array}{ll}
n_1 (\mathbb{S}^3 )=0\,,\qquad  & B =0\,,\quad \;\; \, B =\pm \sqrt{2}\,,\smallskip
\\
n_1 (\mathbb{S}^3 )= 1\,, & B = 1\,,\;\; \,  \quad B =- \dfrac{1}{2}\pm \dfrac{\sqrt{5}}{2}\,,\smallskip  \\
n_1 (\mathbb{S}^3 )=-1\,, & B =-1\,,\quad B =\dfrac{1}{2}\pm \dfrac{\sqrt{5}}{2}\,,\smallskip  \\
|n_1 (\mathbb{S}^3 )|\geq 2\,, & 
B =\dfrac{\Delta ^{1/3}}{6}+4\; \Delta
^{-1/3},
\end{array}
\end{equation*}
where $\Delta = 12(-9n_1 +\sqrt{81 n_1^2-96})$. 
For example, the topological sector with $\tilde{\Xi}=0$ contains pure gauge solutions $B =\pm 1$ classified by the winding numbers $n_1 =\pm 1$. As another particular case, the topological sector with $\tilde{\Xi}\neq 0$ and $B =0$, $\pm \sqrt{2}$ has trivial topology, $n_1 =0$. In a generic case, the sector with $\tilde{\Xi}\neq 0$ and nontrivial topology classified by the integer $|n_1 |\geq 2$, corresponds to non-pure gauge solutions.\medskip

On the projective plane, we find:
\begin{equation*}
\begin{array}{ll}
n_1(\mathbb{RP}^3 )=0\,, & B =1\,,\; \; \,\quad B =-2\,,\smallskip \\
n_1(\mathbb{RP}^3 )=1\,, & B =-1\,,\quad B =2\,,\smallskip \\
n_1(\mathbb{RP}^3 )\neq 0\,,1\,,\quad  & B =\tilde{\Delta}^{1/3}+\tilde{\Delta} ^{-1/3},
\end{array}
\end{equation*}
where $\tilde{\Delta}=-1+2n_1+2\sqrt{n_1^2-n_1}$. For positive index $n_1 \geq 2$, it can also be written as $\tilde{\Delta}=\left(\sqrt{n_1}+\sqrt{n_1-1}\right)^2$.
The absence of the reflection symmetry allows now to have  the same $|n_1 |\leq 1$ for two different solitonic configurations. When $|n_1 |\geq 2$, both $\mathbb{RP}^3 $ and $\mathbb{S}^3 $ have a unique soliton for a given topology.

The topological sector with $\tilde{\Xi}=0$ contains only pure gauge solutions where the non-Abelian field is $\mathcal{A}=g\diff g^{-1}$. In that case,  when $B =1$, the field is topologically trivial because $\mathcal{A}=0$ and $g=1$. However,  when $B =-1$, it becomes topologically nontrivial. 

When $\tilde{\Xi}\neq 0$, there is always only one soliton with given $B$ and nontrivial topology characterized by the integer $n_1\neq 0,1$.

Since the topological charge is conserved and the energy is finite, the evolution cannot change a solution from one topological sector to another.

\subparagraph{$SU(2)_D$ soliton coupled to the black hole.}

Consider now an internal symmetry soliton of amplitude $B$ in the spacetime with axial torsion $C$ and a black hole with the mass parameter $\mu $. The $U(1)_c $ field again does not contribute to the topological charges, with the same argument as discussed above. In this case, there is the quantized topological charge defined in the second line of eq.~\eqref{n1n2}, the Pontryagin index associated to $\widehat{SU}(2)_D$. It is computed similarly to \eqref{non-Abelian P}, with the result
\begin{equation}
n_2 (\mathbb{S}^3 )=-C\left( 2-C^2 \right) \,,\qquad n_2 (\mathbb{RP}^3 )=-\frac 14 \,  (1-C)^2 (2+C)\,.  
\label{hat P}
\end{equation}
Again, the sphere index only changes the sign under a discrete symmetry $C\to -C$, that is $n_2 \to -n_2$, whereas the projective hyperplane index has an additional shift, $n_2 \to -n_2-1$. It describes an independent $\widehat{SU}(2)_D$ soliton of the amplitude $C$ produced by the axial torsion. Thus, there are two solitons, whose charges $B$ and $C$ are related to the topological numbers $n_1$ and $n_2$, respectively, and  with the total $\widehat{SU}(2)_D\times SU(2)_{D}$ Pontryagin index given by \eqref{n},
\begin{equation}
n=n_1 +n_2 =\left\{
\begin{array}{ll}
-C(2-C^2 )+B (2-B ^2)\,, & \text{on } \mathbb{S}^3 \,,\medskip  \\
-\frac 14 \, (1-C)^2 (2+C)+\frac 14 \, (1-B )^2(2+B )\,, & \text{on }\mathbb{RP}^3 \,.
\end{array}
\right.   \label{Pontryagin n}
\end{equation}
When the BPS conditions \eqref{BPSstate} are fulfilled, the two solitons satisfy $B =C$, becoming a soliton and anti-soliton system that unwinds, summing up to a topologically trivial configuration from the point of view of gauge fields,
\begin{equation}
n_{\mathrm{BPS}}=0\,.
\end{equation}
The above relation can be seen as a  `balance of strengths' between the two oppositely charged solitons, and it reflects the  `balance of forces' associated with gravity and internal symmetry which is usual for BPS states in standard supergravity.
However, the solution does not possess only $\widehat{SU}(2)_D\times SU(2)_{D}$ isometries but the extended $SO(2,4)\times SU(2)_{D}$ ones, that means that the total topological charge acquires an additional contribution coming from the AdS field. Using the auxiliary expressions for the AdS field on $\Sigma $,
\begin{eqnarray}
\left\langle \mathbf{\hat{A}\hat{F}}\right\rangle  &=&  C p^2 \epsilon _{ijk}\tilde{e}^{i}\tilde{e}^j\tilde{e}^k-\frac{\Xi }{2}\,\epsilon _{ijk}\,\tilde{e}^{i}\tilde{e}^j\hat{\mathcal{A}}^k,  \notag \\
\mathbf{\hat{A}}^2  &=&\frac{1}{2}\,\epsilon _{ij}^{\ \ k}\,\left( p^2 \tilde{e}^{i}\tilde{e}^j-\hat{\mathcal{A}}^{i}\hat{\mathcal{A}}^j\right) \mathbf{J}_k-\epsilon _{ij}^{\ \ k}\,\tilde{e}^{i}\hat{\mathcal{A}}^j\left( \frac{r}{\ell }\,\mathbf{P}_{k}-f\,\mathbf{J}_{1k}\right) \,,  \notag \\
\left\langle \mathbf{\hat{A}}^3 \right\rangle  &=&-\frac{3}{2}\,p^2 \epsilon_{ijk}\,\tilde{e}^{i}\tilde{e}^j\hat{\mathcal{A}}^k+\frac{1}{2}\,\epsilon _{ijk}\hat{\mathcal{A}}^{i}\hat{\mathcal{A}}^j\hat{\mathcal{A}}^k\,,
\end{eqnarray}
where all divergences cancel out as the $r$-dependence always appears in the combination $\frac{r^2 }{\ell ^2 }-f^2 =p^2 $, eq.~\eqref{hatP} yields 
\begin{equation}\label{psr}
\hat{P}(\mathbb{S}^3 )=n_2 -\frac{3}{2}\,Cp^2 \,,\qquad \hat{P}(\mathbb{RP}^3 )=n_2 +\frac{3}{4}\,Cp^2 \,.
\end{equation}
The total topological charge of the solution is 
\begin{equation}
P(\mathbb{S}^3 )=n_1 +\hat{P}=n-\frac{3}{2}\,Cp^2 \,,\qquad P(\mathbb{RP}^3 )=n+\frac{3}{4}\,Cp^2 \,.  
\label{total P}
\end{equation}
In the above expression, the $SU(2)$ charges are quantized by $n$ and the mass parameter $p^2=\mu-1$ is not. This result is similar to the one in standard gravity where $U(1)$ charges are also quantized and the mass is not.

The expressions \eqref{total P} also shows that the Pontryagin charge $P$ is quantized only in the extremal limit, otherwise it is a real quantity.

The BPS state associated with the full $SU(2,2|4)$ CS gauge connection, with the parameters satisfying \eqref{BPSstate}, is topologically trivial, 
\begin{equation}
P_{\mathrm{BPS}}=0\,.
\end{equation}
Since $P$ is a conserved quantity, the above relation reflects an exact balancing of the contribution from the topological charge in the Pontryagin term for the AdS and from the internal symmetry parts,
\begin{equation}
  \mu=1\,, \qquad n_1=-n_2\,,  
\end{equation}
in the CS AdS action at fixed time. This result is non trivial because, as we saw before, the $SO(2,4)$ gauge fields in CS AdS supergravity are identified with the {\it geometry} of spacetime, that is, the spin connection and vielbein of the 5D space-time over which the CS action is integrated. Thus, thinking of the obtained result from the point of view of a black hole in AdS space, it can be read as a no-force condition, as is usual in BPS solutions, 
 reflecting the balance of forces between the geometry and the internal symmetry, similarly as in standard gravity.

As we will see later, the BPS state has all conserved charges equal to zero, that is, it is $U(1)_q$-neutral with zero energy (see, for instance, eqs.~\eqref{Ebps} and \eqref{Ebps}). It is expected that the obtained BPS condition would saturate the supercharge algebra bound $\{\mathbf{Q},\mathbf{Q}^{\dag }\}\geq 0$, since a linear combination of the conserved charges appears on the r.h.s. of the inequality.

The results of this subsection also hold for the $\Xi$-neutral black holes after setting $p^2=C^2-1$.

\subsection{Conserved charges \label{CC}}

CS theory is invariant under  two sets of local symmetries: the gauge ones with the local parameters $\Lambda ^{\Lambda }(x)$ and spacetime diffeomorphisms with the parameters $\delta x^{\mu }=\xi ^{\mu }(x)$. Because they are not all independent, the corresponding charges are difficult to find.

In $D=3$, the charges and their algebra have been computed in \cite{Banados:1994tn}. Since it is on-shell satisfied $\mathbf{\Lambda =\pounds }_{\xi }\mathbf{A}$, where $\pounds_\xi$ denotes Lie derivative, treating $\mathbf{\Lambda }$ as the only independent parameter leads to the affine (Kac-Moody) asymptotic charge algebra, whereas treating $\xi$ as an independent parameter gives rise to the asymptotic Virasoro algebra. They both describe symmetries of a two-dimensional boundary, since the Virasoro algebra can be obtained from two Kac-Moody ones through the Sugawara construction.

In odd $D\geq 5$, CS gravity is nonlinear in the curvature tensor, and its number of independent local symmetries  can change depending on the phase space region. It means that its number of degrees of freedom and the number of physical fields also varies throughout the phase space, taking values from zero to some maximal number in so-called generic CS theories. In addition, in CS theories it is difficult to separate first and second class constraints and prove which ones are linearly independent. This makes a computation of conserved charges very difficult and intrinsically dependent on the background. 

There are at least two sets of charges: $Q[\mathbf{\Lambda }]$ associated to gauge symmetries and $H[\xi ]$ associated to spatial diffeomorphisms. In \cite{Olea-PhD,Mora:2004kb}, these charges have been constructed in CS gravity in asymptotically locally AdS spaces using the Noether theorem. On the other hand, thermodynamic mass for torsionless black holes in AdS space has been found in \cite{Crisostomo:2000bb} using Hamiltonian method.

The Nester’s formula \cite{Nester:1991yd,Chen:2015vya} is useful for finding conserved charges when the torsional degrees of freedom are present. In \cite{Cvetkovic:2016ios},  conserved charges of the five-dimensional BTZ black rings in first order formalism have been worked out, and they in special case reduce to CS gravity. However, a general formula applicable to any CS gravity solution remains unknown.

\subparagraph{Bulk degrees of freedom.}

Hamiltonian formalism provides a systematic way to identify independent local symmetries, determine the number of degrees of freedom and compute the charges in CS theory \cite{Banados:1995mq,Banados:1996yj}. It has a particular feature that allows for appearance of \textit{accidental symmetries}, additional to gauge symmetries and diffeomorphisms, which turn physical fields into unphysical ones and, in that way, eliminate physical degrees of freedom from the theory. The quantity that counts accidental symmetries is the symplectic matrix, $\Omega (x)$. In $D=5$, the symplectic matrix is defined by \cite{Banados:1995mq,Banados:1996yj}
\begin{equation}
\Omega _{MN}^{\mu \nu }=\ii \epsilon ^{t \mu \nu \lambda \rho }\left\langle \mathbf{G}_M\mathbf{G}_{N}\mathbf{F}_{ \lambda\rho
} \right\rangle \,, \quad \mu,\nu,\ldots=1,\dots,4\,, \label{symplectic}
\end{equation}
on the four-dimensional spatial section $\Gamma $ corresponding to the slice $t=const.$, and all  components along $t$ vanish, $\Omega _{MN}^{t\mu}=-\Omega _{MN}^{\mu t}\equiv 0$.
In \eqref{symplectic},  $\mathbf{G}_M$ are the generators of the $SU(2,2|4)$ supergroup  introduced in Section 2.  Due to its dependence on $\mathbf{F}(x)$, the rank of $\Omega (x)$  varies throughout the phase space \cite{Saavedra:2000wk}. The zero modes of $\Omega $ contain four always-present spatial diffeomorphisms. Any additional zero mode, if it exists, leads to a new accidental symmetry. As a consequence, CS gravity has different number of degrees of freedom in different sectors of the phase space, depending on the number of accidental symmetries present.
If $\mathfrak{D}$ is the dimension of the gauge group, then \textit{generic} CS theory has the maximal rank of $\Omega $, that is $4\mathfrak{D}$, and minimal number of local symmetries: four spatial diffeomorphisms and zero accidental symmetries. It this case, as we will show later in this section, according to eq.~\eqref{dof}, the number of degrees of freedom is maximal, $\mathfrak{D}-2$. Most of them are torsional degrees of freedom. The time-like diffeomorphism (the one that defines total energy) is not an independents symmetry, as it is on-shell equal to a gauge transformation. An example of a generic 5D CS theory invariant under $G\times U(1)$ has been worked out in \cite{Banados:1995mq,Banados:1996yj}. On the other hand, an example of non-generic 5D spherically symmetric and static CS AdS supergravity with torsion and the accidental symmetry $U(1)\times U(1)\times U(1)$, has been discussed in \cite{Giribet:2014hpa}. It has zero degrees of freedom. Another important example of a non-generic CS theory is a special case of torsion-free Lovelock gravity. Remarkably, a generic sector within this class of non-generic theories has the same number of degrees of freedom as General Relativity \cite{Teitelboim:1987zz}, that is, $D(D-3)/2$ in $D$ dimensions, even when the Einstein-Hilbert term is absent and the gravitational dynamics is described by only one higher-order curvature polynomial \cite{Dadhich:2015ivt}.

Another effect to take into account is when two existing local symmetries become linearly dependent on certain backgrounds, decreasing a number of local symmetries and, thus, producing more degrees of freedom. The quantity that accounts these \textit{regularity conditions} (conditions of linear independence of the symmetries) is the Jacobian $\mathcal{J}(x)$. When the gauge constraints become dependent in some backgrounds, the theory is said to be \textit{irregular} \cite{Chandia:1998uf,Miskovic:2003ex}. An example of irregular CS AdS$_{5}$ gravity based on $SU(4)\times U(1)$ has been discussed in \cite{Chandia:1998uf}. In five dimensions, the Jacobian matrix $\mathcal{J}_{MN}(x)$ of the canonical gauge generators $\mathcal{G}_M(A,\pi )$,  is defined by
\begin{equation}
\mathcal{J}_{MN}=\ii \left\langle \mathbf{G}_M\mathbf{G}_{N}\mathbf{F}\right\rangle \,.  \label{Jacobian}
\end{equation}
The generators, satisfying $\delta \mathcal{G}_M\propto \mathcal{J}_{MN}\,D\delta A^{N}$, are linearly independent if the corresponding Jacobian matrix has maximal rank  on shell, equal to the dimension $\mathfrak{D}$ of the gauge group. Even though the symplectic matrix and the Jacobian matrix components are related by $\Omega_{MN}^{\mu \nu }=\epsilon ^{t \mu \nu \lambda \rho}\mathcal{J}_{MN,\lambda \rho }$, their ranks are independent because they act in different spaces. Indeed, there are examples of four different backgrounds corresponding to different phase space sectors in one CS theory (generic/regular, non-generic/regular, generic/irregular and non-generic/irregular).

In our particular case, $\mathfrak{D}=\dim SU(2,2|4)=63$ but, as shown in Appendix \ref{PhaseSpace}, the fermionic part of the phase space decouples in the backgrounds considered in this text and, because we set all fermions equal to zero in the solution, there are no fermionic  degrees of freedom left. On the other hand, the bosonic gauge fields are associated to the group $SO(2,4)\times SU(2)_{D}\times U(1)_{c}\times U(1)_{q}$ of dimension $\mathfrak{D}_{\mathrm{B}}=20$. From the solutions at our disposal, we do not discuss the non extremal cases when $\Xi =0\neq \tilde{\Xi}$ or $\Xi \neq 0=\tilde{\Xi}$ because they do not possess BPS states and, in addition, we are not interested in the backgrounds of pure gauge solutions (such as $\Xi $-neutral solution \eqref{neutral} with $\tilde\Xi$, $\Xi =0$) because we know that they do not propagate in the bulk. It remains to analyse only the phase space sector of the $\Xi $-charged solutions ($\tilde\Xi$, $\Xi \neq 0$) using the field strength $\mathbf{F}$ given by eq.~\eqref{Laura solution}, which has all components on $\Gamma $. Then a non-trivial sub-matrix of $\mathcal{J}_{MN}$ has components in AdS group, $\mathcal{J}_{[AB][CD]}$, and the interacting terms with the $U(1)_{q}$ field, $\mathcal{J}_{1M}$, where $\mathcal{J}_{11}=0$. In addition, although $SU(4)\simeq SO(6)$ also possesses a rank-three invariant tensor, the solution breaks this symmetry to $SU(2)_{D}\times U(1)_{c}$, in which case the symmetric rank-3 tensor of the internal group identically vanishes, implying that the $SU(4)$ sub-matrix vanishes as well, $\mathcal{J}_{[IJ][KL]}=0$. All the components of the bosonic Jacobian are explicitly written in in Appendix \ref{PhaseSpace}. 

The rank of the Jacobian matrix is determined by examining its zero modes $V^M$, 
 \begin{equation}
\mathcal{J}_{MN}V^{N}=0\,.
\end{equation}
If the vector $V^{M}=(V^{AB},V^{\hat{\imath}\hat{\jmath}},V^{c},V^{1})$ (with 20 components) vanishes, the matrix has maximal rank, i.e., $20$. Each zero mode decreases the rank for one. In Appendix \ref{PhaseSpace}, we showed that there are four non-vanishing components $V^{c},V^{i5}\in \mathbb{R}$, thus the rank of the Jacobian is $\mathrm{rank}(\mathcal{J})=16$.

As regards the $4\mathfrak{D}_{\mathrm{B}}\times 4\mathfrak{D}_{\mathrm{B}}$ symplectic matrix \eqref{symplectic}, its maximal possible rank on $\Gamma$ is $76$ (when 4 spatial diffeomorphisms are subtracted). The zero modes $V_{\mu }^{M}$ on $\Gamma $ have $80$ components and they are non-trivial solutions of
\begin{equation}
\Omega _{MN}^{\mu \nu }V_{\nu }^{N}=0\,,\qquad \mu \in \{r,m\}\,.
\end{equation}
In Appendix \ref{PhaseSpace} it is shown that there are $48$ zero modes $V_r^c$, $V_m^c$, $V^{\hat{\imath}\hat{\jmath},\,k}-V^{[\hat{\imath}\hat{\jmath},\,k]}$, $V_r^{i5}$, $V_m^{05}$,$\,V^{0(ij)_{\mathrm{T}}}$, $V^{1[ij]}$, $V^{1(ij)_{\mathrm{T}}}$, $V_m^{i5}$, $V^{ijk}-V^{[ijk]}$, and the rank of the symplectic matrix is $\mathrm{rank}(\Omega)=32$. 

Knowing the ranks of these two matrices that determine the symmetry structure of any CS  theory, the number of degrees of freedom (d.o.f.) is computed in Dirac's canonical formalism according to the formula
\begin{equation}
\mathrm{d.o.f.}\,=\frac 12 \, \mathrm{rank}(\Omega )-\mathrm{rank}(\mathcal{J}) \,,  \label{dof}
\end{equation}
that will be shown in the next paragraph. In generic and regular theories, both ranks are maximal, that is, $\mathrm{rank}(\Omega )=4(\mathfrak{D}-1)$ and $\mathrm{rank}(\mathcal{J})=\mathfrak{D}$, reproducing the known result $\mathfrak{D}-2$ for the number of degrees of freedom \cite{Banados:1995mq,Banados:1996yj}.

In our case, the solution with $\Xi$, $\tilde\Xi \neq 0$ belongs to the non-generic ($\mathrm{rank}(\Omega )<76$) and irregular ($\mathrm{rank}(\mathcal{J})<20$) sector of CS AdS supergravity, such that
\begin{equation}
\mathrm{d.o.f.}\,=\frac 12 \,32-16=0\,,
\end{equation}
meaning that there are no gauge fields propagating in the bulk geometry around the background of the form \eqref{Laura solution}.

Finally, let us comment that our ansatz on the metric \eqref{metric} and torsion \eqref{torsion}, \eqref{nu_chi} considers only four radial functions, $f(r)$, $N(r)$, $\nu (r)$ and $\varphi (r)$, and the internal non-Abelian gauge field represented by the scalar function $\tilde\varphi(r)$, which becomes constant on-shell. Abelian fields are pure gauge. In the branches discussed in Subsections \ref{Xi-neutral} and \ref{LS}, we have $\nu =0$, leaving four functions $f$, $N$, $\varphi $, $\tilde\varphi$ when $\Xi \neq 0$, or three functions $N$, $\varphi$, $\tilde\varphi$ in the $\Xi$-neutral case, whose radial evolution is determined by given initial conditions in first order formalism (satisfying first order differential equations). In addition, we choose boundary conditions such that $N=1$. In that way, we remain with three ($\Xi \neq 0$) and two ($\Xi =0$) functions in the AdS sector, whose radial evolution is determined by the integration constants $(\mu,B,C)$ and $(B,C)$, respectively. 

However, the constants $B$ and $C$ are quantized by the integers $n_1$ and $n_2$ due to topological considerations, so they are not arbitrary. This leaves the parameter $\mu$ as the only integration constant, which is not enough to describe a full degree of freedom because, in the d.o.f.~count, two integration constants usually describe one degree of freedom.
This result is similar to BTZ black hole in three dimensions, where the mass parameter has topological origin. Furthermore, there is also a similarity with standard supergravity, where the BPS states share the feature that they look like `a half' of a degree of freedom, according to the fact that the BPS conditions are first-order differential equations that, however, imply the second-order field equations of motion. This means that, in our ansatz, we indeed have zero physical degrees of freedom, thus our solutions belong to a topological sector of CS AdS$_5$ supergravity.

\subparagraph{Hamiltonian charge formula.}

In CS AdS supergravity, the $SU(2,2|4)$ charge can be obtained by Hamiltonian formalism. Separating the time and space coordinates, $x^{\mu }=(t,x^{\alpha })\in \mathbb{R}\times \Gamma $, the Hamiltonian form of CS action \eqref{L(A)}, up to a boundary term, is \cite{Banados:1996yj}
\begin{equation}
I=\int\limits_{\mathcal{M}}\diff^{5}x\,\left( \mathcal{L}_M^{\alpha }\dot{A}_{\alpha }^M-A_{t}^M\chi _M\right),
\end{equation}
where the $A_{t}^M$-independent quantities are
\begin{eqnarray}
\mathcal{L}_M^{\alpha } &=&-\frac{k}{3}\,\epsilon ^{\alpha \beta \gamma
\delta }\left( g_{MNK}A_{\beta }^{N}F_{\gamma \delta }^K-\frac{1}{4}\,g_{MNL}\,f_{KS}^{\ \ \ L}\,A_{\beta }^{N}A_{\gamma }^K A_{\delta }^{S}\right) \,,  \notag \\
\chi _M &=&\frac{k}{4}\,\epsilon ^{\alpha \beta \gamma \delta }g_{MNK}F_{\alpha \beta }^{N}F_{\gamma \delta }^K\,.
\end{eqnarray}
Passing to the phase space with  canonical variables ($A_{\alpha }^M,\pi _M^{\alpha }$), where $\pi _M^{t}=0$ and $A_{t}^M$ is a Lagrange multiplier, canonical analysis shows that there are two kinds of constraints (canonical equations of motion that do not involve time derivatives). The first ones are
\begin{equation}
\phi _M^{\alpha }=\pi _M^{\alpha}-\mathcal{L}_M^{\alpha}\,, \qquad [\phi _M^{\alpha}(x),\phi _{N}^{\beta}(x')]=\Omega _{MN}^{\alpha \beta }(x)\, \delta^{(4)}(x-x')\,,
\end{equation} 
where the Poisson brackets are taken at the same time, $t=t'$. Zero modes of $\Omega$ correspond to first class constraints that are accidental symmetries.
For us, the relevant first class constraint is the one that generates gauge symmetry,
\begin{equation}
\mathcal{G}_M=-\chi _M+D_{\alpha }\phi _M^{\alpha }\,, \qquad [\mathcal{G}_M(x),\mathcal{G}_{N}(x')]=f_{MN}^{\ \ \ K}\,\mathcal{G}_{K}(x)\, \delta^{(4)}(x-x')\,.
\end{equation}

With this constraint content at hand, a number of d.o.f. in 5D CS theory is computed in Dirac's canonical formalism in the following way. There are $\mathfrak{D}$ canonical gauge generators $\mathcal{G}_M$ (first class constraints), but only $\mathrm{rank}(\mathcal{J})$ of them are linearly independent.  There are also $4\mathfrak{D}$ constraints $\phi_M^{\alpha }$ on $\Gamma $, and $\mathrm{rank}(\Omega )$ of them are not gauge generators, but they are so-called second class constraints that eliminate redundant components of the fields. The rest, $4\mathfrak{D}-\mathrm{rank}(\Omega )$, are gauge generators (diffeomorphisms and accidental symmetries). In total we have $n_{_i}=\mathrm{rank}(\mathcal{J})+4\mathfrak{D}-\mathrm{rank}(\Omega )$ first class constraints and $n_{_{II}}=\mathrm{rank}(\Omega )$ second class constraints. Among the $4\mathfrak{D}$ gauge fields $A_{\alpha }^M$ on $\Gamma $ ($A_{t}^M$ does not count because it is a Lagrange multiplier), not all correspond to the physical fields. Each local symmetry generator (first class constraint) eliminates one field component, and a nature of second class constraints is such that two of them eliminate only one field component (because a second class constraint can be imagined as first class constraint plus its corresponding gauge fixing function in the phase space). What remains is the number of physical gauge field components $\mathrm{d.o.f.}=4\mathfrak{D}-n_{_i}-\frac{1}{2}n_{_{II}}$, leading to the final formula \eqref{dof}.

We will compute a formula for conserved charges associated to gauge symmetries. The smeared generator,
\begin{equation}
G[\Lambda ]=\int\limits_{\Gamma }\diff ^{4}x\,\Lambda ^M\mathcal{G}_M+Q[\Lambda]\,,
\end{equation}
produces gauge transformations via the Poisson brackets, $\delta _{\Lambda }A_{\alpha }^M=\left[ A_{\alpha }^M,G[\Lambda ]\right]
=-D_{\alpha }\Lambda ^M$.
The boundary term, $Q[\Lambda ]$, has to be chosen so that the generator has well-defined functional derivatives \cite{Regge:1974zd}. 

In order to identify its non-differentiable part, we vary canonical fields, keeping the gauge parameter field-independent in the asymptotic limit ($\delta \Lambda \rightarrow 0$ on $\partial \Gamma =\Sigma$). The bulk terms are constraints and they  vanish on-shell, implying that the smeared generator becomes equal to its boundary term, i.e., the charge $G[\Lambda]=Q[\Lambda]$. Thus, to compute $Q[\Lambda]$, it is enough to focus on the boundary terms that arise from making the above expression free of derivatives acting on $\delta A$ and $\delta \pi $. Up to explicitly written global sign, we obtain on-shell
\begin{equation}
\delta Q[\Lambda ]=\pm \ii\int\limits_{\Gamma }\diff\left(
2k\left\langle \mathbf{\Lambda F}\delta \mathbf{A}\right\rangle
 +\frac{2k}{3}\,\left\langle D\mathbf{\Lambda A}\delta \mathbf{A}\right\rangle \right) .
\end{equation}
Then we apply the Stoke's theorem, noticing that the term containing $D \mathbf{\Lambda }$ also vanishes since $D\mathbf{\Lambda }\rightarrow 0$ on $\Sigma$ (see the conditions on the local parameters, eq.~\eqref{parameters} below).

The result is the on-shell Hamiltonian charge  whose variation is given by the functional
\begin{equation}
\delta Q\left[ \Lambda \right] =2 \ii k \theta \int\limits_{\Sigma}\left\langle \mathbf{\Lambda F}\delta \mathbf{A}\right\rangle \,,  \label{charge}
\end{equation}
where $\mathbf{\Lambda }$ is an asymptotically covariantly constant parameter that does not vary on the boundary, and $\theta $ is a global sign chosen as $\theta = 1$ when $\Xi \neq 0$ and $\theta =-1$ when $\Xi =0$. This choice, $\theta =1-2\delta_{0\,\Xi}$, ensures that the energy, which is always a bounded quantity, becomes bounded from below. The above formula is not integrable in general, i.e., not possible to write as a total variation in general, given  boundary conditions for $\mathbf{A}$.

One way to integrate out \eqref{charge} is to use boundary conditions $\mathbf{F} \to \mathbf{\bar{F}}$, where $\mathbf{\bar{F}}$ is a Lie-algebra valued 2-form that does not vary on $\Sigma$. The background-dependent charge is then solved as $Q\left[ \Lambda; \mathbf{\bar{F}} \right] =2\ii k\theta \int_{\Sigma}\left\langle\mathbf{\Lambda \bar{F}A}\right\rangle $. We can always add an $\mathbf{\bar{A}}$-dependent `integration constant' to this charge. This approach has been adopted in \cite{Miskovic:2006ei,Banados:1995mq}.

The background-dependent method is not suitable when the boundary field depends on the parameters $\Xi, \tilde{\Xi} \neq 0$ because it would introduce background charges. Instead, we will look at a class of solutions where $\delta Q$ becomes integrable.

The parameter $\mathbf{\Lambda }$ is a Lie-algebra valued function that has to be fixed and covariantly constant on the asymptotic boundary, namely
\begin{equation}
\Sigma:\quad D\mathbf{\Lambda }=0\,,\qquad \delta \mathbf{\Lambda
}\to 0\,,  \label{parameters}
\end{equation}
where $D=D(\mathbf{A})$ is the group covariant derivative, $D\mathbf{\Lambda
}=\diff \mathbf{\Lambda }+\left[ \mathbf{A},\mathbf{\Lambda }\right] $. In our case, non-vanishing gauge fields are the bosonic ones and since they all commute, the expression becomes a sum of independent terms, $D\mathbf{\Lambda } =\hat{D}\mathbf{\Lambda}_{\mathrm{AdS}} +\mathcal{D}\mathbf{\Lambda }\,_{SU(4)}+\diff \mathbf{\Lambda }_{U(1)}$. This means that, because of independence of generators, each term has to vanish independently. The same has to be true for each bosonic subgroup in $DD\mathbf{\Lambda }=[\mathbf{F,\Lambda }]=0$.

On the other hand, there is the charge $H[\xi ]$ associated to diffeomorphisms, or more precisely to asymptotic Killing vectors $\xi =\xi^{\mu }\,\partial _{\mu }$ describing isometries of the asymptotic sector. Since in $D>3$ we have in general $\mathbf{F}\neq 0$, which implies $\mathbf{\pounds }_{\xi }\mathbf{A}=D(i_{\xi }\mathbf{A})+i_{\xi }\mathbf{F}\neq D\Lambda $ and two set of charges ($Q\left[ \Lambda \right] $ and $H[\xi ]$) are independent. The exception is the time-like charge that defines the total energy and where $\mathbf{\pounds }_{\xi }\mathbf{A}$ and $D(i_{\xi }\mathbf{A})$ become related on-shell. In that case, the parameter $\mathbf{\Lambda }$ turns field-dependent, still satisfying the conditions \eqref{parameters}. The total energy is $E=H[\partial _t]=Q[\Lambda _0] $, where the last equality holds on-shell and the function  $\Lambda _0(A)$ is yet to be determined.

The solution of eq.~\eqref{parameters} in the $U(1)_q \times SU(2)_D \times U(1)_c$ sector is  
\begin{equation}
\Lambda^1=1\,, \qquad \Lambda^i=0\,,\qquad \Lambda^c=1\,.    \label{Lambda1}
\end{equation}
It means that we can have Abelian conserved charges. In particular, even if the $U(1)_q$ field is absent, an effective $U(1)_q$ charge might exist due to interaction.

As regards the asymptotically covariantly constant AdS parameter, it has the form
\begin{eqnarray}
\mathbf{\Lambda } &=&u\left( f+\frac{r}{\ell }\right) \left( \mathbf{J}_{01}+\mathbf{P}_{0}\right) +\dfrac{p^2 u}{f+\frac{r}{\ell }} \left( \mathbf{J}_{01} -\mathbf{P}_{0}\right)  \notag \\
&= &u\left( \frac{2r}{\ell }-\frac{\ell p^2 }{2r}\right)
\,\left( \mathbf{J}_{01}+\mathbf{P}_{0}\right) +\frac{\ell p^2 u}{2r}\,\left( \mathbf{J}_{01}-\mathbf{P}_{0}\right) +\mathcal{O}\left( \frac{1}{r^3 }\right) \,.  \label{Lambda_AdS}
\end{eqnarray}
It is associated to the asymptotic isometries $\xi =\xi ^{\mu }\partial _{\mu }$ as
\begin{equation}
\pounds _{\xi }\mathbf{A}=D\mathbf{\Lambda }+i_{\xi }\mathbf{F}\,,\qquad \mathbf{\Lambda }=i_{\xi }\mathbf{A\,.}
\end{equation}
Since the mass corresponds to the time-like diffeomorphisms whose asymptotic Killing vector is $\xi =\partial _{t}$, we have $i_{\xi }\mathbf{F}=0$ on the solution, and the result is
\begin{equation}
\mathbf{\Lambda}_0=\mathbf{A}_{t}=\frac{r}{\ell ^2 }\,\mathbf{J}_{01}+\frac{f}{\ell }\,\mathbf{P}_{0}=\frac{r}{\ell ^2 }\,\mathbf{J}_{01}+\left[ \frac{r%
}{\ell ^2 }-\frac{p^2 }{2r}+\mathcal{O}\left( \frac{1}{r^3 }\right) %
\right] \,\mathbf{P}_{0}\,.  \label{Killing t}
\end{equation}
This behavior is consistent with the one found in eq.~\eqref{Lambda_AdS}, determining the normalization as
\begin{equation}
\partial _{t}:\quad u=\frac{1}{2\ell }\,,\qquad v=\frac{p^2 }{2\ell }\,.
\end{equation}

We focus now on explicit computation of conserved charges $Q_q$ and $E$ associated to the only two non-vanishing local parameters, $\Lambda^1$ given by \eqref{Lambda1} and $\mathbf{\Lambda }_0$ given by \eqref{Killing t}, respectively.

\subparagraph{Total energy.}

The total energy of the black hole-soliton system is computed using the formula \eqref{charge},
\begin{equation}
\delta E=\delta Q[\mathbf{\Lambda} _{0}] =2\ii k\theta \int\limits_{\Sigma }\left\langle \mathbf{\Lambda }_{0}\mathbf{\hat{F}}\delta \mathbf{\hat{A}}\right\rangle \,.  \label{Eb}
\end{equation}
Since the energy density $\varepsilon$ is the same for any $\Sigma $ which is locally a sphere, the total energy will be of the form $\varepsilon \mathrm{Vol}(\Sigma )$.
The internal symmetry soliton and the $U(1)_c$ gauge field do not contribute to the total energy because they do not interact with the AdS field directly.

The asymptotic parameter \eqref{Killing t} and the solution
\eqref{Laura solution} have the form
\begin{eqnarray}
\mathbf{\Lambda }_{0} &=&\frac{r}{\ell ^2 }\,\mathbf{J}_{01}+\frac{f}{\ell } \,\mathbf{P}_{0}\,,  \notag \\
\mathbf{\hat{F}}|_{\Sigma} &=&\epsilon ^{ijk}\,\tilde{e}_i\tilde{e}_j\left( C\mathbf{S}_{k}+\frac{\Xi }{2}\,\mathbf{J}_{k}\right) ,  \notag \\
\mathbf{\hat{A}}|_{\Sigma} &=&\tilde{e}^{i}\mathbf{S}_i+\left( \tilde{\omega}^{i}-C\,\tilde{e}^{i}\right) \mathbf{J}_i \,,
\end{eqnarray}
where we denoted non-$\widehat{SU}(2)_D$-term by $\mathbf{S}_i\equiv \frac{r}{\ell }\,\mathbf{P}_i-f\,\mathbf{J}_{1i}$ to easier compute the traces. The variations are taken in the boundary parameters such that $\delta r=0$. 

There are several ambiguities in the charge formula that we have to fix. For instance, it is always possible to add an arbitrary variation of an $r$ -dependent function to the above expressions. But if we keep track of all $\delta r$ arising from the supertraces (instead of setting them to zero), we will obtain a natural cancellation of all divergences in the charge. Thus, the $\delta r$-freedom is fixed by the finiteness of the charge when $r \to \infty$. 
Lagrangian version of this method requires addition of local counterterms to the action, that has been done in CS AdS gravity without torsion in any odd dimension in \cite{Miskovic:2007mg} by addition of the Euler topological invariant to the bulk action, but it still has an unknown form in a general case.

Another ambiguity lies in the parameter choice, e.g.,  $\delta p^2=\delta \mu $ and $\delta (C-1)=\delta C$. The additive constant will be chosen so that the charge $Q[\mathbf{\Lambda}]$ vanishes when its source is absent.\footnote{Similar normalization has been chosen in \cite{Crisostomo:2000bb}, where the additive constant related with the mass has been fixed requiring that the horizon shrinks to a point when $\mu-1 \to 0$, and the electrically charged solution reduces to the uncharged one when the parameter $q \to 0$.} For example, on the sphere, the solitons are absent when $C=B =0$, but on the projective space they are absent when $C=B =1$ (because the corresponding topological numbers are zero). As respect to the black hole, the charge is absent when the horizon \eqref{extremality} contracts to a point, $p^2=0$.

With this method at hand, we compute the following supertraces,
\begin{eqnarray}
\ii \left\langle \mathbf{J}_{01}\mathbf{S}_i\mathbf{J}_j\right\rangle &=&\frac{r}{2\ell }\,\delta _{ij}\,,\qquad \ii \left\langle \mathbf{J}_{01}\mathbf{J}_i\delta \mathbf{S}_j\right\rangle =\frac{\delta r}{2\ell }\,\delta _{ij}\,,  \notag \\
\ii \left\langle \mathbf{P}_{0}\mathbf{S}_i\mathbf{J}_j\right\rangle &=&-\frac{f}{2}\,\delta _{ij}\,,\qquad \ii \left\langle \mathbf{P}_{0}\mathbf{J}_i\delta \mathbf{S}_j\right\rangle
=-\frac{\delta f}{2}\,\delta _{ij}\,.
\end{eqnarray}
Plugging them in into  eq.~\eqref{Eb} and recognizing the integral of $\epsilon ^{ijk}\, \tilde{e}_i\tilde{e}_j \tilde{e}_{k}$ as $6\mathrm{Vol}(\Sigma)$, we obtain
\begin{equation}
\delta E=\frac{3k\mathrm{Vol}(\Sigma)\theta}{\ell }\lim_{r\rightarrow \infty }\left[ \Xi \left( \frac{r\delta r}{\ell ^2 }-f\delta f\right) -\delta C^2 \,\left( \frac{r^2 }{\ell ^2 }-f^2 \right) \right],
\end{equation}
where we used that the AdS field varies as $\delta \hat{\mathcal{A}}^{i}=-\delta C\,\tilde{e}^{i}$ independently on the choice of the topology $\Sigma $.
The second term is finite thanks to $\frac{r^2 }{\ell ^2 }-f^2 =p^2 $.
In the first term, we use also $\delta r/r\rightarrow 0$ to evaluate
\begin{equation}
\lim_{r\rightarrow \infty }\left( \frac{r\delta r}{\ell ^2 }-f\delta f\right) =\frac{\delta p^2 }{2}\,,  \label{var_r}
\end{equation}
arriving to the result
\begin{equation}
\delta E=\frac{3k\mathrm{Vol}(\Sigma)\theta} {2\ell}\,\left[ \rule{0pt}{13pt}\left( p^2 +1-C^2 \right) \,\delta p^2 -2p^2 \delta C^2 \right] . \label{var_E}
\end{equation}
We also applied $\Xi=p^2+1-C^2 $. The finiteness of the above expression is a non-trivial result, as the charges in AdS spaces are usually divergent in the asymptotic sector and have to be regularized.

In the $\Xi $-charged case, when $p^2 $ and $C^2 $ are independent parameters, the above expression is not integrable. To circumvent this problem, we will assume that the axial soliton has fixed topology, $\delta C=0$, so that it becomes a background configuration. This can be understood from the fact that the Pontryagin index $n_2(C)$ is quantized, because then it cannot vary infinitesimally, and therefore $\delta n_2=\frac 34 \, (1-C^2) \delta C=0$ implies $\delta C=0$ when $C \neq \pm 1$, but we will always keep it zero. As a result, also setting $\theta =1$, we obtain for the total energy
\begin{eqnarray}
E &=&\frac{3k\mathrm{Vol}(\Sigma)}{2\ell }\,\left( \frac{p^2}{2}+1-C^2 \right)\,p^2  \notag \\
&=&\frac{3k\mathrm{Vol}(\Sigma)}{4\ell }\,(\mu +1-2C^2) (\mu-1) \geq -\frac{3k\mathrm{Vol}(\Sigma)}{4\ell }\,(1-C^2)^2 \,. \label{E_rp3}
\end{eqnarray}
We notice that it is always bounded from below for fixed $C$.

Obtained energy includes the mass of the black hole, interaction energy between the black hole and the axial soliton, as well as the vacuum energy of the AdS space. Indeed, we can write the above result as
\begin{equation}
E=M+E_{\mathrm{int}}+E_{\mathrm{AdS}}\,,
\end{equation}
where
\begin{eqnarray}
M &=&\frac{3k\mathrm{Vol}(\Sigma)}{4\ell }\,\mu^2\,,  \notag \\
E_{\mathrm{int}} &=&-\frac{3k\mathrm{Vol}(\Sigma)}{2\ell }\,\mu C^2\,,  \notag \\
E_{\mathrm{AdS}} &=&-\frac{3k\mathrm{Vol}(\Sigma)}{4\ell}\,(1-2C^2)\,\,. \label{MEE}
\end{eqnarray}
The first term ($M$) is the black hole mass, and the second term ($E_{\mathrm{int}}$) corresponds to an interaction energy between the black hole and the axial soliton background. The last term ($E_{\mathrm{AdS}}$) describes the vacuum energy of the AdS space with torsion. When the torsion vanishes, the result matches the known one of the torsionless asymptotically
AdS$_5$ black hole \cite{Crisostomo:2000bb}, where the gravitational energy (which includes the black hole mass and vacuum energy) has been computed using the Regge-Teitelboim method,\footnote{To translate the notation from \cite{Crisostomo:2000bb} and prove $m=E_{C=0}$, first we choose the dimension $D=2n+1=5$, $n=2$, and fix the volume of the transversal space $\Omega _{3} =\mathrm{Vol}(\Sigma)$. The gravitational coupling constant in the paper is $k_{0}=\frac{1}{12\Omega _{3}G_2 }$, and it is related to our gravitational constant as $k_{0}=\frac{k}{8\ell }$. Finally, the gravitational energy $m=M+E_{\mathrm{AdS}}$ can be obtained from the mass parameter $\mu =\sqrt{2G_2 m+1}$.} 
\begin{equation}
E_{C=0}=\frac{3k\mathrm{Vol}(\Sigma)}{4\ell }\,\left( \mu ^2 -1\right) \,. 
\end{equation}

Note that the internal symmetry soliton with the charge $B$ does not curve nor torsion spacetime and it does not contribute to the total energy, as we already observed since the internal non-Abelian field does not interact with the black hole directly (only through the $U(1)_q $ term). It is a consequence of the fact that $SU(2)$ does not possess a symmetric rank-3 invariant tensor.

The BPS limit of the configuration corresponds to the extremal black hole with $\mu =1$, whose energy is
\begin{equation}
E_{\mathrm{BPS}}=0\,. \label{Ebps}
\end{equation}
The same is true for dimensionally continued black holes \cite{Banados:1993ur}, where the smallest horizon black hole  has zero energy. In addition, it is worth pointing out the similarities with the spectrum of supersymmetric black holes in 3D supergravity in AdS \cite{Coussaert:1993jp}, which is itself a CS theory. In 3D, the BTZ black hole geometry exhibits maximal supersymmetry in the global AdS$_3$ vacuum, which is the vacuum of the Neveu-Schwarz sector. The vacuum of the Ramond sector is given by the so-called massless BTZ, where the origin represents the topological obstruction. Also in the BTZ, as it is usual in supergravity, the BPS configurations represent extremal black holes. Another similarity with the 3D case is the mass gap between the global AdS and the black hole spectrum, a gap that is filled with solutions exhibiting naked singularities. The gap becomes more evident for solutions with weak torsion field ($\mu >>C^2$), when the interaction energy can be neglected.

For the $\Xi $-neutral solution, we have $p^2=C^2-1$. Since the axial charge is related to the black hole mass, it cannot describe a fixed background. Then the energy \eqref{var_E} can be exactly integrated out, leading to
\begin{equation}
\delta E=\frac{3k\mathrm{Vol}(\Sigma)}{2\ell }\,\delta p^{4},
\end{equation}
where we set $\theta =-1$, that further yields
\begin{equation}
E =\frac{3k\mathrm{Vol}(\Sigma)}{2\ell }\,p^{4}=\frac{3k\mathrm{Vol}(\Sigma)}{2\ell }\,(\mu -1) ^2 \geq 0\,,  \label{Eneutral}
\end{equation}
and also
\begin{equation}
E_{\mathrm{AdS}} =\frac{3k\mathrm{Vol}(\Sigma)}{2\ell }\,, \qquad 
E_{\mathrm{BPS}} =0\,. \label{EneutralAdS}
\end{equation}
Again, the energy is bounded from below. In this case, the black hole mass and the interaction between the axial soliton and the black hole cannot be distinguished. Furthermore, the
vacuum energy is not computed in a similar way as before, with $\mu =0$ and $C\neq 0$. In the extremal case, the black hole has zero energy, as expected.

The quantized topological charges $n_1$ and $n_2$ can always be added to the total energy because they satisfy $\delta n_1=0$ and $\delta n_2=0$. This is a more general property funded on the fact that boundary conditions modify conserved quantities. For instance, in AdS$_4$ gravity,
imposing (anti)self-dual boundary conditions on the AdS tensor (on-shell equivalent to the Weyl tensor), permits existence of solitonic solutions with `magnetic mass', computed as a Noether charge when the Pontryagin topological invariant is added to Einsten-Hilbert action \cite{Araneda:2016iiy,Araneda:2018orn}. In the next step, we will see that there is another conserved charge, $Q_q$, of topological origin.\medskip

To conclude, the $\Xi $-neutral solutions have discrete energy spectrum, related to their  topological properties. This is also true for $\Xi $-charged solutions in the BPS limit while, out of the limit, the energy spectrum is continuous. These conclusions are similar to the ones in superstring theory where the black $p$-brane charges \cite{Horowitz:1991cd,Townsend:1995gp} appearing in  supergravity are associated with the number of coincident D-branes \cite{Polchinski:1995mt} in the underlying string-theory configuration
\cite{Gubser:1996de,Strominger:1996sh,Maldacena:1996ky,Maldacena:1997re} and, therefore, quantized. In the BPS limit the mass, thus, becomes quantized.

\subparagraph{Abelian charges.} 

We have two Abelian parameters $\Lambda^1$ and $\Lambda^c$ associated to the symmetry $U(1)_q \times U(1)_c$. It is straightforward to check that the pure gauge field leads to $Q_c=0$.

On the other hand, as we discussed before, the $U(1)_q$ sector of the theory is strongly coupled when $\mathcal{N}=4$. Thus,  even though the $U(1)_q $ gauge field vanishes, its coupling to both AdS and $SU(4)$ fields leads to the effective interaction between them, and produces an effective $U(1)_q $ charge associated to the gauge parameter $\Lambda ^1 =1$. Using the formula \eqref{charge} and the invariant tensor \eqref{inv tensor} and \eqref{susy}, we find
\begin{equation}
\delta Q_q =\frac{k\theta }{2}\int\limits_{\Sigma}\left( \frac{1}{2}\,\hat{F}_{AB}\,\delta \hat{A}^{AB}-\mathcal{F}_i\,\delta \mathcal{A}^{i}\right) \,. \label{Qq}
\end{equation}
Note that $\ii\left\langle \mathbf{G}_1  \mathbf{T}_i\mathbf{T}_j\right\rangle =\frac{1}{4}\,\mathrm{Tr}\left( \tau _i\tau _j\right)
=-\frac{1}{4}\,\delta _{ij}$ (see eq.~\eqref{inv}). Replacing the solution \eqref{solutionSigma} in the variation and using eq.~\eqref{var_r}, we obtain for $\Sigma \in \{\mathbb{S}^2 ,\mathbb{RP}^3 \}$,
\begin{equation}
\delta Q_q = \frac{k\theta }{2} \int\limits_{\Sigma}\epsilon_{ijk}\, \tilde{e}^{i}\tilde{e}^j\tilde{e}^k\lim_{r\rightarrow \infty }\left[ C\,\left( f\delta f-\frac{r\delta r}{\ell ^2 }\right)-\frac 12 \,\Xi \,\delta C+\frac 12 \,\tilde{\Xi}\,\delta B \right],
\end{equation}
or after taking the limit
\begin{equation}
\delta Q_q = -\frac{3k\theta\mathrm{Vol}(\Sigma)}{2}\left[ \rule{0pt}{13pt} \delta (Cp^2) - (1-C^2) \,\delta (1-C)+(1-B ^2) \,\delta B  \right] \,. \label{var_Qq}
\end{equation}
In this case it is always possible to integrate out the charge for any $\Xi$.

On $\mathbb{RP}^3 $, since $\tilde{\omega}=\tilde{e}$, natural charges in $\hat{\mathcal{A}}^i$ and $\mathcal{A}^i$ are $1-C$ and $1-B $ (or equivalently the integration constant of the variation is fixed so that the solitons vanish when $C=1$ and $B =1$), so we can write the above variation as
\begin{equation}
\delta Q_q =-\frac{3k \pi^2  \theta}{2}\left[ \rule{0pt}{13pt}\delta
(Cp^2 )-(1-C^2 )\,\delta (1-C)-(1-B ^2 )\,\delta (1-B )\right] \,,
\end{equation}
and the result after integration is
\begin{equation}
Q_q (\mathbb{RP}^3 )=-\frac{3k\pi ^2 \theta }{2}\left( Cp^2 +\frac{4}{3}%
\,n_2 +\frac{4}{3}\,n_1 \right) =-2k\pi ^2 \theta \,P\, , \label{Q_q}
\end{equation}
where $P \in \mathbb{R}$ is the total Pontryagin index \eqref{total P}. 
 The proportionality between $Q_q$ and $P$ is a consequence of the identity $g_{1MN}=-\frac 14 \,g_{MN}$. The latter is true only for the
bosonic components of the Cartan-Killing metric for the $\mathcal{N}=4$ superalgebra (see eq.~\eqref{CK(N)} in Appendix \ref{Representation} and the comment after eq.~\eqref{susy}). 
 Thus, the effective charge \eqref{Q_q} does not contribute with any new conserved quantity, but itshows a topological origin, similarly as the electric charge of Reissner-Nordstr\"om  black holes in standard supergravity which is related to the central charge of the corresponding algebra.

On $\mathbb{S}^3 $, natural charges are $C$ and $B $ (since the solitons vanish when $C=0$, $B =0$), and the variation \eqref{var_Qq} is 
\begin{equation}
Q_q(\mathbb{S}^3 )=-\theta \pi^2 \left(3Cp^2+n_1 +n_2 +B-C \right)  \,.
\end{equation}
The charge is, therefore, proportional to $3n-2P+B-C$. Because $P$ and $n$ are topological numbers, $B-C$ is a conserved quantity directly related to $U(1)_q$ symmetry on the 3-sphere. 

Since the result is exact, it is valid for both $\Xi $-charged  and $\Xi $-neutral solutions.  The BPS states in all cases satisfy 
\begin{equation}
Q_q ^{\mathrm{BPS}}=0 \,. \label{Qbps}
\end{equation}

\subsection{Right and left non-Abelian solitons}\label{RLsol}

Properties of CS supergravity depend on invariant tensors associated to its gauge group. We saw in Appendix \ref{SU(4)} that the subgroups $SU(2)_{D}$ and $SU(2)_{\pm }$, when coupled to $U(1)_c$, have different invariant tensors (compare eqs.~\eqref{rank3_diagonal} and \eqref{rank3_LR}). Namely, the left and right groups have an additional non-vanishing component of the rank-3 tensor, that is $g_{c[\pm i][\pm j]}=\pm \frac{1}{4}\,\delta_{ij}$. This means that previously discussed solutions, based on the diagonal subgroup, could have different physical features if it is replaced by the left of right subgroup. A purpose of this subsection is to explore this possibility.

Let us consider a solution \eqref{Laura solution} mapped to a nonequivalent one when the non trivial gauge subgroup $SU(2)\subset SU(4)$ in the BPS solution is changed as $SU(2)_{D}\rightarrow SU(2)_{+}$.\footnote{There is an additional minus sign in the mapping $\mathbf{T}_i\leftrightarrow -\mathbf{T}_{+i}$ to adjust the structure constants in the algebra $[\mathbf{T}_i,\mathbf{T}_j]=-\epsilon _{ij}^{\ \ k}\mathbf{T}_{k}$ $\leftrightarrow $ $[\mathbf{T}_{+i},\mathbf{T}_{+j}]=\epsilon _{ij}^{\ \ k}\mathbf{T}_{+k}$.}
A new solution on $\Sigma \in \{\mathbb{S}^3 ,\mathbb{RP}^3 \}$  reads
\begin{eqnarray}
\mathbf{A} &=&\left( \dfrac{r}{\ell }\,\mathbf{J}_{01}+f\,\mathbf{P}_{0}\right) \frac{\diff t}{\ell }+\dfrac{\diff r}{\ell f}\,\mathbf{P}_1  +\tilde{e}^{i}\left( \frac{r}{\ell }\,\mathbf{P}_i-f\,\mathbf{J}_{1i}\right) +\hat{\mathcal{A}}^{i}\mathbf{J}_i+\mathcal{A}_{+}^{i}\mathbf{T}_{+i}+\diff \Omega\,\mathbf{T}_c \,,  \notag \\
\mathbf{F} &=&C\, \epsilon _{ij}^{\ \ k} \tilde{e}^{i}\tilde{e}^j\,\left( \frac{r}{\ell }\,\mathbf{P}_{k}-f\,\mathbf{J}_{1k}\right) +\frac{1}{2}\,\epsilon _{ij}^{\ \ k}\,\tilde{e}^{i}\tilde{e}^j\,\left( \Xi \,\mathbf{J}_{k}-\tilde{\Xi}\,\mathbf{T}_{+k}\right) \,,
\end{eqnarray}
where $\hat{\mathcal{A}}^{i}=\tilde{\omega}^i -C \tilde{e}^{i}$, $\mathcal{A}_{+}^i=\tilde{\omega}^i-B \, \tilde{e}^{i}$ and $\mathcal{A}_{-}^{i}=0$. In the old basis, the components of the internal non-Abelian gauge field are
\begin{equation}
\mathcal{A}^{i}=-\mathcal{A}^{0i}=-\frac{1}{2}\,(\tilde{\omega}^i-B \, \tilde{e}^{i})\,,\qquad \mathcal{F}^{i}=-\mathcal{F}^{0i}=-\frac{\tilde{\Xi}}{4}\,\epsilon _{\ jk}^{i}\,\tilde{e}^j\tilde{e}^k\,.
\end{equation}
The BPS state is reached when
\begin{equation}
\mu =1\,,\qquad B =C\,.
\end{equation}
The corresponding Killing spinor has the same form as before, and it satisfies the projection conditions
\begin{equation}
\Gamma _1 \epsilon =-\epsilon \,,,\qquad \mathbf{J}_i\epsilon =-\mathbf{T}_{+i}\epsilon \,.
\end{equation}
While many features, due to formal equivalence of the solutions, remain the same, an essential difference is the non-vanishing coupling $g_{c[+i][+j]}$ in $SU(2)_{+}\times SU(2)_{-}\times U(1)_c $, so that the $U(1)_c $ component could become relevant.

For that reason, and because the total energy, as well as $Q_q $, remain essentially unchanged by introduction of the right soliton, we will focus only on the new charge $Q_c $ unrelated to the AdS field, associated to the parameter $\Lambda^c=1$, obtained from \eqref{charge} as
\begin{equation}
\delta Q_c =\frac{k\theta}{2}\int\limits_{\Sigma }\mathcal{F}_{+i}\delta \mathcal{A}_{+}^i=-\frac{3k\theta}{2}\,\mathrm{Vol}(\Sigma) (1-B ^2) \delta B  \,.
\end{equation}
We can integrate out the charge exactly, arriving to the result
\begin{equation}
Q_c =\frac{k\theta }{2}\,\mathrm{Vol}(\Sigma )(B ^3 -3B )+const\,,
\end{equation}
where the additive constant is $3k\theta \,\pi^2$ on $\mathbb{RP}^3 $
(corresponding to $Q_c \rightarrow 0$ when $B \rightarrow 1$)  and it is zero on $\mathbb{S}^3 $ (corresponding to $Q_c \rightarrow 0$ when $B \rightarrow 0$). This charge is, therefore, on $\mathbb{RP}^3 $ proportional to the $SU(2)$ Pontryagin index \eqref{non-Abelian P}, and on $\mathbb{S}^3 $ it has more complicated dependence, but in both cases conservation of $n_1$ implies a conservation of $Q_c$. Explicitly, we have
\begin{equation}
Q_c (\mathbb{RP}^3 )=2k\theta \pi ^2 \,n_1 \,,\qquad Q_c (\mathbb{S}^3 )=-k\theta \pi ^2 \,\left(\rule{0pt}{12pt}n_1 +B (n_1)\right)\,.
\end{equation}

Therefore, the new solution is not equivalent to the old one, as it relates the Hamiltonian conserved charge to the quantized topological one associated to the $SU(2)_+$ soliton.

Another solution can be obtained in an analogous way using the mapping $SU(2)_{D}\rightarrow SU(2)_{-}$, with similar conclusions.


\section{Conclusions}

Chern-Simons AdS$_5$ supergravity presents many differences with respect to standard supergravity. The gravitational sector of the former takes the form of Einstein-Gauss-Bonnet AdS theory, with the Gauss-Bonnet coupling constant being fixed so that the global AdS vacuum is 2-fold degenerate. With addition of torsional fields, local symmetries of this gravity theory become enhanced from the Lorentz to AdS gauge group $SO(2,4)$. Furthermore, the addition of $U(1)_q \times SU(\mathcal{N})$ internal gauge fields and $\mathcal{N}$ supersymmetry generators close the supersymmetry algebra \textit{off-shell}, without necessity of the auxiliary fields \cite{Troncoso:1997me}, such that the number of bosonic and fermionic fields is not the same in general. This is related to the topological origin of the CS theory. Also related to that, the number of propagating degrees of freedom depends on background and, in general, it is not the same as in standard gravity. Still, these two gravity theories have a surprising number of common features, such as existence of black holes \cite{Banados:1993ur} respecting the laws of thermodynamics \cite{Crisostomo:2000bb}.

We show in this paper that more similarities arise in the supersymmetric extension of the theory. We focus on the special case with $\mathcal{N}=4$ supersymmetries, in which the $U(1)_q $ field becomes strongly coupled, fermions turn $U(1)_q $-neutral and the $U(1)_q $ sector converts to a central extension. We choose the $U(1)_q $ field locally vanishing, but because it interacts with both the geometry and the internal symmetry, it leads to non-trivial global effects. 

{It is worth mentioning that most results obtained  in this manuscript can be generalized to any $\mathcal{N} \geq 3$, when $SU(\mathcal{N})$ contains an $SU(2) \times U(1)_c$ subgroup, and the $U(1)_q$ field is pure gauge.}

\medskip
\newpage

\textit{New black hole solutions}\smallskip 

First we solve the gravitational sector of the theory and find a new class of black hole solutions with the mass parameter $\mu $ and the axial torsion charge $C$. Namely, the torsion tensor restricted to the 3D spatial section, which is locally a sphere, admits the maximal number of Killing vectors. Thus, the 3D space embedded in higher-dimensional geometry is characterised by the axial torsion strength $C$, whereas the intrinsic torsion of the 3-sphere remains zero. The black holes also possess the gravitational hair that modifies the asymptotics of the metric functions $g_{tt}$, $g_{rr}$ compared to standard AdS fall-off. Such black hole has two horizons, and it becomes extremal, with zero temperature, when they coincide.

Other relevant AdS gauge fields in the solution are associated  to $\widehat{SU}(2)_{D}\subset SO(2,4)$. We find two inequivalent branches of the black holes that differ according to the  $\widehat{SU}(2)_{D}$ field strengths $\Xi $. One of them has $\Xi \neq 0$, and another is a pure gauge. When $\Xi =0 $, the mass parameter is not independent from the axial torsion, being $\mu =C^2 $. In both cases the $\widehat{SU}(2)_{D}$ field is topologically non-trivial and it describes a solitonic solution. Namely, in absence of the gravitational hair ($b=b_{0}=0$), the metric manifold is asymptotically AdS in the standard sense, and the axial torsion can be seen as a scalar field $\varphi (x)$ on the curved space satisfying a boundary condition $\varphi (x)\rightarrow C$. It has properties of a solitonic solution, which we call the axial soliton. In \cite{Canfora:2007xs}, it was shown that the axial soliton preserves $1/2$ of the original supersymmetries, and it is therefore a BPS state.
\medskip

\textit{BPS states and conserved charges}\smallskip 

In this paper, we also investigate other BPS states and their underlying topological origin for the black hole solutions of CS AdS supergravity. To this end, we consider a solution with a non vanishing internal $SU(2)_{D}\times U(1)_c \subset SU(4)$ gauge field, which describes an independent solitonic solution, with amplitude $B$. We show that the internal and axial solitons have nontrivial winding numbers $n_1 (B)$, $n_2 (C)\in \mathbb{Z}$, meaning that the constants $B$ and $C$ cannot take arbitrary values, but only the ones fixed by  $n_1$ and $n_2$. The total Pontryagin number  $P\in \mathbb{R}$, which depends on $n_1$, $n_2$ and on the extremality parameter of the black hole, is a conserved quantity. A way to understand why $P$ is not an integer is that it has an additional term that depends on the mass parameter $\mu $ that appears in the extremality parameter of the black hole.

We compute the conserved charges in the theory, both Noether (due to time-like asymptotic space-time isometries) and topological charges (due to existence of $\widehat{SU}(2)_{D}\times SU(2)_{D}$ soliton). The computations are performed for $\mathbb{S}^3 $ and $\mathbb{RP}^3 \simeq \mathbb{S}^3 /\mathbb{Z}_2 $ spatial topologies on the horizon, motivated by the fact that, in absence of internal symmetries and the axial torsion, the Killing spinor becomes locally constant, so it would in principle not be globally well-defined on (simply connected) $\mathbb{S}^3 $. It could be so on $\mathbb{RP}^3 $. This crucial difference motivated us to investigate the effects of both topologies on the physical properties of the solution. Nevertheless, as the spinors carry also the $U(1)_c $ charge finally, when all gauge fields are added, both cases discussed in the paper present similar behavior, with the difference being that expressions on $\mathbb{S}^3 $ possesses an additional reflection symmetry $C\leftrightarrow -C$, absent on $\mathbb{RP}^3 $ due to identifications.

Another interesting result is that the topological charge $P$ is also an effective charge $Q_q $ for $U(1)_q $ symmetry on $\mathbb{RP}^3 $ and, in the case of $\mathbb{S}^3 $, it leads to a conserved quantity $B-C$. 
As respect to the unbroken supersymmetries, we find that the BPS states are reached in the limit when the internal soliton and axial soliton form a soliton-antisoliton system that finally unwinds ($n_1 =-n_2 $), which is achieved by the soliton amplitude matching, $B=C$, and when the black hole becomes extremal. Thus, similarly as in standard supergravity, and as expected, BPS states are
extremal states. In addition, in both CS and standard supergravity, the space of extremal solutions is larger than the space of BPS solutions -- namely, the state $B=-C$ is extremal, but it is not a BPS state.

It is also worthwhile to emphasize important differences between the $\Xi $-charged and $\Xi $-neutral cases, corresponding to $\frac{1}{16}$-BPS state and $\frac 12$-BPS state, respectively.
The $\Xi $-neutral solutions have fully discrete energy spectrum because $\mu =C^2 $ and $C$ is a discrete function of $n_2 \in \mathbb{Z}$. This is
also true for $\Xi $-charged solutions in the BPS limit, otherwise the energy spectrum is continuous. The last statement resembles the case of the
Reissner–Nordstr\"{o}m black holes in $\mathcal{N}=2$ supergravity (at least when $\Lambda =0$), where the $U(1)_q $ charge is also a central extension in the superalgebra. Another example are Taub-NUT-AdS and Taub-Bolt-AdS solutions in AdS$_4$ gravity, whose Noether charge acquires the `magnetic mass' term that is the Pontryagin number \cite{Araneda:2016iiy,Araneda:2018orn}. Furthermore, it satisfies (anti) self-dual asymptotic conditions in the field strength, frequently associated to solitonic solutions. 

Finally, motivated by possible (anti) self-duality between the $SU(2)_{+}$ and $SU(2)_{-}$ internal sectors of the theory, we re-interpret the solution by replacing the $SU(2)_{D}$ internal soliton by the right $SU(2)_{+}\subset SU(4)$ soliton, where the left one identically vanishes. Although we do not find a duality relation in this simple setting, we prove that in this case the obtained solution is physically inequivalent from the one studied in detail in this paper, as it has a new conserved quantity, that is the $U(1)_c $ charge $Q_c $, related to the topological number $n_1 (B)$. In this way, the charge $n_1 $ becomes conserved, whereas in the previous solution it was conserved only when combined with $n_2$ and the extremality parameter in $P$. \medskip

\textit{Discussion and outlook}\smallskip

An appearance of (anti) self-duality   and solitonic solutions in the context of gauge theories is not sursprising. The well-known example is the Belavin-Polyakov-Schwarz-Tyupkin (BPST) instanton in the $SU(2)$ Yang-Mills theories, where the Euclidean action is bounded by its pure-gauge value $I_{\mathrm{YM}}[g\diff g^{-1}]$. The bound is saturated for the (anti) self-dual Yang-Mills instantons, where $I_{\mathrm{YM}}[g\diff g^{-1}]$ becomes proportional to the winding number $P_{\mathrm{YM}}\in \mathbb{Z}$ (Pontryagin index). This is why we expect that, in principle, it should be possible to extend the static $SU(2)_+$ solution sketched in Section \ref{RLsol} to the full non-static $SU(2)_+\times SU(2)_- $ one, and realize the (anti) self-duality between the left and right sectors at the level of the CS action, similarly as in Yang-Mills theories. In this case, the likely topological charge that would play a role in the bound is not the Pontryagin index, but the invariant obtained from $I_{\mathrm{CS}}[g\diff g^{-1}]$. The same invariant leads to the quantization of the CS coupling constant $k$ in the quantum theory (by requiring the invariance of $\mathrm{e}^{\ii I_{\mathrm{CS}}}$ also globally), because the CS theory changes under finite gauge transformations, $\mathbf{A}\rightarrow g^{-1}(\mathbf{A}+\diff)g$, as \cite{Troncoso:1998ng}
\begin{equation}
I_{\mathrm{CS}}\rightarrow I_{\mathrm{CS}}+\frac{\ii k}{30}\int\limits_{\mathcal{M}}\left\langle (g^{-1}\diff g)^{5}\right\rangle \,,
\end{equation}
where we discarded a (topologically trivial) boundary term.

In addition, a more general study of the spin structure on non-Riemannian manifolds in the framework of CS supergravity could give some insight about possible more general BPS state solutions, similar to the one mentioned above.

Another task to be carried out is to compute the superalgebra of charges and evaluate its central extension, as shown in \cite{Banados:1994tn} in a simple toy model of generic, regular CS theory. Evaluation of the central charge is important to identify a dual boundary theory, in the framework of AdS/CFT correspondence, studied in higher-dimensional CS gravity in \cite{Banados:2005rz,Banados:2006fe,Mora:2014fba,Cvetkovic:2017fxa,Gallegos:2020otk}. In particular, the anticommutator of supercharges would lead to the Bogomol'nyi bound, which is saturated on the BPS states in standard supergravity \cite{Gibbons:1982fy}.
This method has been applied to global AdS solutions with non-Abelian fields (not black holes) in AdS$_5$ CS supergravity in \cite{Miskovic:2006ei}.

Finally, we cannot talk about black holes without knowing their entropy and studying the Hawking radiation via the black hole thermodynamics. The problem is particularly interesting because the found black hole is coupled to two solitons and the configuration is described by three integration constants ($\mu $, $B$, $C$) but only two conserved charges ($E$, $Q_q $) and several topological numbers ($n_1 $, $n_2 $, $P$), where only one of them is conserved ($P$). In the case of the right internal soliton, $Q_c \sim n_1$ is an additional conserved quantity.

Related to the entropy, we would also like to understand whether some well-known facts about the black holes in standard supergravity are still valid in the CS theory. For instance, whether the area of the extremal black hole horizon is proportional to (some power of) the black hole entropy. Without torsion, the entropy in gravity with $k$-fold degenerate AdS vacuum is polynomial in the horizon  \cite{Crisostomo:2000bb}, and this also holds in Lovelock gravity in general \cite{Cai:2003kt}. 

Then, the extremal limit in standard gravity is asymptotically reached at the zero Hawking temperature, where all the energy is lost by thermal Hawking radiation. In absence of topological charges in the gauge field sector, the minimal energy is zero, thus the black hole totally evaporates. When a topological charge is present, the minimal energy is determined by the topological charge, and  the extremal limit gives the no-force condition analogous to $M=|Q_q |$ for the Reissner–Nordstr\"{o}m black holes. Extremal CS AdS black holes also have vanishing temperature and they do not Hawking-radiate, but  in our case the presence of additional solitons singles out only the ones with zero total topological charge as the configuration  of equilibrium. These elements suggest that the usual interpretation of stable BPS states should also hold in the CS case. However, in order to prove it, we still have to work out explicitly the Bogomol'nyi bound of the BPS states by diagonalizing the supercharges part of the algebra that is positive definite, $\{\mathbf{Q}^\alpha_s,(\mathbf{Q}^{\dagger })_\beta^u\}\geq 0$. The topological charge is expected to appear in the superalgebra, such us $Q_q$, $Q_c$ or  -- as discussed above -- the winding number proportional to $I_{\mathrm{CS}}[g\diff g^{-1}]$ of the $SU(2)_+\times SU(2)_-$ soliton. 

Some of the ideas mentioned above are already work in progress.


\section*{Acknowledgments}

We would like to thank to Jorge Zanelli for useful discussions. This work was funded in part by CONICET and ANPCyT through the grants PIP-1109-2017, PICT-2019-00303; ANID through the grant FONDECYT N°1190533; and VRIEA-PUCV grant N°123.764. D.L.D. is supported by the National Agency for Research and Development (ANID) scholarship Doctorado Nacional N°21170779.

\appendix

\section{Representation of the generators of \texorpdfstring{$SU(2,2|\mathcal{N})$}{SU(2,2|N)} }\label{Representation}

The bosonic sector of the supersymmetric extension of AdS$_{5}$ is AdS$_{5}\times SU(\mathcal{N})\times U(1)$ for $\mathcal{N}>1$, where the $SU(\mathcal{N})$ part is absent
when $\mathcal{N}=1$. Representation of all generators can be found in \cite{Chandia:1998uf}, as follows.

\begin{enumerate}
\item[$\bullet $] AdS$_{5}$ generators $\mathbf{J}_{AB}=(\mathbf{J}_{ab},
\mathbf{J}_{a5}\equiv \mathbf{P}_{a})$, $A=(a,5)$ ($a=0,\ldots 4$):
\begin{eqnarray}
\mathbf{J}_{AB} &=&\left[
\begin{array}{cc}
\frac{1}{2}\,(\Gamma _{AB})_{\beta }^{\alpha } & 0 \\
0 & 0
\end{array}
\right] ,  \notag \\
\Gamma _{ab} &=&\frac 12 \,[\Gamma_a,\Gamma_b]\,,\quad
\Gamma _{a5}=\Gamma _a\,, \notag \\
\left\{ \Gamma _{a},\Gamma _{b}\right\} &=&2\eta _{ab}\,,\;\;\quad \eta _{ab}=(-,+,+,+,+)\,.
\end{eqnarray}

\item[$\bullet $] $SU(\mathcal{N})$ generators:
\begin{equation}
\mathbf{G}_{\Lambda }=\left[
\begin{array}{cc}
0 & 0 \\
0 & (\tau _{\Lambda })_{s}^{u}
\end{array}
\right] ,\qquad \tau _{\Lambda }^{\dagger }=-\tau _{\Lambda }\,.
\end{equation}

\item[$\bullet $] $U(1)$ generator:
\begin{equation}
\mathbf{G}_1 =\left[
\begin{array}{cc}
\frac{\ii}{4}\,\delta _{\beta }^{\alpha } & 0 \\
0 & \frac{\ii}{\mathcal{N}}\,\delta _{s}^{u}
\end{array}
\right] .
\end{equation}

\item[$\bullet $] SUSY generators:
\begin{equation}
\mathbf{Q}_{s}^{\alpha }=\left[
\begin{array}{cc}
0 & 0 \\
-\delta _{\beta }^{\alpha }\delta _{s}^{u} & 0
\end{array}
\right] ,\qquad \mathbf{\bar{Q}}_{\alpha }^{s}=\left[
\begin{array}{cc}
0 & \delta _{u}^{s}\delta _{\alpha }^{\beta } \\
0 & 0%
\end{array}
\right] .
\end{equation}
\end{enumerate}
The Cartan-Killing metric of the supergroup is defined by
\begin{equation}
g_{MN}=\left\langle \mathbf{G}_M\mathbf{G}_{N}\right\rangle =\mathrm{STr}\left( \mathbf{G}_M\mathbf{G}_{N}\right) \,,  \label{CK}
\end{equation}
and it has non-vanishing components
\begin{equation}
\begin{array}{llll}
g_{[AB][CD]} & =-\eta _{[AB][CD]}\,,\qquad  & g_{11} & =-q \,, \medskip\\
g_{\Lambda _1 \Lambda _2 } & =-\gamma _{\Lambda _1 \Lambda _2 }\,, & g_{\binom{\alpha }{u}\binom{s}{\beta }} & =-\delta _{\beta }^{\alpha }\delta_{u}^{s}\,,
\end{array}
\label{CK(N)}
\end{equation}
where $q=\frac{1}{4}-\frac{1}{\mathcal{N}}$ and $\gamma _{\Lambda_1\Lambda_2}=\mathrm{Tr}(\tau _{\Lambda_1}\tau_{\Lambda_2})$.

\section{Five-dimensional gamma matrices \label{Gamma}}

The five-dimensional gamma matrices $\Gamma _{a}\ (a=0,\ldots ,4)$ satisfy
the Crifford algebra
\begin{equation}
\{\Gamma _{a},\Gamma _{b}\}=2\eta _{ab}\,,\qquad \eta _{ab}=(-,+,+,+,+)\,.
\label{Clifford}
\end{equation}
They can be constructed from the $4\times 4$ four-dimensional gamma matrices
$^{(4)}\Gamma _{\bar{a}}$, such that $\Gamma _{a}=(\Gamma _{\bar{a}},\Gamma _4)=(^{(4)}\Gamma _{\bar{a}},^{(4)}\Gamma_5)$, where the last matrix is
\begin{equation}
\Gamma_{4}=\,^{(4)}\Gamma _{5}=\ii \,\Gamma _{0}\Gamma _1 \Gamma _2 \Gamma _{3}\,,\quad
\Gamma_4^{\dagger }=\Gamma_4\,,\quad (\Gamma_4)^2=1\,. \label{gamma5}
\end{equation}
In our conventions, their hermicity can be summarized as
\begin{equation}
\Gamma _{a}^{\dagger }=\Gamma _{0}\Gamma _{a}^{\dagger }\Gamma _{0}\,.
\end{equation}
Introducing the Levi-Civita as given by eqs.~\eqref{LeviCivita}, we can prove that
\begin{equation}
\text{Tr}\left( \Gamma _{a}\Gamma _{b}\Gamma _c \Gamma _{d}\Gamma_{e}\right) =-4\ii\,\epsilon _{abcde}\,.  \label{Trace-5}
\end{equation}
Other traces are
\begin{eqnarray}
\mathrm{Tr}\left( \Gamma _{a}\right)  &=&0\,,  \notag \\
\mathrm{Tr}\left( \Gamma _{a}\Gamma _{b}\right)  &=&4\eta _{ab}\,,  \notag \\
\mathrm{Tr}\left( \Gamma _{a}\Gamma _{b}\Gamma _c \right)  &=&0\,,  \notag
\\
\mathrm{Tr}\left( \Gamma _{a}\Gamma _{b}\Gamma _c \Gamma _{d}\right)
&=&4\left( \eta _{ab}\eta _{cd}-\eta _{ac}\eta _{bd}+\eta _{ad}\eta
_{bc}\right) \,.
\end{eqnarray}
The basis of the gamma matrices is
\begin{equation}
\{\mathbb{I}_{4},\Gamma _{a},\Gamma_{ab}\}=\{\mathbb{I}_{4},\Gamma
_{AB}\}\,,\qquad \Gamma _{ab} \equiv \frac 12 \,[\Gamma _{a},\Gamma_{b}]\,,
\end{equation}
where $\Gamma _{AB}=\{\Gamma _{ab},\Gamma _{a5}=\Gamma _{a}\}$, and $\eta
_{AB}=(\eta _{ab},-1)$ ($A,B=0,\ldots ,5$). They satisfy the orthogonality relations
\begin{equation}
\frac{1}{2}\,\left( \Gamma ^{AB}\right) _{\ \beta }^{\alpha }\left( \Gamma _{AB}\right) _{\ \delta }^{\gamma }=\delta _{\beta }^{\alpha }\delta _{\delta }^{\gamma }-4\delta _{\delta }^{\alpha }\delta _{\beta }^{\gamma }\,.
\end{equation}
Any matrix $\mathbb{M}_{\ \beta }^{\alpha }$ can be expanded in this basis as
\begin{equation}
\mathbb{M}=\frac{1}{4}\,\mathrm{Tr}\left( \mathbb{M}\right) +\frac{1}{4}\,\mathrm{Tr}\left( \mathbb{M}\Gamma ^{a}\right) \Gamma _{a}-\frac{1}{8}\,
\mathrm{Tr}\left( \mathbb{M}\Gamma ^{ab}\right) \Gamma _{ab}\,.
\end{equation}

As a consequence of \eqref{Trace-5} and the Levi-Civita conventions \eqref{LeviCivita}, we get the identity
\begin{equation}
\frac{\ii}{2}\,\epsilon _{ijk}\,\Gamma ^{jk}=\Gamma _{0}\Gamma _1 \Gamma _i\,.  \label{epsilonG}
\end{equation}

Let us write out a particular representation of five-dimensional gamma matrices decomposed as $\Gamma _{a}=(\Gamma _{0},\Gamma _1 ,\Gamma _i)$, with $\eta _{ab}=(-,+,\delta _{ij})$, in terms of three-dimensional matrices $\gamma _i$. 
They close the Clifford algebra
\begin{equation}
\left\{ \gamma _i,\gamma _j\right\} =2\delta _{ij}\,,
\end{equation}
and they can be represented in terms of the Pauli's matrices as $\gamma _{i+1}=\sigma _i$. They also satisfy the following identities,
\begin{equation}
\gamma _i\gamma _j=\delta _{ij}\,\mathbb{I}_2 +\ii\epsilon
_{ijk}\gamma ^k\,,\qquad \gamma _{ij}=\ii\epsilon _{ijk}\gamma
^k\,,  \label{id}
\end{equation}
where the second one is three-dimensional version of the identity \eqref{epsilonG}. 
Then the representation of five-dimensional $\Gamma$-matrices reads
\begin{eqnarray}
\Gamma _{0} &=&\mathrm{i\,}\sigma _2 \otimes \mathbb{I}_2 =\left(
\begin{array}{cc}
0 & \mathbb{I}_2  \\
-\mathbb{I}_2  & 0
\end{array}%
\right) ,\quad \Gamma _1 =\sigma _{3}\otimes \mathbb{I}_2 =\left(
\begin{array}{cc}
\mathbb{I}_2  & 0 \\
0 & -\mathbb{I}_2 
\end{array}
\right) ,  \notag \\
\Gamma _i &=&\sigma _1 \otimes \gamma _i=\left(
\begin{array}{cc}
0 & \gamma _i \\
\gamma _i & 0
\end{array}
\right),  \label{Repr5D}
\end{eqnarray}
and also for $\Gamma_{ab}=-\Gamma_{ba}$,
\begin{eqnarray}
\Gamma _{01} &=&-\sigma _1 \otimes \mathbb{I}_2 =\left(
\begin{array}{cc}
0 & -\mathbb{I}_2  \\
-\mathbb{I}_2  & 0
\end{array}
\right) ,\quad \Gamma _{0i}=\sigma _{3}\otimes \gamma _i=\left(
\begin{array}{cc}
\gamma _i & 0 \\
0 & -\gamma _i
\end{array}
\right) ,  \notag \\
\Gamma _{1i} &=&\ii \,\sigma _2 \otimes \gamma _i=\left(
\begin{array}{cc}
0 & \gamma _i \\
-\gamma _i & 0
\end{array}
\right) ,\qquad \Gamma _{ij}=\mathbb{I}_2 \otimes \gamma _{ij}=\left(
\begin{array}{cc}
\gamma _{ij} & 0 \\
0 & \gamma _{ij}
\end{array}
\right) .  \label{Repr5D_ij}
\end{eqnarray}

\section{Generators and invariant tensors of  \texorpdfstring{$SU(4)$}{SU(4)} \label{SU(4)}}

In Appendix \ref{Representation} we reviewed a representation of the algebra $\mathfrak{su}(2,2|N)$ used in this text. The basic properties and identities of $SU(N)$ generators have been summarized in Appendix of \cite{Greiner-etal}. 

Here we focus on the special case $N=4$. Because of the isomorphism $SU(4)\simeq SO(6)$, the generators of $SU(4)$ can be represented in terms of the gamma matrices $\tilde{\Gamma}_{IJ}$. Let us write $\mathbf{G}_{\Lambda}=\mathbf{T}_{IJ}$, where $\Lambda =[IJ]$ and
\begin{equation}
\mathbf{T}_{IJ}=\left[
\begin{array}{cc}
0 & 0 \\
0 & \frac{1}{2}\,(\tilde{\Gamma}_{IJ})_{u}^{s}
\end{array}
\right] ,\quad I,J=\hat{0},\ldots ,\hat{5}\,,\quad s,u=1,\ldots 4\,.
\end{equation}
The signature of the group indices $I,J$ is flat and the numeration from $\hat{0}$ to $\hat{5}$ is for convenience, stemming from the identification \eqref{i}. Since the Euclidean and Lorentzian indices cannot be confused in this section, we drop writing the hats.\footnote{\label{T=J}
The mapping of the indices $I,J=\left( \hat{0},\ldots ,\hat{5}\right) \leftrightarrow A,B=\left( 0,\ldots ,5\right) $ and the signatures $\eta_{AB}\leftrightarrow \delta _{IJ}$ means that the AdS$_{5}$ generators with the particular indices $0$ and $5$ acquire the factor $\ $`$-\ii $' with respect to the $SO(6)$ generators with the indices $\hat{0}$ and $\hat{5}$, so the isomorphism is established via the mapping:
\begin{equation*}
\begin{array}{llllllll}
\mathbf{T}_{\hat{0}\hat{1}} & \leftrightarrow -\ii\,\mathbf{J}_{01}\,,\quad  & \mathbf{T}_{\hat0\,\hat\imath} & \leftrightarrow -\ii\,\mathbf{J}_{0i}\,,\quad  & \mathbf{T}_{\hat0\hat5} &
\leftrightarrow -\mathbf{J}_0\,,\quad  & \mathbf{T}_{\hat\imath\hat\jmath} & \leftrightarrow \mathbf{J}_{ij}\,, \\
\mathbf{T}_{\hat1\hat\imath} & \leftrightarrow \mathbf{J}_{1i}\,, &
\mathbf{T}_{\hat1\hat5} & \leftrightarrow -\ii\,\mathbf{J}_{15}\,,
& \mathbf{T}_{\hat\imath\hat5} & \leftrightarrow -\ii\,\mathbf{J}_i\,. &  &
\end{array}
\end{equation*}}

The $\mathfrak{so}(6)$ algebra reads
\begin{equation}
[\mathbf{T}_{IJ},\mathbf{T}_{KL}]=\delta _{IL}\mathbf{T}_{JK}-\delta
_{JL}\mathbf{T}_{IK}-\delta _{IK}\mathbf{T}_{JL}+\delta _{JK}\mathbf{T}%
_{IL}\,,
\end{equation}%
and the gamma matrices $(\tilde{\Gamma}_i)_{s}^{u}$
are the $4\times 4$ matrices satisfying the Euclidean Clifford algebra
\begin{equation}
\{\tilde{\Gamma}_i,\tilde{\Gamma}_j\}=2\delta _{IJ}\,,\qquad
(I,J=1,\ldots ,6)\,.
\end{equation}
All gamma matrices are Hermitean and the generators are anti-Hermitean,
\begin{equation}
\tilde{\Gamma}_i^{\dag}=\tilde{\Gamma}_i\text{\thinspace },\qquad
\tilde{\Gamma}_{IJ}=\frac{1}{2}\,[\tilde{\Gamma}_i,\tilde{\Gamma}_j]=-
\tilde{\Gamma}_{IJ}^{\dag }\,.
\end{equation}

Using the definition  \eqref{inv tensor}, and the fact that the supertrace of $\mathbf{T}_{IJ}$ and the trace of $\tilde{\Gamma}_{IJ}$ have a relative minus sign, the supergroup symmetric invariant tensor is
\begin{eqnarray}
g_{[IJ][KL][MN]} &=&\gamma _{[IJ][KL][MN]}\,,  \notag \\
g_{1[IJ][KL]} &=&\frac{1}{\mathcal{N}} \,\gamma _{[IJ][KL]}\,, \\
g_{[IJ]\left( _{s}^{\alpha }\right) \left( _{\beta }^{u}\right) } &=&-\frac{\ii}{2}\, \delta _{\beta }^{\alpha }(\tilde{\Gamma}_{IJ})_{s}^{u}\,, \notag
\end{eqnarray}
where the particular $SO(6)$ symmetric invariant tensors are defined by
\begin{eqnarray}
\gamma _{[IJ][KL]} &=&\frac{1}{4}\,\mathrm{Tr}(\tilde{\Gamma}_{IJ}\tilde{\Gamma}_{KL}) ,  \notag \\
\ii\gamma _{[IJ][KL][MN]} &=&\frac{1}{16}\,\mathrm{Tr}(\{ \tilde{\Gamma}_{IJ}, \tilde{\Gamma}_{KL}\}\tilde{\Gamma}_{MN} ) .
\label{Cartan and inv su(4)}
\end{eqnarray}

It is useful to write the following traces of the gamma matrices, 
\begin{eqnarray}
\mathrm{Tr}(\tilde{\Gamma}_i\tilde{\Gamma}_j)  &=&4\delta
_{IJ}\,,  \notag \\
\mathrm{Tr}(\tilde{\Gamma}_i\tilde{\Gamma}_j\tilde{\Gamma}_{K}\tilde{\Gamma}_{L})  &=&4\left( \delta _{IJ}\delta _{KL}-\delta
_{IK}\delta _{JL}+\delta _{IL}\delta _{JK}\right) \,,  \notag \\
\mathrm{Tr}(\tilde{\Gamma}_i\tilde{\Gamma}_j\tilde{\Gamma}_{K}\tilde{\Gamma}_{L}\tilde{\Gamma}_m\tilde{\Gamma}_{N})  &=&4\delta
_{IJ}\,\left( \delta _{KL}\delta _{MN}-\delta _{KM}\delta _{LN}+\delta
_{KN}\delta _{LM}\right)   \notag \\
&&-4\delta _{IK}\left( \delta _{JL}\delta _{MN}-\delta _{JM}\delta
_{LN}+\delta _{JN}\delta _{LM}\right)   \notag \\
&&+4\delta _{IL}\left( \delta _{JK}\delta _{MN}-\delta _{JM}\delta
_{KN}+\delta _{JN}\delta _{KM}\right)   \notag \\
&&-4\delta _{IM}\left( \delta _{JK}\delta _{LN}-\delta _{JL}\delta
_{KN}+\delta _{JN}\delta _{KL}\right)    \\
&&+4\delta _{IN}\left( \delta _{JK}\delta _{LM}-\delta _{JL}\delta
_{KM}+\delta _{JM}\delta _{KL}\right) +4\ii\,\epsilon _{IJKLMN}\notag\,,
\end{eqnarray}
where the trace of six gamma matrices, with all indices different, is totally antisymmetric, and therefore proportional to the Levi-Civita  tensor $\epsilon _{IJKLMN}$. Its coefficient is evaluated using the identity \eqref{Trace-5} and the isomorphism with AdS$_{5}$, as explained in Footnote \ref{T=J}.  To compute the invariant tensor, we need also 
\begin{eqnarray}
\mathrm{Tr}(\tilde{\Gamma}_{IJ})  &=&0\,, \qquad
\frac{1}{4}\,\mathrm{Tr}(\tilde{\Gamma}_{IJ}\tilde{\Gamma}_{KL})=-\delta _{[IJ][KL]}\,,  \notag \\
\frac{1}{4}\,\mathrm{Tr}(\tilde{\Gamma}_{IJ}\tilde{\Gamma}_{KL}\tilde{\Gamma}_{MN})  &=&-\delta _{[IJ][KN]}\delta _{LM}+\delta_{[IJ][LN]}\delta _{KM}  \notag \\
&&+\delta _{[IJ][KM]}\delta _{LN}-\delta _{[IJ][LM]}\delta_{KN}+\ii\,\epsilon _{IJKLMN}\,,
\end{eqnarray}
where we defined
\begin{equation}
\delta _{[IJ][KL]}=\delta _{IK}\delta _{JL}-\delta _{IL}\delta_{JK}\,,\qquad \epsilon _{123456}=1\,.
\end{equation}
Now we can evaluate the symmetric tensors of rank two and three using \eqref{Cartan and inv su(4)},
\begin{eqnarray}
\mathrm{Tr}\left( \tau_{IJ}\tau_{KL}\right)  &=&\gamma _{[IJ][KL]}=-\delta _{[IJ][KL]}\,,  \notag \\
\mathrm{Tr}\left( \tau_{IJ}\tau_{KL}\tau_{MN}\right) 
&=&\ii\gamma _{[IJ][KL][MN]}=\frac{\ii}{2}\,\epsilon_{IJKLMN}\,.  \label{invSU(4)}
\end{eqnarray}
The corresponding supergroup invariant tensors are
\begin{eqnarray}
\ii\left\langle \mathbf{T}_{IJ}\mathbf{T}_{KL}\mathbf{T}_{MN}\right\rangle  &=&g_{[IJ][KL][MN]}=\frac 12\,\epsilon
_{IJKLMN}\,\,,  \notag \\
\ii\left\langle \mathbf{G}_1 \mathbf{T}_{IJ}\mathbf{T}_{KL}\right\rangle  &=&g_{1[IJ][KL]}=-\frac{1}{\mathcal{N}} \,\delta _{[IJ][KL]}\,,
\label{inv}
\end{eqnarray}
and for the Cartan-Killing metric
\begin{equation}
\left\langle \mathbf{T}_{IJ}\mathbf{T}_{KL}\right\rangle
=g_{[IJ][KL]}=\delta _{[ IJ][KL]}\,.  \label{CK-SU(4)}
\end{equation}

In our settings, we are focused on $SU(2)$ group, which does not possess a rank three symmetric invariant tensor (see Appendix of \cite{Greiner-etal}). In order to acquire a non-trivial rank three tensor, the $SU(2)$ fields have to interact with $U(1)$ fields. This can be achieved in a few nonequivalent ways, which we summarize below.

\begin{quote}
\textit{a}) \textbf{Subgroup} $U(1)_c\times SU(2)_{D}\times U(1)_q $
\end{quote}
This subgroup is generated by $\mathbf{T}_c=\mathbf{T}_{15}$, $\mathbf{T}_i=\frac 12 \,\epsilon_i^{\ jk}\,\mathbf{T}_{jk}$\ $(i,j,k=2,3,4)$ and $\mathbf{G}_1 $. The Levi-Civita tensor associated to this subgroup is identically zero (because the generator $\mathbf{T}_0$ is missing), giving that the rank-three symmetric tensor is also zero. All non-vanishing components are due to the Cartan-Killing metric that appears in either \eqref{inv} or \eqref{CK-SU(4)}, and they are
\begin{equation}
\begin{array}{llllll}
g_{1cc} & = & -\dfrac{1}{\mathcal{N}}\,,\qquad  & g_{1[ij][kl]} & = & -\dfrac{1}{\mathcal{N}}\,\delta _{[ij][kl]}\,,\medskip  \\
g_{cc} & = & 1\,, & g_{[ij][kl]} & = & \delta _{[ij][kl]}\,.
\end{array}
\label{rank3_diagonal}
\end{equation}
In terms of the generators $\tau_i=\frac{1}{2}\, \epsilon _i^{\ jk}\,\tau_{jk}$, useful relations used in the text are 
\begin{equation}
\begin{array}{llllll}
[ \tau_i, \tau_j] & = & -\epsilon _{ij}^{\ \ k} \tau_{k}\,,\qquad  & \mathrm{Tr}\left( \left[  \tau_i, \tau_j\right]  \tau_{k}\right)  & = & \epsilon _{ijk}\,,\medskip  \\
\mathrm{Tr}\left(  \tau_i \tau_j\right)  & = & -\delta _{ij}\,,
& \mathrm{Tr}\left( \{ \tau_i,  \tau_j\} \tau_{k}\right)
& = & 0\,.\end{array} 
\end{equation}
For the supergenerators, these relations become
\begin{equation}
\begin{array}{llllll}
[ \mathbf{T}_i, \mathbf{T}_j] & = & -\epsilon _{ij}^{\ \ k} \mathbf{T}_{k}\,,\qquad  & \left\langle \left[  \mathbf{T}_i, \mathbf{T}_j\right]  \mathbf{T}_{k}\right\rangle  & = & -\epsilon _{ijk}\,,\medskip  \\
\left\langle  \mathbf{T}_i \mathbf{T}_j\right\rangle  & = & \delta _{ij}\,,
& \left\langle \{ \mathbf{T}_i,  \mathbf{T}_j\} \mathbf{T}_{k}\right\rangle 
& = & 0\,.\end{array} \label{SU(2)all}
\end{equation}

\begin{quote}
\textit{b}) \textbf{Subgroup} $U(1)_c\times SU(2)_{-}\times SU(2)_{+}\times U(1)_q $
\end{quote}
This subgroup is an extension of the previous one, with the generators $\mathbf{T}_c=\mathbf{T}_{15}$, $\mathbf{T}_{\pm i}$ and $\mathbf{G}_1 $. Now the non-Abelian generators cover the full range of the indices and they can form non-trivial Levi-Civita tensor. Using the invariant tensor \eqref{inv} with the Levi-Civita conventions \eqref{LeviCivita}, and also \eqref{CK-SU(4)}, we find the following non-vanishing components,
\begin{equation}
g_{c[\pm i][\pm j]}=\pm \frac{1}{4}\,\delta _{ij}\,,\qquad g_{1cc}=-\frac{1}{\mathcal{N}}\,, \qquad  g_{1[\pm i][\pm j]}=-\frac{1}{2\mathcal{N}}\,\delta _{ij}\,,  \label{rank3_LR}
\end{equation}
as well as
\begin{equation}
g_{cc}=1\,,\qquad g_{[\pm i][\pm j]}=\frac 12 \,\delta _{ij}\,.
\label{rank2_LR}
\end{equation}
For completeness, we summarize the definition of the left and right generators \eqref{Tipm} and their algebra as
\begin{eqnarray}
\mathbf{T}_{\pm i} &=&\frac 12\,(\pm \mathbf{T}_{0i}-\mathbf{T}_i)\,,  \notag \\
\left[ \mathbf{T}_{\pm i},\mathbf{T}_{\pm j}\right]  &=&\epsilon _{ij}^{\ \ k}\,\mathbf{T}_{\pm k}\,,\quad \left[ \mathbf{T}_{+i},\mathbf{T}_{-j}\right]=0\,.
\end{eqnarray}

Particular results corresponding to the broken gauge symmetry $SU(2)_{\pm} \to U(1)_{\pm }$ can be easily deduced from the above text by taking $\mathbf{T_{2\pm}}$ as the generator of $U(1)_{\pm}$.


\section{Two branchings of  
\texorpdfstring{$SU(4)$}{SU(4)}}
\label{two_branchings}

The internal symmetry of interest is $SU(4)$, which is the double covering of $SO(6)$.  On one hand, the fundamental representation  $\mathbf{4}$ of $SU(4)$ is complex and we label its indices by  $s,u,\ldots=1,\ldots,4$. Then the 15 generators  are traceless matrices $T_s^{\ u}$ (with $T_s^{\ s}=0$)  in the $\mathbf{4}$ (lower index) times  $\mathbf{\bar{4}}$ (upper index). On the other hand, the fundamental representation  $\mathbf{6}$ of $SO(6)$ is real and we label its indices by  $I,J,\ldots=1,\ldots,6$.
The 15 generators in this representation are  $T_{IJ}=-T_{JI}$.\medskip

 $SU(4)$ admits two inequivalent branchings  $SU(4) \to SO(4)$ (see \cite{Slansky:1981yr}, page 117), which are associated with different black hole solutions. These are:
\begin{enumerate}
\item[{\bf 1)}] $SU(4) \to SO(4)$ irreducibly or, equivalently, $SO(6)\to SO(3)\times SO(3)$;
\item[{\bf 2)}]  $SU(4) \to SU(2)_+\times SU(2)_- \times U(1)_c$, that is, $SO(6)\to SO(4)\times SO(2)_c$.
\end{enumerate}

 In the first branching, the (complex) fundamental  representation $\mathbf{4} \in SU(4)$ goes into the (real) $\mathbf{4} \in SO(4)\simeq SU(2)_+\times SU(2)_-$ irreducibly, that is,
\begin{equation}
\mathbf{4} \to (\mathbf{2},\mathbf{2} )\,,    
\end{equation}
which implies in terms of both the representation and the corresponding indices,
\begin{eqnarray}
\mathbf{6} &\rightarrow &(\mathbf{3},\mathbf{1})+(\mathbf{1},\mathbf{3})\,,
\nonumber \\
I &\rightarrow &(i,\tilde{\imath})\,,\quad i,\tilde{\imath}=1,2,3\,.
\end{eqnarray}
The adjoint representation, similarly, decomposes as
\begin{equation}
   \mathbf{15} \to (\mathbf{3},\mathbf{1} )+  (\mathbf{1},\mathbf{3} ) + (\mathbf{3},\mathbf{3} )\,. 
\end{equation}
From the point of view of the generators, the above decomposition is realized in the $\mathbf{4}$ as $T_s^{\ u}\to T_{[su]}$, which are the generators of $SO(4)$, and  in the $\mathbf{6}$ we have instead $T_{IJ}\to T_{ij}+ T_{\tilde\imath\tilde\jmath}$, which are the generators of $SU(2)_\pm$ respectively.\medskip

In the second, different, branching of $SU(4)$, there is an additional Abelian group that assigns the $U(1)_c$ weight to each representation, which we denote as a subscript. Then the $\mathbf{4} \in SU(4)$ decomposes into
\begin{equation}\mathbf{4} \to (\mathbf{2},\mathbf{1})_{+1} +(\mathbf{1},\mathbf{2})_{-1} \,,\label{spindec}
\end{equation}
where the first item  in the brackets  is the dimension of the $SU(2)_+$ representation and the second one is the dimension of the $SU(2)_-$ representation.
 In components, we have that the (complex) index $u=1,\dots ,4$ decomposes into $u =(\alpha ,\tilde \alpha)$, where $\alpha,\dot\alpha=1,2$ label the fundamental representations of  $SU(2)_+$ and of $SU(2)_-$ respectively, carrying opposite  charges $\pm 1$ with respect to $U(1)_c$.
 
 For the $\mathbf{6}$, we have
 \begin{equation}
\mathbf{6} \to  (\mathbf{2},\mathbf{2})_{0} + (\mathbf{1},\mathbf{1})_{+2} +(\mathbf{1},\mathbf{1})_{-2} \,,     
 \end{equation}
 which can be realized by the decomposition (as we do in the main body of the paper) $I \to (0,1,i,5)$, with $i=2,3,4$, where the fundamental representation of $SO(4)$ is spanned by the indices $(0,i)$ while the indices  $(1,5)$ span $SO(2)_c$.
 For  the adjoint, similarly, we obtain
\begin{equation}
\mathbf{15} \to  (\mathbf{3},\mathbf{1})_{0} + (\mathbf{1},\mathbf{3})_{0}+ (\mathbf{1},\mathbf{1})_{0}+ (\mathbf{2},\mathbf{2})_{+2} +(\mathbf{2},\mathbf{2})_{-2} \,.    
\end{equation}
Finally, for the generators in the $\mathbf{4}$ we now find $T_s^{\ u}\to T_\alpha^{\ \beta}+T_{\tilde\alpha}^{\ \tilde{\beta}} + T$, generating $SU(2)_+\times SU(2)_-\times U(1)$,  where $T_\alpha^{\ \beta}$ and $T_{\tilde\alpha}^{\ \tilde{\beta}}$ are symmetric and traceless  and where the $U(1)_c$ generator is $T\equiv T_\alpha^{\ \alpha}=-T_{\tilde\alpha}^{\ \tilde{\alpha}}$. For the generators in the $\mathbf{6}$ we have, instead, $T_{IJ}\to T_{0i}+T_{ij}+ T_{15}$, 
 where the combinations
 \begin{equation}
\mathbf{T}_{\pm i}=\frac 12 \left( \pm\mathbf{T}_{0i}-\frac 12 \, \epsilon_{ijk} \mathbf{T}_{jk}\right)\,,\quad i=2,3,4, 
\end{equation}
introduced in \eqref{Tipm}, generate $SU(2)_\pm$, while $T_{15}$ generates $SO(2)_c\simeq U(1)_c$. 
 
\section{Counting of supersymmetries \label{Counting}}

Here we prove that the Killing spinor for the $\Xi $-charged solution \eqref{A_BPS} has $1/16$ of unbroken supersymmetries.

We focus on the constant part of the spinor, in the text denoted by $\eta_0$, but we drop the index $0$ in this appendix for the sake of simplicity. Using the representation \eqref{Repr5D} of the $\Gamma _{1}$ matrix, the projection $\Gamma _{1}\eta_{s}=-\eta_{s}$ leaves, for each $s$, just a half of spinor components,
\begin{equation}
\eta _{s}=\left(
\begin{array}{c}
0 \\
\tilde{\eta}_{s}
\end{array}
\right) ,\quad \Gamma _i\,\eta _{s}=\left(
\begin{array}{c}
\gamma _i\tilde{\eta}_{s}^{\alpha } \\
0
\end{array}
\right) ,\qquad s=1,\ldots ,4\,,
\end{equation}
where we wrote $\Gamma _i\eta _{s}$ represented in (\ref{Repr5D_ij})
because we will need it in the next step.

Now we have to check, when $\Xi \neq 0$, how many supersymmetries breaks the condition \eqref{iso} written as
\begin{equation}
\left( \Gamma _i\right) _{\beta }^{\alpha }\,\eta _{s}^{\beta }=(\tilde{\Gamma}_i)_{s}^{u}\,\eta _{u}^{\alpha }\,.
\end{equation}
Because the spatial and spinorial indices are mixed up, we will write $\eta _{s}^{\alpha }\equiv \eta _{\ s}^{\alpha }$\ as a $4\times 4$ matrix $\mathbb{H}$, where each column has fixed group index $s$, namely
\begin{equation}
\mathbb{H}=[\eta _{\ s}^{\alpha }]=\left(
\begin{array}{ll}
0 & 0 \\
\mathbb{A} & \mathbb{B}
\end{array}
\right) ,
\end{equation}
and $\mathbb{A}=\left( \tilde{\eta}_{1}\,\tilde{\eta}_{2}\right) $ and $\mathbb{B}=\left( \tilde{\eta}_{3} \, \tilde{\eta}_{4}\right) $ are $2\times 2$
matrices with the spinorial indices $\alpha =3,4$, which are exactly the left chiral $(0,1/2)$ indices that can also be denoted by $\dot{a}=\dot{1},\dot{2}$. Then the matrices
can be written explicitly as
\begin{equation}
\mathbb{A}=\left(
\begin{array}{ll}
\tilde{\eta}_{\ 1}^{\dot{1}} & \tilde{\eta}_{\ 2}^{\dot{1}} \\
\tilde{\eta}_{\ 1}^{\dot{2}} & \tilde{\eta}_{\ 2}^{\dot{2}}%
\end{array}
\right) ,\qquad \mathbb{B}=\left(
\begin{array}{ll}
\tilde{\eta}_{\ 3}^{\dot{1}} & \tilde{\eta}_{\ 4}^{\dot{1}} \\
\tilde{\eta}_{\ 3}^{\dot{2}} & \tilde{\eta}_{\ 4}^{\dot{2}}%
\end{array}
\right) .
\end{equation}
Thus $\left( \Gamma _i\right) _{\ \beta }^{\alpha }\,\eta _{\ s}^{\beta }=\left( \Gamma _i\mathbb{H}\right) _{\ s}^{\alpha }$ and $(\tilde{\Gamma}_i)_{\ s}^{u}\,\eta _{\ u}^{\alpha }=(\mathbb{H}^{T}\tilde{\Gamma}_i)_{\ s}^{\alpha }$ can be expressed just as products of matrices, leading to
\begin{equation}
\Gamma _i\mathbb{H=H}^{T}\tilde{\Gamma}_i\quad \Leftrightarrow \quad
\left(
\begin{array}{cc}
0 & \sigma _i \\
\sigma _i & 0
\end{array}
\right) \left(
\begin{array}{ll}
0 & 0 \\
\mathbb{A} & \mathbb{B}
\end{array}
\right) =\left(
\begin{array}{ll}
0 & \mathbb{A} \\
0 & \mathbb{B}%
\end{array}%
\right) \left(
\begin{array}{cc}
0 & \tilde{\sigma}_i \\
\tilde{\sigma}_i & 0
\end{array}
\right) ,
\end{equation}
where we represented $\tilde{\Gamma}_i$ analogously to \eqref{Repr5D_ij}, that is $\tilde{\Gamma}_i=\tilde{\sigma}_{1}\otimes \tilde{\gamma}_i$.
The above condition becomes
\begin{equation}
\left(
\begin{array}{ll}
\sigma _i\mathbb{A} & \sigma _i\mathbb{B} \\
0 & 0%
\end{array}%
\right) =\left(
\begin{array}{ll}
\mathbb{A}\tilde{\sigma}_i & 0 \\
\mathbb{B}\tilde{\sigma}_i & 0%
\end{array}%
\right) \quad \Rightarrow \quad \sigma _i\mathbb{A=A}\tilde{\sigma}%
_i\,,\quad \mathbb{B}=0\,.
\end{equation}%
The fact that $\mathbb{B}=0$ means that a half of the remaining components are further projected out, $\tilde{\eta}_{3}^{\dot{a}}=0$ and $\tilde{\eta}_{4}^{\dot{a}%
}=0$, leaving only $1/4$ of the original components denoted by $\eta _{\ \dot{s}}^{\dot{a}}$, with $\dot{a}=\dot{1},\dot{2}$ and  $\dot{s}=1,2$, which are 4 complex numbers.

The last set of the conditions to be analyzed is $(\sigma _i)_{\ \dot{b}}^{\dot{a}}\ \mathbb{A}_{\ \dot{s}}^{\dot{b}}\ =\mathbb{A}_{\ \dot{s}}^{\dot{a} }\ (\tilde{\sigma}_i)_{\ \dot{u}}^{\dot{s}}$. Here tilde just means that the Pauli matrices $\tilde{\sigma}_i$ act by the right, and $\sigma _i$
by the left, so this statement is the condition that $\mathbb{A}$ has to commute with all Pauli matrices. Using the Schur's lemma that says that the only matrices commuting with a finite-dimensional irreducible representation of a group, $SU(2)$ in this case, are the scalar matrices, implies that $\mathbb{A}$ is a scalar, i.e., proportional to $\mathbb{I} _{2}$. Therefore, only one component, $\tilde{\eta}_{\ 1}^{\dot{1}}=\tilde{\eta}_{\ 2}^{\dot{2}}$ survives the last condition,  while $\tilde{\eta}_{\ 2}^{\dot{1}}=\tilde{\eta}_{\ 1}^{\dot{2}}=0$, cutting further the number of supersymmetries to $1/4$. As a result, a pair of dependent spinors $\eta_1^\alpha$ and $\eta _2^\alpha$ remains non-vanishing, where either both are different than zero or both are zero, because they are proportional to the same complex Grassmann number $\tilde{\eta}_{\ 1}^{\dot{1}}$. Note that a different choice of the $SU(4)$ basis could transform it to only one spinor. 

In sum, accounting all projection conditions in the considered representation leaves $1/16$ of the original supersymmetries unbroken, given by the Killing spinors of the form
\begin{equation}
\eta _{1}^{\alpha }=\left(
\begin{array}{c}
0 \\
0 \\
1 \\
0
\end{array}
\right) \tilde{\eta}_{\ 1}^{\dot{1}}\ ,\qquad \eta _{2}^{\alpha
}=\left(
\begin{array}{c}
0 \\
0 \\
0 \\
1
\end{array}
\right) \tilde{\eta}_{\ 1}^{\dot{1}}\,,
\end{equation}
and the corresponding conjugated ones, $\bar{\eta}_{\alpha }^1 $ and $\bar{\eta}_{\alpha }^{2}$.

\section{Coordinates on the 3-sphere \label{Sphere}}

The 3-sphere can be described in the Hopf's fibration, with the angles $y^m=(\theta ,\varphi _1 ,\varphi _2 )\in \left[ 0,\frac{\pi }{2}\right] \times \left[ 0,2\pi \right] \times \left[ 0,2\pi \right] $, where $\mathbb{S} ^3 $ is embedded in $\mathbb{R}^{4}$ through \cite{Wilson}
\begin{eqnarray}
Y^1 &=&\cos \varphi _1 \sin \theta \,,\qquad Y^2=\sin \varphi _1 \sin \theta \,,  \notag \\
Y^3 &=&\cos \varphi _2 \cos \theta \,,\qquad Y^4=\sin \varphi _2 \cos \theta \,.  \label{X^s}
\end{eqnarray}
It satisfies the unit 3-sphere equation $\delta_{pq} Y^p Y^q=1$, whose metric
is
\begin{eqnarray}
\diff \Omega ^2 &=&\delta_{pq} \,\diff Y^p \diff Y^q=\diff \theta ^2 +\sin ^2 \theta \,\diff \varphi _1 ^2 +\cos
^2 \theta \,\diff \varphi _2 ^2 \,.  \label{Hopf}
\end{eqnarray}
For any fixed value of $\theta $, the coordinates $(\varphi _1 ,\varphi _2 )$ parametrize a 2-dimensional torus. When $\theta =0$ and $\pi /2$, these tori reduce to circles.

The vielbein and the spin connection of the 3-sphere in the Hopf's coordinates are
\begin{equation}
\begin{array}[b]{ll}
\tilde{e}^2 =\diff \theta \,, & 
\tilde{\omega}^2 =0\,, \\
\tilde{e}^3 =\sin \theta \,\diff \varphi _1 \,,\qquad  & \tilde{\omega}^3 =-\sin \theta \,\diff \varphi _2 \,, \\
\tilde{e}^{4}=\cos \theta \,\diff \varphi _2 \,, & 
\tilde{\omega}^{4}=-\cos \theta \,\diff \varphi _1 \,.
\end{array}
\label{tilde e,w}
\end{equation}
What we call the volume of the above three-dimensional geometry is
\begin{equation}
\mathrm{Vol}(\mathbb{S}^3 ) =\int\limits_{0}^{2\pi }\diff \varphi
_1 \int\limits_{0}^{2\pi }\diff \varphi _2 \int\limits_{0}^{\pi /2}\diff \theta \sin \theta \cos \theta =2\pi ^2 \,,
\end{equation}
that corresponds to the known expression for the area of the unit 3-sphere, whose four-dimensional volume is $\pi ^2 /2$.
Other integrals that we use in the text are
\begin{equation}
\begin{array}{llll}
\int\limits_{\mathbb{S}^3 }\epsilon_{ijk}\,\tilde{e}^{i}\tilde{e}^j\tilde{e}^k & =12\pi ^2 \,,\qquad \medskip  & \int\limits_{\mathbb{S}^3 }\epsilon_{ijk}\tilde{e}^{i}\tilde{\omega}^j\tilde{\omega}^k &=4\pi ^2 \,, \\
\int\limits_{\mathbb{S}^3 }\epsilon_{ijk}\,\tilde{\omega}^{i}\tilde{\omega}^j\tilde{\omega}^k & =0\,, & \int\limits_{\mathbb{S}^3 }\epsilon_{ijk}\,\tilde{e}^{i}\tilde{e}^j\tilde{\omega}^k & =0\,.
\end{array}
\label{eee}
\end{equation}

\section{Real projective 3-space \label{RProjective}}

The real projective 3-space, $\mathbb{RP}^3 $, is the quotient space $%
\mathbb{S}^3 /\mathbb{Z}_2 $ of a 3-sphere, obtained by identification of its antipodal points, $\mathbb{S}^3 /(x\sim -x)$. Equivalently, it is the quotient space of a half-sphere, i.e., solid 3-disc $\mathbb{D}^3 $ with antipodal points of $\partial \mathbb{D}^3 \simeq \mathbb{S}^2 $
identified \cite{Hatcher}. In that way, each open hemisphere in $\mathbb{S}^3 $ becomes disjoint from its antipodal image.

A double covering of the sphere, used for the above identification, can be written as the quotient $\mathbb{S}^3 \simeq G_{\mathrm{isom}}/SU(2)_{\mathrm{hol}}$ of the isometry group $G_{\mathrm{isom}}\simeq SU(2)_{+}\times SU(2)_{-}$ and the holonomy group $SU(2)_{\mathrm{hol}}\subset G_{\mathrm{isom}}$. A choice of the latter defines the topology and the connection defined on it.   In what follows we will construct  $\mathbb{S}^3 $ and $\mathbb{RP}^3 $. 

The Killing vectors generating $G_{\mathrm{isom}}$ satisfy
\begin{equation}
\left[ \xi _{\pm i},\xi _{\pm j}\right] =-\epsilon _{ij}^{\ \ k}\,\xi _{\pm k}\,,\qquad \left[ \xi _{+i},\xi _{-j}\right] =0\,,
\label{Killing V algebra}
\end{equation}
that can also be expressed in terms of the left-invariant 1-forms $\omega
_{\pm }^{i}$ of the isometry group, defined by their action on the
generators,
\begin{eqnarray}
&&\left. \omega _{+}^{i}\left( \xi _{+j}\right) =\delta _j^{i}\,,\quad
\omega _{-}^{i}\left( \xi _{-j}\right) =\delta _j^{i}\,,\quad \omega _{\pm}^{i}\left( \xi _{\mp j}\right) =0\,,\right.   \notag \\
&&\left. \diff \omega _{\pm }^k\left( \xi _{i\pm },\xi _{\pm j}\right) =-\omega _{\pm }^k\left( \left[ \xi _{\pm i},\xi _{\pm j}\right] \right) =-\epsilon _{ij}^{\ \ k}\,.\right.
\end{eqnarray}
This is equivalent to saying that $\omega _{\pm }^{i}$ satisfy the
Maurer-Cartan equations,
\begin{equation}
\diff \omega _{\pm }^{i}-\frac{1}{2}\,\epsilon _{\ \ jk}^{i}\,\omega
_{\pm }^j\wedge \omega _{\pm }^k=0\,.  \label{opm}
\end{equation}

To represent $G_{\mathrm{isom}}$ in terms of its Killing vectors, we embed the 3-sphere of radius $a$ in $\mathbb{R}^{4}$ using the radial foliation $(a,y^m)$,
\begin{equation}
\diff \ell ^2 =\diff a^2 +a^2 \diff \Omega ^2 \,,\qquad
\diff \Omega ^2 =\gamma _{mn}(y)\,\diff y^m\diff y^{n}\,,
\end{equation}
with the vielbein $e^p=(e^1 ,e^{i})$ and the corresponding Levi-Civita spin
connection given by
\begin{equation}
\begin{array}{llll}
e^1  & =\diff a\,, & e^{i} & =a\,\tilde{e}^{i}(y)\,, \\
\omega ^{ij} & =\epsilon _{\ \ k}^{ij}\,\tilde{\omega}^k(y)\,,\qquad  &
\omega ^{1i} & =-e^{i}\,.%
\end{array}
\label{radfol}
\end{equation}%
Here $\tilde{e}^{i}$ and $\tilde{\omega}^{i}$ are intrinsic quantities of $\mathbb{S}^3 $ in the coordinates $y^m$, with the intrinsic curvature
2-form $\tilde{R}^{i}=\epsilon _{\ jk}^{i}\,\tilde{e}^j\wedge \tilde{e}^k $ and the intrinsic torsion 2-form $\tilde{T}^{i}=0$. The unit 3-sphere $\mathbb{S}^3 $ is obtained from $\mathbb{R}^{4}$ by setting $\diff a=0$, $a=1$.

In this basis, the isometries of the 3-sphere are described by the six
Killing vectors $\xi _{pq}=-\xi _{qp}=\xi _{pq}^m\partial _m$ which
satisfy the $\mathfrak{so}(4)$ Lie-bracket algebra,
\begin{equation}
\left[ \xi _{pq},\xi _{p'q'}\right] =\delta _{pq'}\,\xi _{qp'}-\delta _{qq'}\,\xi _{pp'}-\delta_{pp'}\,\xi _{qq'}+\delta _{qp'}\,\xi_{pq'}\,.
\end{equation}
We can decompose and redefine these Killing vectors as
\begin{equation}
\xi _{pq}=\left( \xi _{1i},\xi _{ij}\right) \rightarrow \left( \xi _{+i},\xi
_{-i}\right) \,,
\end{equation}
to establish the isomorphism $G_{\mathrm{isom}}\simeq SO(4)\simeq
SO(3)_{+}\times SO(3)_{-}$. Then the $SO(3)_{\pm }$ isometries are described
by the Killing vectors,
\begin{equation}
\xi _{\pm i}=\frac{1}{2}\,(\xi _i\pm \xi _{i1})\,,\qquad \xi _{ij}\equiv \epsilon _{ij}^{\ \ k}\xi _{k}\,,  \label{xi_pm}
\end{equation}
which satisfy the algebra \eqref{Killing V algebra}.

To obtain  $\mathbb{S}^3 $, we choose $SU(2)_{\mathrm{hol}}\simeq
SU(2)_{D}$ as the diagonal subgroup of $G_{\mathrm{isom}}$, whose connection and the vielbein are associated to the ones of the 3-sphere,
\begin{equation}
\tilde{\omega}^{i}=\frac{1}{2}(\omega _{+}^{i}+\omega _{-}^{i})\,,\qquad
\tilde{e}^{i}=\frac{1}{2}(\omega _{-}^{i}-\omega _{+}^{i})\,,  \label{opm2}
\end{equation}
or equivalently
\begin{equation}
\omega _{\pm }^{i}=\omega ^{i}\pm \omega ^{1i}=\tilde{\omega}^{i}\mp \tilde{e%
}^{i}\,.  \label{w_pm}
\end{equation}%
Indeed, in terms of $\tilde{\omega}^{i}$ and $e^{i}$, eqs.~\eqref{opm}
become the conditions of constant curvature $\tilde{R}^{i}$ and vanishing torsion $\tilde{T}^{i}$ of the 3-manifold. Unlike $\omega _{\pm }^{i}$, the connection $\tilde{\omega}^{i}$ has a non vanishing curvature, corresponding to the fact the $SU(2)_{D}$ is a local symmetry on $\mathbb{S}^3 $. This choice reproduces the radial foliation described in \eqref{radfol}.

We can choose, instead, one of the $SU(2)_{\pm }$ factors as a holonomy
group, say $SU(2)_{+}$, such that $\mathbb{S}^3 \simeq SU(2)_{-}$. With
this choice, only a subgroup $SU(2)_{-}\subset SO(4)$ of the isometry group is manifest, so that only $\xi _{-i}$ acts on the quotient space $\mathbb{RP}^3 $, while $\xi _{+i}$ has the corresponding gauge connection set to zero
\begin{eqnarray}
\omega _{+}^{i} &=&0\quad \Rightarrow \quad \tilde{\omega}^{i}=\tilde{e}%
^{i}\,,  \notag \\
\omega _{-}^{i} &=&2\tilde{\omega}^{i}=2\tilde{e}^{i}\,.
\end{eqnarray}
To show that obtained topological space is indeed $\mathbb{RP}^3 $, note
that the left-invariant 1-form $\tilde{e}^{i}$ are dual to the generators $\tilde{\xi}_i$ such that $\tilde{e}^{i}(\tilde{\xi}_j)=\delta _j^{i}$. Their algebra can be read off from the structure constants in the Maurer-Cartan equations satisfied be the left-invariant forms $\tilde{\omega}^{i}$, finding that the isometry group is indeed restricted to $SO(3)=SU(2)_{+}/\mathbb{Z}_2 $, that is precisely the one of $\mathbb{RP}^3 $, when it is written as a parallelized manifold \cite{Chavel-Chern}.

It remains to find an explicit coordinate frame for this qotient space.
Since $\omega _{-}^{i}$ is a pure gauge connection, $\tilde{e}^{i}(y)$ can be constructed using $\omega _{-}^{i}\tilde{\mathbf{J}}_{-i}=g\diff %
g^{-1}$, where $g(y)\in SO(3)_{-}$ and $\tilde{[\mathbf{J}}_{-i},\tilde{%
\mathbf{J}}_{-j}]=-\epsilon _{ij}^{\ \ k}\tilde{\mathbf{J}}_{-k}$. One
possible parametrization of the $SO(3)_{-}$ group element is in terms of the
Euler angles $y^m=(\psi ,\theta ,\varphi )\sim (\psi +2\pi ,\theta +\pi
,\varphi +2\pi )$ such that $g=\mathbf{R}_{z}(\theta )\mathbf{R}_{y}(\varphi
)\mathbf{R}_{z}(\psi )$, leading to the vielbein
\begin{eqnarray}
\tilde{e}^2  &=&\frac{1}{2}\,\left( \sin \theta \,\diff \varphi -\sin
\varphi \cos \theta \,\diff \psi \right) \,,  \notag \\
\tilde{e}^3  &=&-\frac{1}{2}\,\left( \cos \theta \,\diff \varphi +\sin
\varphi \sin \theta \,\diff \psi \right) \,,  \label{e_RP3} \\
\tilde{e}^{4} &=&\frac{1}{2}\,\left( \diff \theta +\cos \varphi \,%
\diff \psi \right) \,,  \notag
\end{eqnarray}%
and the projective 3-space metric
\begin{equation}
\diff \Omega ^2 =\frac{1}{4}\,\left( \diff \psi ^2 +\diff %
\theta ^2 +2\cos \varphi \,\diff \theta \diff \psi +\diff %
\varphi ^2 \right) \,,
\end{equation}%
with $\sqrt{\gamma }=\frac{1}{8}\,\sin \varphi $. This enables to compute
the volume of the manifold as
\begin{equation}
\mathrm{Vol}(\mathbb{RP}^3 )=\int \sqrt{\gamma }\,\diff ^3 y=\frac{1}{8%
}\int\limits_{0}^{2\pi }\diff \theta \int\limits_{0}^{2\pi }\diff %
\psi \int\limits_{0}^{\pi }\diff \varphi \,\sin \varphi =\pi ^2 \,.
\label{VolumeRP3}
\end{equation}
We see that it is a half of the volume of the 3-sphere, due to its identification. For completeness, we write the Killing vectors describing isometries of the above metric as
\begin{eqnarray}
\xi _{+2} &=&-\partial _{\psi }\,,  \notag \\
\xi _{+3} &=&\cot \varphi \sin \psi \,\partial _{\psi }-\frac{\sin \psi }{%
\sin \varphi }\,\partial _{\theta }-\cos \psi \,\partial _{\varphi }\,,
\notag \\
\xi _{+4} &=&\cot \varphi \cos \psi \,\partial _{\psi }-\frac{\cos \psi }{%
\sin \varphi }\,\partial _{\theta }+\sin \psi \,\partial _{\varphi }\,,
\notag \\
\xi _{-2} &=&-\partial _{\theta }\,,  \notag \\
\xi _{-3} &=&\frac{\cos \theta }{\sin \varphi }\,\partial _{\psi }-\cot
\varphi \cos \theta \,\partial _{\theta }-\sin \theta \,\partial _{\varphi
}\,,  \notag \\
\xi _{-4} &=&-\frac{\sin \theta }{\sin \varphi }\,\partial _{\psi }+\cot
\varphi \sin \theta \,\partial _{\theta }-\cos \theta \,\partial _{\varphi
}\,.
\end{eqnarray}

\section{Degrees of freedom count \label{PhaseSpace}}

In this Appendix we evaluate the rank of the Jacobian matrix 2-form
\begin{equation}
\mathcal{J}_{MN}=\ii\left\langle \mathbf{G}_M\mathbf{G}_{N}\mathbf{F}\right\rangle =g_{MNK}\,F^K\,,  \label{J}
\end{equation}
and the symplectic matrix
\begin{equation}
\Omega _{MN}^{\mu \nu }=\ii\epsilon ^{t\mu \nu \lambda \rho
}\left\langle \mathbf{G}_M\mathbf{G}_N\mathbf{F}_{\lambda \rho
}\right\rangle =\epsilon ^{t\mu \nu \lambda \rho }\,g_{MNK}\,F^K_{\lambda \rho}\,,  \label{O}
\end{equation}
for the $\Xi $-charged solution \eqref{Laura solution}, for purpose of
computing the number of d.o.f. in the theory around a certain background. The matrix elements of these two tensors are related by
\begin{equation}
\Omega _{MN}^{\mu \nu }=\epsilon ^{t\mu \nu \rho \sigma }\mathcal{J}_{MN,\rho \sigma }\,,
\end{equation}
but they act in different spaces, so the rank of one matrix is not related
to the rank of another matrix.

We use the $\Xi$-charged solution given by eq.~\eqref{Laura solution},
\begin{equation}
\mathbf{F}=C\,\epsilon _{\ jk}^{i}\,\left( \frac{r}{\ell }\,\mathbf{P}%
_i-f\,\mathbf{J}_{1i}\right) \tilde{e}^j\tilde{e}^k+\frac{1}{2}\,\Xi \,\tilde{e}^{i}\tilde{e}^j\,\mathbf{J}_{ij}+\frac{1}{2}\,\tilde{\Xi}%
\,\epsilon _{ij}^{\ \ k}\tilde{e}^{i}\tilde{e}^j\mathbf{T}_{k}\,,\quad
\tilde{\Xi},\Xi \neq 0\,.  \label{Background}
\end{equation}
We do not discuss the cases when $\Xi =0\neq \tilde{\Xi}$ or $\Xi \neq 0=%
\tilde{\Xi}$ because they do not possess BPS states, and we also omit the $\Xi $-neutral solution \eqref{neutral} with $\tilde{\Xi},\Xi =0$ because it is a pure gauge. In addition, the matrices \eqref{J} and \eqref{O} do not have mixed bosonic-fermionic terms so that the purely fermionic sub-matrix decouples from the bosonic one, and because at the end we take all fermions equal to zero, the fermionic degrees of freedom vanish. This is why we will consider only the bosonic submatrices of \eqref{J} and \eqref{O}, corresponding to the gauge group $SO(2,4)\times SU(2)_{D}\times U(1)_{c}\times U(1)_{q}$ whose dimension is $\mathfrak{D}_{\mathrm{B}}=15+3+1+1=20$.

Let us evaluate first the bosonic components of the 2-form \eqref{J}. In the pure AdS sector, it corresponds to the symmetric $15\times 15$ Jacobian matrix,
\begin{equation}
\mathcal{J}_{[AB][CD]}=\dfrac{1}{2}\,C\,\epsilon _{\ jk}^{i}\,\left( \frac{r}{\ell }\,\epsilon _{ABCDi5}-f\epsilon _{ABCD1i}\right) \tilde{e}^j\tilde{e}^k+\frac{1}{4}\,\epsilon _{ABCDij}\Xi \,\tilde{e}^{i}\tilde{e}^j\,.
\end{equation}
Decomposing the group indices as $A=(0,1,i,5)$, its non-vanishing components are
\begin{eqnarray}
\mathcal{J}_{[01][ij]} &=&\mathcal{J}_{[1i][0j]}=\dfrac{Cr}{\ell }\,\tilde{e}_i\tilde{e}_j\,,  \notag \\
\mathcal{J}_{[i5][0j]} &=&-\mathcal{J}_{[ij][05]}=Cf\,\tilde{e}_i\tilde{e}_j\,,  \notag \\
\mathcal{J}_{[01][i5]} &=&\mathcal{J}_{[1i][05]}=-\mathcal{J}_{[15][0i]}=
\frac{\Xi }{4}\,\epsilon _{ijk}\,\tilde{e}^j\tilde{e}^k\,,
\end{eqnarray}
plus their symmetric counterparts. Next,
the Jacobian matrix with $SU(4)$ indices identically vanishes because this
group is broken to $SU(2)_{D}\times U(1)_{c}$ which, according to \eqref{rank3_diagonal}, does not have a rank-3 symmetric invariant tensor and the $U(1)_{q}$ field strength is zero. As regards the $U(1)_{q}$ interaction, using \eqref{inv tensor} and (\ref{rank3_diagonal}), we find
the components
\begin{equation}
\begin{array}{llll}
\mathcal{J}_{1[i5]} & =-\dfrac{rC}{4\ell }\,\epsilon_{ijk}\, \tilde{e}^j\tilde{e}^k\,,\qquad  & \mathcal{J}_{1[ij]} & =\dfrac{\Xi }{4}\,\tilde{e}_i\tilde{e}_j\,,\medskip  \\
\mathcal{J}_{1[1i]} & =-\dfrac{Cf}{4}\,\epsilon_{ijk}\, \tilde{e}^j\tilde{e}^k\,, & \mathcal{J}_{1[\hat{\imath}\hat{\jmath}]} & =-\dfrac{\tilde{\Xi}}{4}\,\tilde{e}_i\tilde{e}_j\,,
\end{array}
\end{equation}
where we denoted the internal symmetry
indices by $\hat{\imath},\hat{\jmath}$ in the Jacobian, not to be confused with the AdS ones, $i,j$.

Zero modes of the Jacobian are determined from the algebraic equation
\begin{equation}
0=\mathcal{J}_{MN}\,V^{N}\,,
\end{equation}%
where $\mathfrak{D}_{\mathrm{B}}=20$ is the maximal rank of the $20\times 20$ Jacobian matrix $\mathcal{J}_{MN}$, and each non-vanishing component of the vector field $V^m=(V^{AB},V^{\hat{\imath}\hat{\jmath}},V^{c},V^1 )$
decreases this rank for one. It is straightforward to show that, because $\mathcal{J}_{cM}=0$ and $U(1)_{c}$ field is the pure gauge, the component $V^{c}\in \mathbb{R}$ remains arbitrary, presenting one zero mode of the
Jacobian. Keeping this in mind, the above equations can equivalently be
written as
\begin{eqnarray}
0 &=&\frac{1}{2}\,\mathcal{J}_{[AB][CD]}V^{CD}+\mathcal{J}_{[AB]1}V^1 \,,
\notag \\
0 &=&\mathcal{J}_{1[AB]}V^{AB}+\mathcal{J}_{1[\hat{\imath}\hat{\jmath}]}V^{%
\hat{\imath}\hat{\jmath}}\,, \\
0 &=&\mathcal{J}_{[\hat{\imath}\hat{\jmath}]1}V^1 \,.  \notag
\end{eqnarray}
Due to $\mathcal{J}_{[\hat{\imath}\hat{\jmath}]1}\neq 0$, the last equation yields $V^1 =0$, and the $SU(2)_{D}$ components $V^{\hat{\imath}\hat{\jmath}}$ can be solved in terms of $V^{AB}$ from the second equation, so that they do not correspond to the zero modes
of the matrix. As a result, the system reduces in equations in $V^{AB}$ components, $\mathcal{J}_{[AB][CD]}V^{CD}=0$, which can also be written as
\begin{eqnarray}
0 &=&\dfrac{Cr}{\ell }\,\tilde{e}_i\tilde{e}_jV^{ij}+\frac{\Xi }{2}%
\,\epsilon _{ijk}\,\tilde{e}^j\tilde{e}^kV^{i5}\,,  \notag \\
0 &=&-Cf\,\tilde{e}_i\tilde{e}_jV^{ij}+\frac{\Xi }{2}\,\epsilon _{ijk}\,\tilde{e}^j\tilde{e}^kV^{1i}\,,  \notag \\
0 &=&\frac{\Xi }{4}\,\epsilon _{ijk}\,\tilde{e}^j\tilde{e}^kV^{0i}\,,
\notag \\
0 &=&2C\left( \dfrac{r}{\ell }\,V^{1j}+f\,V^{j5}\right) \tilde{e}_i\tilde{e}_j+\frac{\Xi }{4}\,\epsilon _{ijk}\,\tilde{e}^j\tilde{e}^kV^{15}\,, \\
0 &=&\dfrac{Cr}{\ell }\,\tilde{e}_i\tilde{e}_jV^{0j}+\frac{\Xi }{4}%
\,\epsilon _{ijk}\,\tilde{e}^j\tilde{e}^kV^{05}\,,  \notag \\
0 &=&Cf\,\tilde{e}_i\tilde{e}_jV^{0j}+\frac{\Xi }{4}\,\epsilon _{ijk}\,\tilde{e}^j\tilde{e}^kV^{01}\,,  \notag \\
0 &=&C\left( \dfrac{r}{\ell }\,V^{01}-f\,V^{05}\right) \tilde{e}_i\tilde{e}_j\,.  \notag
\end{eqnarray}
The solution in $V^{AB}$ is
\begin{equation}
\begin{array}{lll}
V^{0i}=0\,,\qquad  & V^{05}=0\,, & V^{01}=0\,,\medskip  \\
V^{15}=0\,, & V^{1i}=-\dfrac{\ell f}{r}V^{i5}\,,\quad  & V^{ij}=-\dfrac{\Xi\ell }{2Cr}\,\epsilon _{\ \ k}^{ij}\,V^{k5}\,,
\end{array}
\end{equation}
where the component $V^{i5}\in \mathbb{R}$ remains arbitrary representing
three independent zero modes. Together with $V^c$, there are 4 zero modes in total and the rank of the bosonic part of the Jacobian is
\begin{equation}
\mathrm{rank}(\mathcal{J})=20-4=16\,,
\end{equation}%
where the result is also valid in the extremal limit $f\rightarrow \frac{r}{\ell }$.

Let us focus now on the symplectic matrix $\Omega _{MN}^{\mu \nu }$ with
non-zero components easily deduced from the known Jacobian matrix,
\begin{eqnarray}
\Omega _{[ 01][ij]}^{rm} &=&\Omega _{[ 1i][0j]}^{rm}=\sqrt{\gamma }\,\frac{2Cr}{\ell }\,\epsilon _{ijk}\,\tilde{e}^{mk}\,,  \notag \\
\Omega _{[ i5][0j]}^{rm} &=&-\Omega _{[ ij][05]}^{rm}=\sqrt{\gamma }\,2Cf\,\epsilon _{ijk}\,\tilde{e}^{mk}\,,  \notag \\
\Omega _{[ 01][i5]}^{rm} &=&\Omega _{[ 1i][05]}^{rm}=-\Omega
_{[ 15][0i]}^{rm}=\sqrt{\gamma }\,\Xi \,\tilde{e}_i^m\,,
\end{eqnarray}
as well as
\begin{equation}
\begin{array}{llll}
\Omega _{1[i5]}^{rm} & =-\sqrt{\gamma }\,\dfrac{rC}{\ell }\,\tilde{e}%
_i^m\,,\qquad  & \Omega _{1[ij]}^{rm} & =\sqrt{\gamma }\,\dfrac{\Xi }{2}\,\epsilon _{ijk}\,\tilde{e}^{mk}\,,\medskip  \\
\Omega _{1[1i]}^{rm} & =-\sqrt{\gamma }\,Cf\,\tilde{e}_i^m\,, & \Omega
_{1[\hat{\imath}\hat{\jmath}]}^{rm} & =-\sqrt{\gamma }\,\dfrac{\tilde{\Xi}}{2}\,\epsilon _{ijk}\,\tilde{e}^{mk}\,.
\end{array}
\end{equation}
Its rank is determined by the following bosonic equations on $\Gamma $,
\begin{equation}
\Omega _{MN}^{\mu \nu }V_{\nu }^{N}=0\,\quad \Rightarrow \quad \left\{
\begin{array}{c}
\Omega _{MN}^{rm}V_r^{N}=0\,, \medskip\\
\Omega _{MN}^{rm}V_m^{N}=0\,.
\end{array}
\right.
\end{equation}
There are $4\times 20$ components of the bosonic vector field $V_{\mu
}^m=\left( V_r^m,V_m^m\right) $ and the antisymmetric symplectic form $\Omega_{MN}^{\mu \nu }$ is $80\times 80$ matrix. Since $M\in \left\{ c,[\hat{\imath}\hat{\jmath}],[AB],1\right\} $, we find
\begin{eqnarray}
0 &=&\Omega _{[\hat{\imath}\hat{\jmath}]1}^{rm}V_r^1 \,,  \notag \\
0 &=&\Omega _{[\hat{\imath}\hat{\jmath}]1}^{rm}V_m^1 \,,  \notag \\
0 &=&\Omega _{1[CD]}^{rm}V_r^{CD}+\Omega _{1[\hat{\imath}\hat{\jmath}]}^{rm}V_r^{\hat{\imath}\hat{\jmath}}\,,  \notag \\
0 &=&\Omega _{1[CD]}^{rm}V_m^{CD}+\Omega _{1[\hat{\imath}\hat{\jmath}]}^{rm}V_m^{\hat{\imath}\hat{\jmath}}\,, \\
0 &=&\frac{1}{2}\,\Omega _{[AB][CD]}^{rm}V_r^{CD}+\Omega
_{1[AB]}^{rm}V_r^1 \,,  \notag \\
0 &=&\frac{1}{2}\,\Omega _{[AB][CD]}^{rm}V_m^{CD}+\Omega
_{1[AB]}^{rm}V_m^1 \,.  \notag
\end{eqnarray}
Because $\Omega _{[\hat{\imath}\hat{\jmath}]1}^{rm}\neq 0$, first two equations lead to $V_r^1 ,V_m^1 =0$, and the third and fourth equations give
\begin{eqnarray}
V_r^{\hat{\imath}\hat{\jmath}} &=&\frac{1}{\tilde{\Xi} \sqrt{\gamma }}\,\epsilon_{\ \ k}^{ij}\tilde{e}_m^k\Omega _{1[CD]}^{rm}V_r^{CD}\,,  \notag \\
\epsilon _{ijk}V^{\hat{\imath}\hat{\jmath},\,k} &=&\frac{2}{\tilde{\Xi}\sqrt{\gamma }}\,\Omega _{1[CD]}^{rm}V_m^{CD}\,,
\end{eqnarray}
where we observe that only one component of $V_m^{\hat{\imath}\hat{\jmath}}$ (the completely antisymmetric one) has been solved, while 8 others remain arbitrary. As a result, we get a decoupled system of equations for the AdS components,
\begin{eqnarray}
0 &=&\Omega _{[ AB][CD]}^{rm}V_r^{CD}\,,  \notag \\
0 &=&\Omega _{[ AB][CD]}^{rm}V_m^{CD}\,.
\end{eqnarray}
The $V_r^{AB}$ components (15 of them) satisfy
\begin{eqnarray}
0 &=&\dfrac{Cr}{\ell }\,\epsilon _{\ jk}^{i}\,V_r^{jk}+\Xi \,V_r^{i5}\,,
\notag \\
0 &=&-Cf\,\epsilon _{\ jk}^{i}\,V_r^{jk}+\Xi \,V_r^{1i}\,,  \notag \\
0 &=&\Xi \,V_r^{0i}\,,  \notag \\
0 &=&2C\epsilon _{\ jk}^{i}\,\left( \dfrac{r}{\ell }\,V_r^{1k}+f\,V_r^{k5}\right) +\Xi \,\delta _j^{i}V_r^{15}\,, \\
0 &=&\dfrac{2Cr}{\ell }\,\epsilon _{\ jk}^{i}\,V_r^{0k}+\Xi \,\delta _j^{i}V_r^{05}\,,  \notag \\
0 &=&2Cf\,\epsilon _{\ jk}^{i}\,V_r^{0k}+\Xi \,\delta _j^{i}V_r^{01}\,,
\notag \\
0 &=&\dfrac{r}{\ell }\,V_r^{01}-f\,V_r^{05}\,,  \notag
\end{eqnarray}
and the solution determines $12$ components only,
\begin{equation}
\begin{array}{ll}
V_r^{0i}=0\,,~~\quad V_r^{05}=0\,,\quad  & V_r^{01}=0\,,~~\quad
V_r^{15}=0\,,\medskip  \\
V_r^{1i}=-\dfrac{\ell f}{r}\,V_r^{i5}\,, & V_r^{ij}=-\dfrac{\Xi \ell }{%
2Cr}\,\epsilon _{\ \ k}^{ij}V_r^{k5}\,,
\end{array}
\end{equation}
leaving three zero modes $V_r^{i5}\in \mathbb{R}$ among the radial components. As regards the $3\times 15$ transversal components $V_m^{AB}$, using the notation $V^{ABi}=V_m^{AB}\tilde{e}_i^m$, they satisfy
equations
\begin{eqnarray}
0 &=&\dfrac{Cr}{\ell }\,\epsilon _{ijk}\,V^{ijk}+\Xi \,V_{\ \ i}^{i5}\,, \notag \\
0 &=&- Cf\,\epsilon _{ijk}\,V^{ijk}+\Xi \,V_{\ \ i}^{1i}\,,  \notag \\
0 &=&\Xi \,V_{\ \ i}^{0i}\,,   \notag \\
0 &=&\dfrac{2Cr}{\ell }\,\epsilon_{ijk}\, V^{1jk}+2Cf\,\epsilon_{ijk}\,V^{j5k}+\Xi \,V_{\ \ i}^{15}\,,
 \\
0 &=&\dfrac{2Cr}{\ell }\,\epsilon _{ijk}\,V^{0jk}+\Xi \,V_{\ \ i}^{05} \,,  
\notag \\
0 &=&2Cf\,\epsilon _{ijk}\,V^{0jk}+\Xi \,V_{\ \ i}^{01}\,,  \notag \\
0 &=&\dfrac{r}{\ell }\,V^{01i}-f\,V^{05i} \,,\notag
\end{eqnarray}
with the solution for $12$ components only,
\begin{equation}
\begin{array}{ll}
V_{\ \ i}^{0i}=0\,, & \epsilon _{ijk}\,V^{ijk}=-\dfrac{\ell \Xi }{rC}\,V_{\ \ i}^{i5}\,,\medskip \\
V_{\ \ i}^{01}=\dfrac{\ell f}{r}\,V_{\ \ i}^{05}\,, & V^{0[ij]}=-\dfrac{\ell \Xi }{4Cr}\,\epsilon ^{ijk}V_{\ \ k}^{05}\,,\medskip  \\
V_{\ \ i}^{1i}=-\dfrac{\ell f}{r}\,V_{\ \ i}^{i5}\,,\quad  & V_{\ \ i}^{15}=-\dfrac{2C}{\Xi }\,\epsilon _{ijk}\,\left( \dfrac{r}{\ell } \,V^{1[jk]}-f\,V^{5[jk]}\right) .
\end{array}
\end{equation}
There are $3+5+3+5+9+8=33$ unsolved components $V_{\ \ i}^{05}$,$\,V^{0(ij)_{\mathrm{T}}}$, $V^{1[ij]}$, $V^{1(ij)_{\mathrm{T}}}$, $V_{\ \ j}^{5i}$, $%
V^{ijk}-V^{[ijk]}\in \mathbb{R}$, where $V^{A(ij)_{\mathrm{T}}}$ denotes a traceless symmetric part of the tensor $V^{Aij}$ in the indices $ij$. This means that, in total, there are $4_{c}+8_{SU(2)}+36_{\mathrm{AdS}}=48$ zero modes, meaning that the rank of the $80\times 80$ symplectic matrix is
\begin{equation}
\mathrm{rank}(\Omega )=80-48=32=2\,\mathrm{rank}(\mathcal{J})\,.
\end{equation}
A number of degrees of freedom in the phase space with the background \eqref{Background} is given by the Dirac's formula  \cite{Banados:1995mq,Banados:1996yj}
\begin{equation}
\mathrm{d.o.f.}\,=\frac{1}{2}\,\mathrm{rank}(\Omega )-\mathrm{rank}(\mathcal{J})=0\,,
\end{equation}
showing that the $\Xi$-charged solution belongs to the topological sector of the CS gravity.



\begin{thebibliography}{99}

\bibitem{Zanelli:2005sa}
J.~Zanelli,
``Lecture notes on Chern-Simons (super-)gravities. Second edition (February 2008),''
[arXiv:hep-th/0502193 [hep-th]].

\bibitem{Brigante:2008gz}
M.~Brigante, H.~Liu, R.~C.~Myers, S.~Shenker and S.~Yaida,
``The Viscosity Bound and Causality Violation,''
Phys. Rev. Lett. \textbf{100}, 191601 (2008)
[arXiv:0802.3318 [hep-th]].

\bibitem{Brigante:2007nu}
M.~Brigante, H.~Liu, R.~C.~Myers, S.~Shenker and S.~Yaida,
``Viscosity Bound Violation in Higher Derivative Gravity,''
Phys. Rev. D \textbf{77}, 126006 (2008)
[arXiv:0712.0805 [hep-th]].

\bibitem{Hofman:2008ar}
D.~M.~Hofman and J.~Maldacena,
``Conformal collider physics: Energy and charge correlations,''
JHEP \textbf{05}, 012 (2008)
[arXiv:0803.1467 [hep-th]].

\bibitem{Hofman:2009ug}
D.~M.~Hofman,
``Higher Derivative Gravity, Causality and Positivity of Energy in a UV complete QFT,''
Nucl. Phys. B \textbf{823}, 174-194 (2009)
[arXiv:0907.1625 [hep-th]].

\bibitem{Troncoso:1998ng}
R.~Troncoso and J.~Zanelli,
``Gauge supergravities for all odd dimensions,''
Int. J. Theor. Phys. \textbf{38} (1999), 1181-1206
doi:10.1023/A:1026614631617
[arXiv:hep-th/9807029 [hep-th]].

\bibitem{Boulware:1985wk}
D.~G.~Boulware and S.~Deser,
``String Generated Gravity Models,''
Phys. Rev. Lett. \textbf{55} (1985), 2656.

\bibitem{Banados:1993ur} M.~Banados, C.~Teitelboim and J.~Zanelli,
``Dimensionally continued black holes,'' 
Phys. Rev. D \textbf{49} (1994),
975-986 [arXiv:gr-qc/9307033 [gr-qc]].

\bibitem{Zegers:2005vx}
R.~Zegers,
``Birkhoff's theorem in Lovelock gravity,''
J. Math. Phys. \textbf{46}, 072502 (2005)
[arXiv:gr-qc/0505016 [gr-qc]].

\bibitem{Deser:2005gr}
S.~Deser and J.~Franklin,
``Birkhoff for Lovelock redux,''
Class. Quant. Grav. \textbf{22}, L103-L106 (2005)
[arXiv:gr-qc/0506014 [gr-qc]].

\bibitem{Dotti:2007az}
G.~Dotti, J.~Oliva and R.~Troncoso,
``Exact solutions for the Einstein-Gauss-Bonnet theory in five dimensions: Black holes, wormholes and spacetime horns,''
Phys. Rev. D \textbf{76}, 064038 (2007)
[arXiv:0706.1830 [hep-th]].

\bibitem{Adams:2008zk}
A.~Adams, A.~Maloney, A.~Sinha and S.~E.~Vazquez,
``1/N Effects in Non-Relativistic Gauge-Gravity Duality,''
JHEP \textbf{03}, 097 (2009)
[arXiv:0812.0166 [hep-th]].

\bibitem{AyonBeato:2009nh}
E.~Ayon-Beato, A.~Garbarz, G.~Giribet and M.~Hassaine,
``Lifshitz Black Hole in Three Dimensions,''
Phys. Rev. D \textbf{80}, 104029 (2009)
[arXiv:0909.1347 [hep-th]].

\bibitem{Dotti:2006cp}
G.~Dotti, J.~Oliva and R.~Troncoso,
``Static wormhole solution for higher-dimensional gravity in vacuum,''
Phys. Rev. D \textbf{75}, 024002 (2007)
[arXiv:hep-th/0607062 [hep-th]].

\bibitem{Chandia:1998uf} O.~Chandia, R.~Troncoso and J.~Zanelli,
``Dynamical content of CS supergravity,'' 
AIP Conf. Proc. \textbf{484}, no. 1, 231 (1999) [hep-th/9903204].

\bibitem{Miskovic:2006ei} O.~Miskovic, R.~Troncoso and J.~Zanelli,
``Dynamics and BPS states of AdS(5) supergravity with a Gauss-Bonnet term,''  Phys. Lett. B \textbf{637}, 317 (2006) [hep-th/0603183].

\bibitem{Canfora:2007xs} F.~Canfora, A.~Giacomini and R.~Troncoso,
``Black holes, parallelizable horizons and half-BPS states for the Einstein-Gauss-Bonnet theory in five dimensions,'' 
Phys. Rev. D \textbf{77}, 024002 (2008) [arXiv:0707.1056 [hep-th]].

\bibitem{Giribet:2014hpa} G.~Giribet, N.~Merino, O.~Miskovic and J.~Zanelli,
``Black hole solutions in CS AdS supergravity,'' 
JHEP \textbf{1408}, 083 (2014) [arXiv:1406.3096 [hep-th]].

\bibitem{Aros:2002rk} R.~Aros, C.~Martinez, R.~Troncoso and J.~Zanelli,
``Supersymmetry of gravitational ground states,''  
JHEP \textbf{05} (2002), 020 [arXiv:hep-th/0204029 [hep-th]].

\bibitem{Brihaye:2013vsa}
Y.~Brihaye and E.~Radu,
``Black hole solutions in d=5 Chern-Simons gravity,''
JHEP \textbf{11} (2013), 049
doi:10.1007/JHEP11(2013)049
[arXiv:1305.3531 [gr-qc]].

\bibitem{Chamseddine:1976bf} A.~H.~Chamseddine and P.~C.~West,
``Supergravity as a Gauge Theory of Supersymmetry,'' Nucl. Phys. B \textbf{129}, 39 (1977).

\bibitem{Andrianopoli:2014aqa}
L.~Andrianopoli and R.~D'Auria,
``N=1 and N=2 pure supergravities on a manifold with boundary,''
JHEP \textbf{08} (2014), 012
doi:10.1007/JHEP08(2014)012
[arXiv:1405.2010 [hep-th]].

\bibitem{Aros:2006qc}
R.~Aros and M.~Contreras,
``Torsion induces gravity,''
Phys. Rev. D \textbf{73} (2006), 087501
doi:10.1103/PhysRevD.73.087501
[arXiv:gr-qc/0601135 [gr-qc]].

\bibitem{Wang} M.~Wang, ``Parallel spinors and parallel forms,'' 
Ann. Global Anal. Geom. \textbf{7} (1989), 59--68.

\bibitem{Bar:1993gpi} C.~Bar, 
``Real Killing Spinors and Holonomy,'' 
Commun. Math. Phys. \textbf{154} (1993) no.3, 509-521 
doi:10.1007/BF02102106.

\bibitem{Baum} H.~Baum, 
``Complete Riemannian manifolds with imaginary Killing spinors,'' 
Ann. Global Anal. Geom. \textbf{7} (1989), 205--215.

\bibitem{Alonso-Alberca:2002wsh}
N.~Alonso-Alberca, E.~Lozano-Tellechea and T.~Ortin,
``Geometric construction of Killing spinors and supersymmetry algebras in homogeneous space-times,''
Class. Quant. Grav. \textbf{19} (2002), 6009-6024
doi:10.1088/0264-9381/19/23/309
[arXiv:hep-th/0208158 [hep-th]].

\bibitem{Lu:1998nu} H.~Lu, C.~N.~Pope and J.~Rahmfeld, 
``A Construction of Killing spinors on S**n,'' 
J. Math. Phys. \textbf{40} (1999), 4518-4515
doi:10.1063/1.532983 [arXiv:hep-th/9805151 [hep-th]].

\bibitem{small1}
H.~Ooguri, A.~Strominger and C.~Vafa,
``Black hole attractors and the topological string,''
Phys. Rev. D \textbf{70} (2004), 106007
doi:10.1103/PhysRevD.70.106007
[arXiv:hep-th/0405146 [hep-th]].

\bibitem{small2}
G.~Lopes Cardoso, B.~de Wit, J.~Kappeli and T.~Mohaupt,
``Asymptotic degeneracy of dyonic N = 4 string states and black hole entropy,''
JHEP \textbf{12} (2004), 075
doi:10.1088/1126-6708/2004/12/075
[arXiv:hep-th/0412287 [hep-th]].

\bibitem{small3}
A.~Sen,
``Black hole entropy function and the attractor mechanism in higher derivative gravity,''
JHEP \textbf{09} (2005), 038
doi:10.1088/1126-6708/2005/09/038
[arXiv:hep-th/0506177 [hep-th]].

\bibitem{Oliva:2009ip}
J.~Oliva, D.~Tempo and R.~Troncoso,
``Three-dimensional black holes, gravitational solitons, kinks and wormholes for BHT massive gravity,''
JHEP \textbf{07} (2009), 011
doi:10.1088/1126-6708/2009/07/011
[arXiv:0905.1545 [hep-th]].



\bibitem{Riegert:1984zz}
R.~J.~Riegert,
``Birkhoff's Theorem in Conformal Gravity,''
Phys. Rev. Lett. \textbf{53} (1984), 315-318
doi:10.1103/PhysRevLett.53.315.

\bibitem{Wheeler:1985qd}
J.~T.~Wheeler,
``Symmetric Solutions to the Maximally \{Gauss-Bonnet\} Extended Einstein Equations,''
Nucl. Phys. B \textbf{273} (1986), 732-748.

\bibitem{Gabadadze:2012xv}
G.~Gabadadze, G.~Giribet and A.~Iglesias,
``New Massive Gravity on de Sitter Space and Black Holes at the Special Point,''
doi:10.1142/9789814623995\_0484
[arXiv:1212.6279 [hep-th]].

\bibitem{Nakahara} M. Nakahara, 
\textit{Geometry, topology, and physics } (Taylor \& Francis Group, 2003).

\bibitem{Brody}
E.~J.~Brody, ``The topological classification of the lens spaces'', Annals of Mathematics (1960), Vol. 71.

\bibitem{Slansky:1981yr}
R.~Slansky,
``Group Theory for Unified Model Building,''
Phys. Rept. \textbf{79} (1981), 1-128
doi:10.1016/0370-1573(81)90092-2.

\bibitem{tHooft:1974kcl}
G.~'t Hooft,
``Magnetic Monopoles in Unified Gauge Theories,''
Nucl. Phys. B \textbf{79} (1974), 276-284
doi:10.1016/0550-3213(74)90486-6.

\bibitem{Polyakov:1974ek}
A.~M.~Polyakov,
``Particle Spectrum in the Quantum Field Theory,''
JETP Lett. \textbf{20} (1974), 194-195
PRINT-74-1566 (LANDAU-INST).

\bibitem{Wilczek}
F.~Wilczek, \textit{Quark  confinement  and  field  theory} (ed.~Stump  and  Weingarte; Wiley-Interscience, New York, 1977).

\bibitem{Charap:1977ww}
J.~M.~Charap and M.~J.~Duff,
``Gravitational Effects on Yang-Mills Topology,''
Phys. Lett. B \textbf{69} (1977), 445-447, doi:10.1016/0370-2693(77)90841-3.

\bibitem{Hawking:1977ab}
S.~W.~Hawking and C.~N.~Pope,
``Generalized Spin Structures in Quantum Gravity,''
Phys. Lett. B \textbf{73} (1978), 42-44
doi:10.1016/0370-2693(78)90167-3.


\bibitem{Gibbons:1982fy}
G.~W.~Gibbons and C.~M.~Hull,
``A Bogomolny Bound for General Relativity and Solitons in N=2 Supergravity,''
Phys. Lett. B \textbf{109} (1982), 190-194,
doi:10.1016/0370-2693(82)90751-1.

\bibitem{Kallosh:1992ii}
R.~Kallosh, A.~D.~Linde, T.~Ortin, A.~W.~Peet and A.~Van Proeyen, ``Supersymmetry as a cosmic censor,'' Phys. Rev. D \textbf{46} (1992), 5278-5302 doi:10.1103/PhysRevD.46.5278 [arXiv:hep-th/9205027 [hep-th]].

\bibitem{Dabrowski:1986en}
L.~Dabrowski and A.~Trautman, ``Spinor Structures in Spheres and Projective Spaces,''
J. Math. Phys. \textbf{27} (1986), 2022
doi:10.1063/1.527021.

\bibitem{Howe:1995zm} P.~S.~Howe, J.~M.~Izquierdo, G.~Papadopoulos and P.~K.~Townsend, ``New supergravities with central charges
and Killing spinors in (2+1)-dimensions,'' Nucl. Phys. B \textbf{467} (1996), 183-214 doi:10.1016/0550-3213(96)00091-0
[arXiv:hep-th/9505032 [hep-th]].

\bibitem{Jackiw:1977hi}
R.~Jackiw, C.~Nohl and C.~Rebbi,
``Classical and Semiclassical Solutions of the Yang-Mills Theory,''
MIT-CTP-675; published in D.~H.~Boal and A.~N.~Kamal (eds.), \textit{Particles and fields} 
(Plenum Press, New York and London, 1978).

\bibitem{ToplogicalSolitons}
N.~Manton and P.~Sutcliffe, \textit{Toplogical Solitons}, Cambridge University Press (2010). 

\bibitem{Chavel-Chern} I.~Chavel and S.~S.~Chern, ``Poincar\'{e} Metrics on Real Projective Space reference,'' Indiana Univ. Math. J. \textbf{23} (1973), 95-101. 

\bibitem{Sochifi} L.~Andrianopoli, D.~L\'opez D\'iaz, O.~Miskovic, ``BPS soliton winding around the black hole in Chern-Simons AdS$_5$ supergravity'', Presented at XXII SOCHIFI (November 2020). To appear in Proceedings.

\bibitem{Banados:1994tn}
M.~Banados,
``Global charges in Chern-Simons field theory and the (2+1) black hole,''
Phys. Rev. D \textbf{52} (1996), 5816-5825
[arXiv:hep-th/9405171 [hep-th]].

\bibitem{Olea-PhD} R. Olea, 
\textit{Conserved charges in anti-de Sitter gravity} (Ph.D. thesis, in Spanish), November 2000.

\bibitem{Mora:2004kb}
P.~Mora, R.~Olea, R.~Troncoso and J.~Zanelli,
``Finite action principle for Chern-Simons AdS gravity,''
JHEP \textbf{06} (2004), 036
doi:10.1088/1126-6708/2004/06/036
[arXiv:hep-th/0405267 [hep-th]].

\bibitem{Crisostomo:2000bb}
J.~Crisostomo, R.~Troncoso and J.~Zanelli,
``Black hole scan,''
Phys. Rev. D \textbf{62} (2000), 084013
doi:10.1103/PhysRevD.62.084013
[arXiv:hep-th/0003271 [hep-th]].

\bibitem{Nester:1991yd}
J.~M.~Nester,
``A covariant Hamiltonian for gravity theories,''
Mod. Phys. Lett. A \textbf{6} (1991), 2655-2661.

\bibitem{Chen:2015vya}
C.~M.~Chen, J.~M.~Nester and R.~S.~Tung,
``Gravitational energy for GR and Poincar\'e gauge theories: A covariant Hamiltonian approach,''
Int. J. Mod. Phys. D \textbf{24} (2015) no.11, 1530026
[arXiv:1507.07300 [gr-qc]].

\bibitem{Cvetkovic:2016ios}
B.~Cvetkovi\'c and D.~Simi\'c,
``5D Lovelock gravity: new exact solutions with torsion,''
Phys. Rev. D \textbf{94} (2016) no.8, 084037
doi:10.1103/PhysRevD.94.084037
[arXiv:1608.07976 [gr-qc]].

\bibitem{Banados:1995mq} M.~Banados, L.~J.~Garay and M.~Henneaux,
``The Local degrees of freedom of higher dimensional pure CS
theories,'' 
Phys. Rev. D \textbf{53}, 593 (1996) 
[hep-th/9506187].

\bibitem{Banados:1996yj} M.~Banados, L.~J.~Garay and M.~Henneaux,
``The Dynamical structure of higher dimensional CS theory,''  Nucl.  Phys.  B \textbf{476}, 611 (1996) [hep-th/9605159].

\bibitem{Saavedra:2000wk}
J.~Saavedra, R.~Troncoso and J.~Zanelli,
``Degenerate dynamical systems,''
J. Math. Phys. \textbf{42} (2001), 4383-4390
doi:10.1063/1.1389088
[arXiv:hep-th/0011231 [hep-th]].

\bibitem{Teitelboim:1987zz}
C.~Teitelboim and J.~Zanelli,
``Dimensionally continued topological gravitation theory in Hamiltonian form,''
Class. Quant. Grav. \textbf{4} (1987), L125
doi:10.1088/0264-9381/4/4/010.

\bibitem{Dadhich:2015ivt}
N.~Dadhich, R.~Durka, N.~Merino and O.~Miskovic,
``Dynamical structure of Pure Lovelock gravity,''
Phys. Rev. D \textbf{93} (2016) no.6, 064009
doi:10.1103/PhysRevD.93.064009
[arXiv:1511.02541 [hep-th]].

\bibitem{Miskovic:2003ex}
O.~Miskovic and J.~Zanelli,
``Dynamical structure of irregular constrained systems,''
J. Math. Phys. \textbf{44} (2003), 3876-3887
doi:10.1063/1.1601299
[arXiv:hep-th/0302033 [hep-th]].

\bibitem{Regge:1974zd}
T.~Regge and C.~Teitelboim,
``Role of Surface Integrals in the Hamiltonian Formulation of General Relativity,''
Annals Phys. \textbf{88} (1974), 286
doi:10.1016/0003-4916(74)90404-7.

\bibitem{Miskovic:2007mg}
O.~Miskovic and R.~Olea,
``Counterterms in Dimensionally Continued AdS Gravity,''
JHEP \textbf{10} (2007), 028
doi:10.1088/1126-6708/2007/10/028
[arXiv:0706.4460 [hep-th]].

\bibitem{Coussaert:1993jp}
O.~Coussaert and M.~Henneaux,
``Supersymmetry of the (2+1) black holes,''
Phys. Rev. Lett. \textbf{72} (1994), 183-186
doi:10.1103/PhysRevLett.72.183
[arXiv:hep-th/9310194 [hep-th]].

\bibitem{Araneda:2016iiy}
R.~Araneda, R.~Aros, O.~Miskovic and R.~Olea,
``Magnetic Mass in 4D AdS Gravity,''
Phys. Rev. D \textbf{93} (2016) no.8, 084022
doi:10.1103/PhysRevD.93.084022
[arXiv:1602.07975 [hep-th]].

\bibitem{Araneda:2018orn}
R.~Araneda, R.~Aros, O.~Miskovic and R.~Olea,
``Pontryagin Term and Magnetic Mass in 4D AdS Gravity,''
J. Phys. Conf. Ser. \textbf{1043} (2018) no.1, 012016
doi:10.1088/1742-6596/1043/1/012016.

\bibitem{Horowitz:1991cd}
G.~T.~Horowitz and A.~Strominger,
``Black strings and P-branes,''
Nucl. Phys. B \textbf{360} (1991), 197-209,
doi:10.1016/0550-3213(91)90440-9.

\bibitem{Townsend:1995gp}
P.~K.~Townsend,
``P-brane democracy,''
[arXiv:hep-th/9507048 [hep-th]].

\bibitem{Polchinski:1995mt}
J.~Polchinski,
``Dirichlet Branes and Ramond-Ramond charges,''
Phys. Rev. Lett. \textbf{75} (1995), 4724-4727,
doi:10.1103/PhysRevLett.75.4724
[arXiv:hep-th/9510017 [hep-th]].

\bibitem{Gubser:1996de}
S.~S.~Gubser, I.~R.~Klebanov and A.~W.~Peet,
``Entropy and temperature of black 3-branes,''
Phys. Rev. D \textbf{54} (1996), 3915-3919,
doi:10.1103/PhysRevD.54.3915
[arXiv:hep-th/9602135 [hep-th]].

\bibitem{Strominger:1996sh}
A.~Strominger and C.~Vafa,
``Microscopic origin of the Bekenstein-Hawking entropy,''
Phys. Lett. B \textbf{379} (1996), 99-104,
doi:10.1016/0370-2693(96)00345-0
[arXiv:hep-th/9601029 [hep-th]].

\bibitem{Maldacena:1996ky}
J.~M.~Maldacena,
``Black holes in string theory,''
[arXiv:hep-th/9607235 [hep-th]].

\bibitem{Maldacena:1997re}
J.~M.~Maldacena,
``The Large N limit of superconformal field theories and supergravity,''
Adv. Theor. Math. Phys. \textbf{2} (1998), 231-252,
doi:10.1023/A:1026654312961
[arXiv:hep-th/9711200 [hep-th]].

\bibitem{Troncoso:1997me}
R.~Troncoso and J.~Zanelli,
``Chern-Simons supergravities with off-shell local superalgebras,''
doi:10.1142/9789812793270\_0007
[arXiv:hep-th/9902003 [hep-th]].

\bibitem{Banados:2005rz}
M.~Banados, R.~Olea and S.~Theisen,
``Counterterms and dual holographic anomalies in CS gravity,''
JHEP \textbf{10} (2005), 067
doi:10.1088/1126-6708/2005/10/067
[arXiv:hep-th/0509179 [hep-th]].

\bibitem{Banados:2006fe}
M.~Banados, O.~Miskovic and S.~Theisen,
``Holographic currents in first order gravity and finite Fefferman-Graham expansions,''
JHEP \textbf{06} (2006), 025,
doi:10.1088/1126-6708/2006/06/025
[arXiv:hep-th/0604148 [hep-th]].

\bibitem{Mora:2014fba}
P.~Mora,
``Gauge Symmetries and Holographic Anomalies of Chern-Simons and Transgression AdS Gravity,''
JHEP \textbf{04} (2015), 090,
doi:10.1007/JHEP04(2015)090
[arXiv:1408.1436 [hep-th]].

\bibitem{Cvetkovic:2017fxa}
B.~Cvetkovi\'c, O.~Miskovic and D.~Simi\'c,
``Holography in Lovelock Chern-Simons AdS Gravity,''
Phys. Rev. D \textbf{96} (2017) no.4, 044027,
doi:10.1103/PhysRevD.96.044027
[arXiv:1705.04522 [hep-th]].

\bibitem{Gallegos:2020otk}
A.~D.~Gallegos and U.~G\"ursoy,
``Holographic spin liquids and Lovelock Chern-Simons gravity,''
JHEP \textbf{11} (2020), 151,
doi:10.1007/JHEP11(2020)151
[arXiv:2004.05148 [hep-th]].

\bibitem{Cai:2003kt}
R.~G.~Cai,
``A Note on thermodynamics of black holes in Lovelock gravity,''
Phys. Lett. B \textbf{582} (2004), 237-242,
doi:10.1016/j.physletb.2004.01.015
[arXiv:hep-th/0311240 [hep-th]].

\bibitem{Greiner-etal} W.~Greiner, S.~Schramm and E.~Stein, 
\textit{Quantum Chromodynamics}, (3rd edition, Springer, 2007).

\bibitem{Wilson} F.~Wesley Wilson, Jr. 
``Some examples of vector fields on the 3-sphere,'' 
Ann. Inst. Fourier, Grenoble, \textbf{20}, 2 (1970), 1-20.

\bibitem{Hatcher}
A.~Hatcher, \textit{Algebraic topology}, Cambridge University Press (2001).











\end{thebibliography}
\end{document}